\documentclass[prd,twocolumn,superscriptaddress,nofootinbib,showpacs,10pt]{revtex4-1}

\usepackage{wrapfig}
\usepackage{lipsum}
\usepackage{amsmath,amssymb,graphicx,bm,float}
\usepackage{pmat}
\usepackage{slashbox}
\usepackage{tikz}

\allowdisplaybreaks

\begin{document}

\title{Yukawa sector of Multi Higgs Doublet Models in the presence of Abelian symmetries}

\author{H.~Ser\^{o}dio}
\thanks{E-mail: Hugo.Serodio@ific.uv.es}
\affiliation{Departament de F\'{\i}sica Te\`{o}rica and IFIC,
		Universitat de Val\`{e}ncia-CSIC,
		E-46100, Burjassot, Spain}

\date{\today}

\begin{abstract}
A general method for classifying the possible quark models of a multi-Higgs-doublet model,
in the presence of Abelian symmetries, is presented.
All the possible sets of textures that can be present in a given sector are shown,
thus turning the determination of the flavor models into a combinatorial problem.
Several symmetry implementations are studied for two and three Higgs doublet models.
Some models implementations are explored in great detail,
with a particular emphasis on models known as Branco-Grimus-Lavoura
and nearest-neighbour-nnteraction.
Several considerations on the flavor changing neutral currents
of multi-Higgs models are also given.
\end{abstract}


\maketitle

\section{Introduction}

The Standard Model (SM) of strong and electroweak interactions is very successful phenomenologically and the discovery of a Higgs-like particle~\cite{Aad:2012tfa} was the missing piece in order to establish it as the best model available. However, there is a general consensus that this should not be the final theory because it does not explain basic issues such as dark matter, neutrino masses, number of families, and many others. 

One possible extension of the SM is the addition of extra copies of the Higgs field, just like in the fermionic sector. The most common scenario is the two Higgs doublet model (2HDM), which has been extensively studied in the literature; for a review see~\cite{Branco:2011iw}. Models with three or more Higgs bosons have also been considered, but the lack of information on these extension is much larger. With the addition of extra scalar doublets the number of parameters, in the scalar and Yukawa sector, increases largely. In these $N$ Higgs doublet models (NHDM) it is very common to add symmetries to help tackle the problem. For the 2HDM, Ivanov~\cite{Ivanov:2006yq} has shown that, no matter what combination of flavor symmetries and/or generalized CP symmetries one imposes on the scalar potential, one always ends up with one of six distinct classes of potentials. Later, this issue was studied further by Ferreira, Haber, and Silva~\cite{Ferreira:2009wh}.  The recent studies of Ivanov and Vdovin~\cite{Ivanov:2012ry} have extended these analyses to the three Higgs doublet models (3HDM). The study of Abelian symmetries in the NHDM scalar sector was done by Ivanov, Keus, and Vdovin~\cite{Ivanov:2011ae}. Despite the extensive general studies of symmetries in the scalar potential of NHDM, the Yukawa sector has been left partially apart. There are several particular flavor models in literature with two, three, or more Higgs fields, but there is a lack of a general approach as the one existing for the scalar sector. 

The study of the Yukawa sector in NHDM tends to be a little involved since, besides the scalar fields, we have three types of fermions ($Q_L$, $n_R$, and $p_R$) repeated 3 times. This enlarges significantly the number of choices for the representations of a given group. Recently, a general study of 3HDM in the presence of $A_4$ and $S_4$ was done~\cite{Felipe:2013ie}. These are two interesting non-Abelian groups since they lead to a scalar potential highly symmetric, allowing the complete determination of the global minimums~\cite{Degee:2012sk}. While, the study of non-Abelian symmetries in the Yukawa sector depends strongly of the irreducible representations (irreps) and the way we attribute them, for the Abelian case we only have one-dimensional irreps. Using this feature, Ferreira and Silva~\cite{Ferreira:2010ir} have presented a general study of Abelian symmetries in the Yukawa sector of the 2HDM. The aim of this work is to extend this study to the NHDM case.

This article is organized as follows. In Sec.~\ref{AbelianYukawa} we introduce our notation and show how the action of Abelian symmetries constrains the Yukawa textures. In Sec.~\ref{Abelianchains} we show the possible combinations of textures, i.e. chains, that can be built in Abelian models, as well as the possible Higgs fields transformations and associated textures. In Sec.~\ref{Cupdown} we explain how to make the connection between the up-quark and down-quark sectors, allowing us to build explicit models for the quark sector. In Sec.~\ref{DPAG} we extend our previous analyses to cases where the Abelian group is a direct product of cyclic groups. In Sec.~\ref{QuaksM} explicit model implementations are studied in detail, in particular, the well-known Branco-Grimus-Lavoura (BGL) and nearest-neighbour-interaction (NNI) models. We draw our conclusions in Sec.~\ref{conclusions}.

\section{Abelian symmetries versus Yukawa textures}\label{AbelianYukawa}
The most general and renormalizable scalar potential constructed with $N$ copies of the $SU(2)_L\otimes U(1)_Y$ doublet $\Phi_a$ $(a=1,\cdots,N)$ is
\begin{equation}\label{Pot}
V=Y_{ab}\left(\Phi_a^\dagger\Phi_b\right)+Z_{abcd}
\left(\Phi_a^\dagger\Phi_b\right)\left(\Phi_c^\dagger\Phi_d\right)
\end{equation}
with
\begin{equation}
Y_{ab}=Y_{ba}^\ast\,,\quad Z_{abcd}=Z_{cdab}=Z_{badc}^\ast\,,
\end{equation}
due to Hermiticity of the Lagrangian. The model also contains three flavors of left-handed quarks $(Q_{L\alpha})$, right-handed down-type quarks $(n_{R\alpha})$, and right-handed up-type quarks $(p_{R\alpha})$, with the Greek letters denoting the fermion flavors. The scalars and fermion fields are connected through the Yukawa Lagrangian
\begin{align}
\begin{split}
-\mathcal{L}_{\text{Yuk}}=&\left(\mathbf{\Gamma}_a\right)_{\alpha\beta}\,\overline{Q_{L\alpha}}\,\Phi_a\, n_{R\beta}+\left(\mathbf{\Delta}_a\right)_{\alpha\beta}\,\overline{Q_{L\alpha}}\,\tilde{\Phi}_a\, p_{R\beta}\\
&+\text{H.c.}\,,
\end{split}
\end{align}
with $\tilde{\Phi}_a\equiv i\tau_2 \Phi_a$. The matrices in flavor space are denoted with bold. When the scalar fields acquire a vacuum expectation value, i.e. $\left<\Phi_a\right>=v_a$, the quarks become massive. Their mass matrix takes the form
\begin{equation}
\mathbf{M}_u=v_a^\ast \mathbf{\Delta}_a\quad\text{and}\quad \mathbf{M}_d=v_a\mathbf{\Gamma}_a\,.
\end{equation}
They are diagonalized through a left and right unitary transformation
\begin{equation}\label{URL}
\begin{array}{l}
\mathbf{U}_{L}^{p\dagger}\mathbf{M}_uU_R^p=\text{diag}(m_u,m_c,m_t)\,,\\
\mathbf{U}_{L}^{n\dagger}\mathbf{M}_dU_R^n=\text{diag}(m_d,m_s,m_b)\,.
\end{array}
\end{equation}
The quark mixing matrix is defined as $\mathbf{V}_{CKM}=\mathbf{U}_{L}^{p\dagger}\mathbf{U}_{L}^{n}$. 
The invariance of Eq.~\eqref{Pot} under
\begin{equation}
\Phi_a\rightarrow \left(\mathcal{S}^\prime_H\right)_{ab}\Phi_b\,, 
\end{equation}
defines a symmetry of the scalar potential. The $\mathcal{S}^\prime_H$ is the generator of the symmetry group (there could be more than one). This requirement of invariance will put constraints on the $Y_{ab}$ and $Z_{abcd}$ couplings. If we want this symmetry to leave the full Lagrangian invariant, then the fermionic fields will also have to transform,
\begin{equation}\label{symT}
Q_{L}\rightarrow\mathcal{S}_L\,Q_L\,,\quad n_{R}\rightarrow \mathcal{S}^n_R\, n_R\,,\quad p_{R}\rightarrow \mathcal{S}^p_R \,p_R
\end{equation}
and leave the Yukawa sector invariant. This requirement on the Yukawa sector leads to the constraints 
\begin{equation}\label{symeq}
\left\{
\begin{array}{l}
\mathcal{S}_L^\dagger\,\mathbf{\Gamma}_b\,\mathcal{S}^{n}_R\,\left(\mathcal{S}^\prime_{H}\right)_{ba}=\mathbf{\Gamma}_a\\\\
\mathcal{S}_L^\dagger\,\mathbf{\Delta}_b\,\mathcal{S}^{p}_R\,\left(\mathcal{S}^{\prime\ast}_{H}\right)_{ba}=\mathbf{\Delta}_a
\end{array}\right.\rightarrow\,
\mathcal{S}_L^\dagger\,\mathcal{A}_b\,\mathcal{S}_R\,\left(\mathcal{S}_{H}\right)_{ba}=\mathcal{A}_a\,,
\end{equation}
with $\mathcal{A}_a=\left\{\mathbf{\Gamma}_a,\,\mathbf{\Delta}_a\right\}$, while $\mathcal{S}_{L}$, $\mathcal{S}_{R}=\left\{\mathcal{S}_{R}^n,\, \mathcal{S}_{R}^p\right\}$, and $\mathcal{S}_{H}=\left\{\mathcal{S}^\prime_H,\,\mathcal{S}_H^{\prime\ast}\right\}$ are the symmetry generators for $Q_L$, $n_R(p_R)$, and $\Phi_a(\Phi^\ast_a)$, respectively.

Abelian symmetries are characterized by the commutativity of all their group elements, leading to the existence of only one-dimensional representations. This in turn implies the existence of a basis where the generators present in Eq.~\eqref{symT} are all diagonal, i.e.
\begin{equation}
\begin{array}{l}
\mathcal{S}_{L}=\text{diag}\left(e^{i\alpha_1}\,,e^{i\alpha_2},\,e^{i\alpha_3}\right)\,,\,\\
\mathcal{S}_{R}=\text{diag}\left(e^{i\beta_1}\,,e^{i\beta_2},\,e^{i\beta_3}\right)\,,\\
\mathcal{S}_{H}=\text{diag}\left(e^{i\theta_1}\,,e^{i\theta_2},\,\cdots,\,e^{i\theta_N}\right)\,.
\end{array}
\end{equation}
In this basis it becomes clear that the Higgs field transformations define only trivial textures, i.e. the full matrix or the null matrix. Therefore, these transformations will not play any role in finding nontrivial textures, and the best way to get rid of them is through the Hermitian combinations $\mathcal{H}^a_L=\mathcal{A}_a\mathcal{A}_a^\dagger$ and $\mathcal{H}^a_R=\mathcal{A}_a^\dagger \mathcal{A}_a$. These combinations have another particularity, they split the left- and right-handed space
\begin{equation}\label{LRsym}
\mathcal{S}_L^\dagger\mathcal{H}^a_L\mathcal{S}_L=\mathcal{H}^a_L\,,\quad \mathcal{S}_R^\dagger\mathcal{H}^a_R\mathcal{S}_R=\mathcal{H}^a_R\,.
\end{equation} 
In order to find the possible textures of $\mathcal{A}_{a}$, we shall solve the equations above. We shall do this for the left-handed space, having in mind that the right-handed space solution can be found in an equivalent way. 

The solution of the first relation in Eq.~\eqref{LRsym} falls into one of three cases:
\begin{itemize}
\item[(1)] $\mathcal{S}_L$ has a full degeneracy,
\item[(2)] $\mathcal{S}_L$ has two-fold degeneracy,
\item[(3)] $\mathcal{S}_L$ is nondegenerate
\end{itemize}

\subsection{Case (1): $\mathcal{S}_L$ has a full degeneracy}

In this case,
the left-handed Hermitian combination has to be of the form
\begin{equation}
\mathcal{H}^a_L
=
\begin{pmatrix}
\bm{\times}&\bm{\times}&\bm{\times}\\
\bm{\times}&\bm{\times}&\bm{\times}\\
\bm{\times}&\bm{\times}&\bm{\times}
\end{pmatrix}\,.
\end{equation}
The $\times$ represents an entry that in general is nonzero. This means that it could be zero in a particular model implementation, but the symmetry itself does not impose it. In this case, looking to the combination matrix $\mathcal{H}_L$ is not of great advantage. However, having in mind how $\mathcal{A}_a$ transforms under the symmetry, see Eq.~\eqref{symeq}, we get for this case 
\begin{equation}
e^{i\gamma}\mathcal{A}_a\mathcal{S}_R=\mathcal{A}_a\,,
\end{equation}
with $\gamma=\theta_a-\alpha$ and $\alpha=\alpha_1=\alpha_2=\alpha_3$. This, in turn, implies the following textures for $\mathcal{A}_a$
\begin{equation}
\begin{pmatrix}
\bm{\times}&\bm{\times}&\bm{\times}\\
\bm{\times}&\bm{\times}&\bm{\times}\\
\bm{\times}&\bm{\times}&\bm{\times}
\end{pmatrix}\,,\,
\begin{pmatrix}
\bm{\times}&\bm{\times}&\text{ }\\
\bm{\times}&\bm{\times}&\\
\bm{\times}&\bm{\times}&
\end{pmatrix}\mathcal{P}\quad\text{and}\quad
\begin{pmatrix}
\bm{\times}&\text{ }&\text{ }\\
\bm{\times}&&\\
\bm{\times}&&
\end{pmatrix}\mathcal{P}\,,
\end{equation}
for $\mathcal{S}_R$ full degenerate, two-fold degenerate and nondegenerate, respectively. The empty entries represent null elements. The matrix $\mathcal{P}$ represents a permutation matrix. There is no permutation on the left since it does not change the textures. The set of $3\times 3$ permutation matrices is
\begin{equation}
\begin{array}{rl}
I=&\begin{pmatrix}
1&&\\
&1&\\
&&1
\end{pmatrix}\,,\quad
\mathcal{P}_{12}=\begin{pmatrix}
&1&\\
1&&\\
&&1
\end{pmatrix}\,,\\\\
\mathcal{P}_{13}=&\begin{pmatrix}
&&1\\
&1&\\
1&&
\end{pmatrix}\,,\quad
\mathcal{P}_{23}=\begin{pmatrix}
1&&\\
&&1\\
&1&
\end{pmatrix}\,,\\\\
\mathcal{P}_{123}=&
\begin{pmatrix}
&&1\\
1&&\\
&1&
\end{pmatrix}\,,\quad
\mathcal{P}_{321}=
\begin{pmatrix}
&1&\\
&&1\\
1&&
\end{pmatrix}\,.
\end{array}
\end{equation}

\subsection{Case (2): $\mathcal{S}_L$ has two-fold degeneracy}

Here,
the left-handed Hermitian combination has to be of the form
\begin{equation}\label{HLtf}
\mathcal{H}_L^a=\mathcal{P}^\prime\left\{
\begin{pmatrix}
\bm{\times}&\bm{\times}&\\
\bm{\times}&\bm{\times}&\\
&&\bm{\times}
\end{pmatrix}\,,\,
\begin{pmatrix}
\bm{\times}&\bm{\times}&\text{ }\\
\bm{\times}&\bm{\times}&\\
&&
\end{pmatrix}\,\,
\begin{pmatrix}
\text{ }&&\\
\text{ }&&\\
&&\bm{\times}
\end{pmatrix}\right\}\mathcal{P}^{\prime T}
\end{equation}
As a default, the two-fold degeneracy was chosen to be in the $(1,2)$ sector. The role of the permutation matrices is to allow this degeneracy to be in one of the other two sectors, i.e. $(1,3)$ and $(2,3)$.
 
This two-fold degeneracy imposes a two-zero texture in $\mathcal{H}_L^a$. However, it does not forbid the nonzero blocks to be zero. If $\mathcal{H}_L^a$ were a completely general Hermitian matrix with no correlations among entries, the first texture would be the only one present. However, since $\mathcal{H}_L^a$ is a combination of $\mathcal{A}_a$, there can be correlations among entries, due to the texture of $\mathcal{A}_a$. Therefore, the second and third textures in Eq.~\eqref{HLtf} become possible. 

In order to find the textures for $\mathcal{A}_a$ we shall work within the two-fold degeneracy in the $(1,2)$ sector, since the others are obtained through some permutation of rows and columns. Since the entries on the $\mathcal{A}_a$ are unrelated, the only way to have zero entries in the Hermitian combination is to have zero entries in $\mathcal{A}_a$. This fact can be easily seen if one writes $\left(\mathcal{A}_a\right)_{ij}=e^{i\gamma_{ij}}a_{ij}$, with $\gamma_{ij}$ and $a_{ij}$ arbitrary and unrelated. The left-handed Hermitian combination is given by $\left(\mathcal{H}_L^a\right)_{ij}=e^{i\left(\gamma_{ik}-\gamma_{jk}\right)}a_{ik}a_{jk}$. Therefore, a zero in the $(i,j)$ position implies
\begin{equation}
e^{i\left(\gamma_{ik}-\gamma_{jk}\right)}a_{ik}a_{jk}=0\,.
\end{equation}
Since $\gamma_{ik}$ and $\gamma_{jk}$ are unrelated for $i\neq j$, the only way to have the sum equal to zero is to have every element of the sum equal to zero. Otherwise, the above condition would imply a relation between $a_{ij}$ elements that is not imposed by the Abelian symmetry. Therefore, phases of elements in $\mathcal{A}_a$ are irrelevant for defining a texture.  

We start by noticing that the Hermitian combination $\mathcal{H}_{L}^a$ is invariant under the transformation $\mathcal{A}_a\rightarrow \mathcal{A}_a \mathbf{U}$, with $\mathbf{U}$ a general unitary matrix. However, as seen above, phases of $\mathcal{A}_a$ do not play any role in defining textures. Therefore, this freedom on the right can be seen as a real orthogonal transformation. Still, since this orthogonal freedom makes part of $\mathcal{A}_a$ its orthogonality can not depend on relations between entries. The only orthogonal matrices where this is fulfilled are the permutation matrices. Therefore, respecting only textures, the unitary freedom on the right is nothing more than the possibility of permuting columns.

The relevant system of equations for the first texture in Eq.~\eqref{HLtf} is given by
\begin{equation}
\mathcal{H}_L^a=
\begin{pmatrix}
\bm{\times}&\bm{\times}&\\
\bm{\times}&\bm{\times}&\\
&&\bm{\times}
\end{pmatrix}\rightarrow
\left\{
\begin{array}{l}
\left(a_{11}+a_{21}\right)a_{31}\\
+(a_{12}+a_{22})a_{32}\\
+(a_{13}+a_{23})a_{33}=0\\
a_{31}^2+a_{32}^2+a_{33}^2\neq 0\\
a_{11}^2+a_{12}^2+a_{13}^2\neq 0\\
a_{21}^2+a_{22}^2+a_{23}^2\neq 0\\
\end{array}\right.
\end{equation}
We now determine the solutions of this system:
\begin{itemize}
\item[(i)] Last line of $\mathcal{A}_a$ with two zeros. There are three possible implementations of this, which just correspond to the freedom to multiply on the right (permutation of columns). We then choose $a_{31}=a_{32}=0$, which in turn implies that $a_{13}=a_{23}=0$ leading to
\begin{equation}
\mathcal{A}_a=
\begin{pmatrix}
\bm{\times}&\bm{\times}&\\
\bm{\times}&\bm{\times}&\\
&&\bm{\times}
\end{pmatrix}\mathcal{P}\, .
\end{equation}  
We still have the freedom to put $a_{ij}$ elements to zero and still get the same $\mathcal{H}_L^a$. However, one should note that additional zeros are determined by the generator $\mathcal{S}_R$ and therefore entire columns are set to zero. Thus $a_{12}=0$ is not allowed but $a_{12}=a_{22}=0$ is, leading to
\begin{equation}
\mathcal{A}_a=
\begin{pmatrix}
\bm{\times}&\text{ }&\\
\bm{\times}&&\\
&&\bm{\times}
\end{pmatrix}\mathcal{P}\, .
\end{equation} 
Additional matrices can be found by column suppression, but they will always lead to the zero block diagonal case, which will be fully studied below.

\item[(ii)] Last line of $\mathcal{A}_a$ with one zero. Again, there are three possible implementations. We choose $a_{33}=0$ and $a_{11}=a_{21}=a_{12}=a_{22}=0$, leading to
\begin{equation}
\mathcal{A}_a=
\begin{pmatrix}
&&\bm{\times}\\
&&\bm{\times}\\
\bm{\times}&\bm{\times}&
\end{pmatrix}\mathcal{P}\, .
\end{equation}
By setting the first column to zero, 
we get
\begin{equation}
\mathcal{A}_a=
\begin{pmatrix}
&&\bm{\times}\\
&&\bm{\times}\\
\text{ }&\bm{\times}&
\end{pmatrix}\mathcal{P}\, ,
\end{equation}
falling into the previous case.
\end{itemize}
Now we study the system when it has a zero block diagonal form
\begin{equation}
\mathcal{H}_L^a=
\begin{pmatrix}
\bm{\times}&\bm{\times}&\\
\bm{\times}&\bm{\times}&\\
&&\text{}
\end{pmatrix}\rightarrow
\left\{
\begin{array}{l}
\left(a_{11}+a_{21}\right)a_{31}\\
+(a_{12}+a_{22})a_{32}\\
+(a_{13}+a_{23})a_{33}=0\\
a_{31}^2+a_{32}^2+a_{33}^2= 0\\
a_{11}^2+a_{12}^2+a_{13}^2\neq 0\\
a_{21}^2+a_{22}^2+a_{23}^2\neq 0\\
\end{array}\right.
\end{equation}
This implies that the last line equals zero, leading to
\begin{equation}
\mathcal{A}_a=\left\{
\begin{pmatrix}
\bm{\times}&\bm{\times}&\bm{\times}\\
\bm{\times}&\bm{\times}&\bm{\times}\\
&&
\end{pmatrix}\,,\,
\begin{pmatrix}
\bm{\times}&\bm{\times}&\text{ }\\
\bm{\times}&\bm{\times}&\\
&&
\end{pmatrix}\,,\,
\begin{pmatrix}
\bm{\times}&\text{ }&\text{ }\\
\bm{\times}&&\\
&&
\end{pmatrix}\right\}\mathcal{P}\, .
\end{equation}
Finally, the last texture in Eq.~\eqref{HLtf} gives the system
\begin{equation}
\mathcal{H}_L^a=
\begin{pmatrix}
\text{ }&\text{ }&\\
&&\\
&&\bm{\times}
\end{pmatrix}\rightarrow
\left\{
\begin{array}{l}
\left(a_{11}+a_{21}\right)a_{31}\\
+(a_{12}+a_{22})a_{32}\\
+(a_{13}+a_{23})a_{33}=0\\
a_{31}^2+a_{32}^2+a_{33}^2\neq 0\\
a_{11}^2+a_{12}^2+a_{13}^2= 0\\
a_{21}^2+a_{22}^2+a_{23}^2= 0\\
\end{array}\right.
\end{equation} 
This forces the first two lines to be zero:
\begin{equation}
\mathcal{A}_a=\left\{
\begin{pmatrix}
&&\\
&&\\
\bm{\times}&\bm{\times}&\bm{\times}
\end{pmatrix}\,,\,
\begin{pmatrix}
&&\\
&&\\
\bm{\times}&\bm{\times}&\text{ }
\end{pmatrix}\,,\,
\begin{pmatrix}
&&\\
&&\\
\text{ }&\text{ }&\bm{\times}
\end{pmatrix}\right\}\mathcal{P}\, .
\end{equation}

\subsection{Case (3): $\mathcal{S}_L$ nondegenerate }

In this case,
the left-handed Hermitian combination has to be of the form
\begin{equation}
\mathcal{H}_L^a=\mathcal{P}^\prime\left\{
\begin{pmatrix}
\bm{\times}&&\\
&\bm{\times}&\\
&&\bm{\times}
\end{pmatrix}\,,\,
\begin{pmatrix}
\text{ }&&\\
&\bm{\times}&\\
&&\bm{\times}
\end{pmatrix}\,,\,
\begin{pmatrix}
\text{ }&\text{ }&\\
&&\\
&&\bm{\times}
\end{pmatrix}
\right\}\mathcal{P}^{\prime T}\, .
\end{equation}
The first texture implies $\mathcal{A}_a\mathcal{A}_a^T=\text{diag}$, since phases play no role. The only matrices that satisfy this relation are monomial matrices, i.e. matrices with the textures of a permutation matrix. To see this, we start with the equation
\begin{equation}
\mathcal{A}_a\mathcal{A}_a^T=\mathbf{d}
\end{equation}
with $\mathbf{d}$ a nonsingular diagonal matrix. We may rewrite the above equation as
\begin{equation}
\mathbf{d}^{-1/2}\mathcal{A}_a\,\mathcal{A}_a^T\mathbf{d}^{-1/2}=\mathbb{I}\, ,
\end{equation}
with $\mathbf{d}^{-1/2}\mathcal{A}_a$ an orthogonal matrix, with no relations between entries, i.e. permutation matrices. Thus $\mathcal{A}_a=\mathbf{d}^{1/2}\mathcal{P}$ is a monomial matrix. 

The second case, with one zero entry in the diagonal $[\text{with no loss of generality } \left(\mathcal{H}_L^a\right)_{11}=0]$, gives the system 
\begin{equation}
\mathcal{H}_L^a=
\begin{pmatrix}
\text{ }&&\\
&\bm{\times}&\\
&&\bm{\times}
\end{pmatrix}\rightarrow
\left\{
\begin{array}{l}
\left(a_{11}+a_{21}\right)a_{31}\\
+(a_{12}+a_{22})a_{32}\\
+(a_{13}+a_{23})a_{33}=0\\
a_{11}a_{21}+a_{12}a_{22}+a_{13}a_{23}=0\\
a_{31}^2+a_{32}^2+a_{33}^2\neq 0\\
a_{11}^2+a_{12}^2+a_{13}^2= 0\\
a_{21}^2+a_{22}^2+a_{23}^2\neq 0\\
\end{array}\right.
\end{equation}
This automatically imposes, to the textures found in case $(2)$ with no zero block diagonal and the monomial matrices, the first line null 
\begin{equation}
\mathcal{A}_a=\left\{
\begin{pmatrix}
&&\\
\bm{\times}&\bm{\times}&\\
&&\bm{\times}
\end{pmatrix}\,,\,
\begin{pmatrix}
&&\\
\text{ }&\bm{\times}&\\
&&\bm{\times}
\end{pmatrix}\right\}\mathcal{P}\,.
\end{equation}
Finally, for the case with two zeros in the diagonal, we get the same matrices as the ones found in the last texture of case $(2)$.

\subsection{Textures and classes}
The case where $\mathcal{H}_L^a$ is zero always leads to $\mathcal{A}_a$ zero and can always be implemented in any of the three cases presented above.

The same analysis could have been done with the Hermitian combination $\mathcal{H}_R^a$, and the transpose textures would be found. However, all matrices have their transpose in this set of allowed textures. We then summarize the set of all possible textures for $\mathcal{A}_a$ allowed by Abelian symmetries:
\begin{widetext}
\begin{align}\label{Textures}
\begin{split}
\mathcal{P}^\prime&\left\{
A_1=
\begin{pmatrix}
\bm{\times}&\bm{\times}&\bm{\times}\\
\bm{\times}&\bm{\times}&\bm{\times}\\
\bm{\times}&\bm{\times}&\bm{\times}
\end{pmatrix}\,,\,
A_2=\begin{pmatrix}
\bm{\times}&\bm{\times}&\text{ }\\
\bm{\times}&\bm{\times}&\text{ }\\
\text{ }&\text{ }&\bm{\times}
\end{pmatrix}\,,\,
A_3=\begin{pmatrix}
\text{ }&\text{ }&\bm{\times}\\
\text{ }&\text{ }&\bm{\times}\\
\bm{\times}&\bm{\times}&\text{ }
\end{pmatrix}\,,\,A_4=\begin{pmatrix}
\text{ }&\text{ }&\bm{\times}\\
\text{ }&\text{ }&\bm{\times}\\
\text{ }&\bm{\times}&\text{ }
\end{pmatrix}\,,\,A_5=
\begin{pmatrix}
\bm{\times}&\bm{\times}&\bm{\times}\\
\bm{\times}&\bm{\times}&\bm{\times}\\
\text{ }&\text{ }&\text{ }
\end{pmatrix}\,,
\right.\\
&\, A_6=
\begin{pmatrix}
\bm{\times}&\bm{\times}&\text{ }\\
\bm{\times}&\bm{\times}&\text{ }\\
\bm{\times}&\bm{\times}&\text{ }
\end{pmatrix}\,,\,
A_7=
\begin{pmatrix}
\bm{\times}&\bm{\times}&\text{ }\\
\bm{\times}&\bm{\times}&\text{ }\\
\text{ }&\text{ }&\text{ }
\end{pmatrix}\,,
A_8=
\begin{pmatrix}
\text{ }&\text{ }&\bm{\times}\\
\text{ }&\text{ }&\bm{\times}\\
\text{ }&\text{ }&\text{ }
\end{pmatrix}\,,\,
A_9=
\begin{pmatrix}
\text{ }&\text{ }&\text{ }\\
\text{ }&\text{ }&\text{ }\\
\bm{\times}&\bm{\times}&\bm{\times}
\end{pmatrix}\,,\,A_{10}=
\begin{pmatrix}
\text{ }&\text{ }&\bm{\times}\\
\text{ }&\text{ }&\bm{\times}\\
\text{ }&\text{ }&\bm{\times}
\end{pmatrix}\,,\\
&\left.\,
A_{11}=
\begin{pmatrix}
\text{ }&\text{ }&\text{ }\\
\text{ }&\text{ }&\text{ }\\
\bm{\times}&\bm{\times}&\text{ }
\end{pmatrix}\,,
A_{12}=
\begin{pmatrix}
\text{ }&\text{ }&\text{ }\\
\text{ }&\text{ }&\text{ }\\
\text{ }&\text{ }&\bm{\times}
\end{pmatrix}\,,\,
A_{13}=
\begin{pmatrix}
\bm{\times}&\text{ }&\text{ }\\
\text{ }&\bm{\times}&\text{ }\\
\text{ }&\text{ }&\bm{\times}
\end{pmatrix}\,,\,
A_{14}=
\begin{pmatrix}
\text{ }&\text{ }&\text{ }\\
\bm{\times}&\bm{\times}&\text{ }\\
\text{ }&\text{ }&\bm{\times}
\end{pmatrix}\,,\,
A_{15}=
\begin{pmatrix}
\text{ }&\text{ }&\text{ }\\
\text{ }&\bm{\times}&\text{ }\\
\text{ }&\text{ }&\bm{\times}
\end{pmatrix}
\right\}\mathcal{P}
\end{split}
\end{align}
\end{widetext}
and the null matrix is denoted by $A_0$. In Table~\ref{classesij}, we present the nine distinct classes that are possible. The null matrix, i.e. $A_0$, can be implemented in any of these classes, and therefore is not presented in the table. In order to simplify the notation,
we shall denote the nine classes as $\mathbf{(i,j)}$, with $\mathbf{i}$ and $\mathbf{j}$ corresponding to the number of different phases for the left and right generators, respectively. The left-handed transformations are connected with the quark mixing and are shared by both sectors. This implies that the three classes $\mathbf{(2,i)}$ are in fact nine, three for each $\mathcal{P}^L=\left\{1\,,\mathcal{P}_{13},\,\mathcal{P}_{23}\right\}$. Since each sector has to share the same left permutation matrix, we shall choose $\mathcal{P}_L=I$ without loss of generality. The total number of models for such classes will be 3 times the cases studied, with the appropriate left permutations. 

\begin{widetext}
\begin{center}
\begin{table}[h]
\begin{tabular}{|@{\hspace{0.2cm}} c @{\hspace{0.2cm}}||@{\hspace{0.2cm}} c @{\hspace{0.2cm}}|@{\hspace{0.2cm}} c @{\hspace{0.2cm}}|@{\hspace{0.2cm}} c @{\hspace{0.2cm}}|}
\hline
\backslashbox{$\mathcal{H}_L^a$}{$\mathcal{H}_R^a$}
&$\begin{pmatrix}
\bm{\times}&\bm{\times}&\bm{\times}\\
\bm{\times}&\bm{\times}&\bm{\times}\\
\bm{\times}&\bm{\times}&\bm{\times}
\end{pmatrix}$ &$\mathcal{P}^R\begin{pmatrix}
\bullet&\bullet&\text{}\\
\bullet&\bullet&\text{}\\
\text{}&\text{}&\bullet
\end{pmatrix}\mathcal{P}^{R}$&$\begin{pmatrix}
\bullet&\text{}&\text{}\\
\text{}&\bullet&\text{}\\
\text{}&\text{}&\bullet
\end{pmatrix}$\\
\hline
\hline
$\begin{pmatrix}
\bm{\times}&\bm{\times}&\bm{\times}\\
\bm{\times}&\bm{\times}&\bm{\times}\\
\bm{\times}&\bm{\times}&\bm{\times}
\end{pmatrix}$&$A_1$&$\left\{A_6\,,\,A_{10}\right\}\mathcal{P}^R$&$\left\{A_{10}\right\}\mathcal{P}$\\
$\mathcal{P}^{L}\begin{pmatrix}
\bullet&\bullet&\text{}\\
\bullet&\bullet&\text{}\\
\text{}&\text{}&\bullet
\end{pmatrix}\mathcal{P}^{L}$&$\mathcal{P}^L\left\{A_5\,,\,A_{9}\right\}$&$\mathcal{P}^L\left\{A_2\,,\,A_3\,,\,A_7\,,\,A_8\,,A_{11}\,,\,A_{12}\right\}\mathcal{P}^{R}$&$\mathcal{P}^L\left\{A_{4}\,,\,A_{8}\,,\,A_{12}\right\}\mathcal{P}$\\
$\begin{pmatrix}
\bullet&\text{}&\text{}\\
\text{}&\bullet&\text{}\\
\text{}&\text{}&\bullet
\end{pmatrix}$&$\mathcal{P}^\prime\left\{A_{9}\right\}$&$\mathcal{P}^\prime\left\{A_{11}\,,\,A_{12}\,,A_{14}\right\}\mathcal{P}^R$&$\mathcal{P}^\prime\left\{A_{12}\,,A_{13}\,,\,A_{15}\right\}\mathcal{P}$\\
\hline
\end{tabular}
\caption{\label{classesij}Different classes of textures, with $\bullet=\bm{\times}\text{ or }0$.}
\end{table}
\end{center}
\end{widetext}

\section{Abelian symmetries, chains, and charge vector}\label{Abelianchains}
In this section, the textures found previously will be grouped into sets that can be simultaneously implemented by a symmetry. In order to exemplify the problem we face, one example is in order. From Table~\ref{classesij}, we see that the class $\mathbf{(2,2)}$ allows for $A_3$ and $A_8$ textures. However, these textures overlap partially, and no symmetry can be found that allows this. 

Our aim is to find all possible texture combinations in each class. For that we introduce two new concepts: 
\begin{itemize}
\item \textbf{Disjoint textures}: two matrices have disjoint textures if and only if they do not share any nonzero entry. For example
\begin{equation}
\begin{pmatrix}
\bm{\times}&&\bm{\times}\\
&&\\
\bm{\times}&&\bm{\times}
\end{pmatrix}\,:\,
\left\{
\begin{pmatrix}
&\bm{\times}&\\
\bm{\times}&&\bm{\times}\\
&\bm{\times}&
\end{pmatrix},\,
\begin{pmatrix}
&&\\
\text{ }&\bm{\times}&\text{ }\\
&&
\end{pmatrix},\cdots
\right\}\,.
\end{equation}

\item \textbf{Chain}: the set of matrices with disjoint textures belonging to the same class, which together build a full matrix. For example
\begin{equation}
\begin{pmatrix}
\bm{\times}&&\bm{\times}\\
&&\\
\bm{\times}&&\bm{\times}
\end{pmatrix}\,,\,
\begin{pmatrix}
&\bm{\times}&\\
\bm{\times}&&\bm{\times}\\
&\bm{\times}&
\end{pmatrix}\,,\,
\begin{pmatrix}
&&\\
\text{ }&\bm{\times}&\text{ }\\
&&
\end{pmatrix}\,.
\end{equation}
The chains will be denoted as $C_n^{\mathbf{(i,j)}}$, which means the $n$th chain of the class $\mathbf{(i,j)}$. The null matrix can be present in a chain by construction or added \textit{a posteriori}, in the last case the chain is denoted as $^0C_n^{\mathbf{(i,j)}}$.
\end{itemize}
In order to find the possible chains and the Abelian groups that may implement them, we introduce the phase transformation matrix
\begin{equation}\label{PTM}
\begin{array}{rl}
\Theta_{\mathcal{A}_a}=&
\begin{pmatrix}
\beta_1-\alpha_1&\beta_2-\alpha_1&\beta_3-\alpha_1\\
\beta_1-\alpha_2&\beta_2-\alpha_2&\beta_3-\alpha_2\\
\beta_1-\alpha_3&\beta_2-\alpha_3&\beta_3-\alpha_3
\end{pmatrix}\\
=&\dfrac{2\pi}{n}
\begin{pmatrix}
k_1&k_2&k_3\\
k_1-k_{L1}&k_2-k_{L1}&k_3-k_{L1}\\
k_1-k_{L2}&k_2-k_{L2}&k_3-k_{L2}
\end{pmatrix}\,.
\end{array}
\end{equation}
This matrix represents the phases of each entry of $\mathcal{A}_a$ when acted by the left and right symmetry generators. The first line of Eq.~\eqref{PTM} is expressed in term of the continuous phases $\alpha_i$ and $\beta_i$, while in the second line we have discretized it. For simplicity we shall work with the last line. The group could take two forms: $Z_{kn}$; $Z_{n\geq k}$. The first case tells us that the order of the group has to belong to $k\mathbb{Z}$ and therefore the group is discrete. The second case just says that the order of the group has to be equal or larger than $k$. Therefore, the group could be a $Z_{k}$, $Z_{k+1}$ or even a $U(1)$. In this discrete notation, the left and right generators are given by 
\begin{equation}\label{discreteS}
\begin{array}{l}
\mathcal{S}_L=\text{diag}\left(1,\omega_n^{k_{L1}},\omega_n^{k_{L2}}\right)\,,\\
\mathcal{S}_R=\text{diag}\left(\omega_n^{k_1},\omega_n^{k_2},\omega_n^{k_3}\right)\,.
\end{array}
\end{equation}
Without loss of generality, we have chosen the first entry of $\mathcal{S}_L$ to have no phase. Since the class $\mathbf{(3,i)}$ contains textures and all their left permutations, we need to redefine the phases $k_{L1}$ and $k_{L2}$ when left permutations are applied in Eq.~\eqref{discreteS}. The redefinitions are: 
\begin{itemize}
\item[(i)] $\mathcal{P}_{12}\mathcal{S}_L\mathcal{P}_{12}^T$: $k_{L1}\rightarrow-k_{L1}, \, k_{L2}\rightarrow k_{L2}-k_{L1}$
\item[(ii)] $\mathcal{P}_{13}\mathcal{S}_L\mathcal{P}_{13}^T$: $k_{L2}\rightarrow-k_{L2}, \, k_{L1}\rightarrow k_{L1}-k_{L2}$
\item[(iii)] $\mathcal{P}_{23}\mathcal{S}_L\mathcal{P}_{23}^T$: $k_{L1}\rightarrow k_{L2}, \, k_{L2}\rightarrow k_{L1}$
\item[(iv)] $\mathcal{P}_{123}\mathcal{S}_L\mathcal{P}_{123}^T$: $k_{L1}\rightarrow k_{L2}-k_{L1}, \, k_{L2}\rightarrow -k_{L1}$
\item[(v)] $\mathcal{P}_{321}\mathcal{S}_L\mathcal{P}_{321}^T$: $k_{L1}\rightarrow -k_{L2}, \, k_{L2}\rightarrow k_{L1}-k_{L2}$
\end{itemize}
Up to this point the number of Higgs fields and their symmetry transformations have not been used. The reason has to do with the fact that they are just a global phase transformation for each Yukawa coupling. Therefore, they have no impact on the determination of the possible textures for each individual matrix. The role of the Higgs fields will be to select the textures from a given chain. Different charges for the scalar fields will lead to disjoint textures of the same chain. We then define \textit{charge vector} as the set of phases associated with the disjoint textures of a given chain. These will be the charges under which the Higgs fields will transform.

The size of a chain is equal to the order of the smallest Abelian group needed to build the chain. This is true since if a texture has $m$ distinct $k's$, we may subtract to the equivalent phase matrix (equivalent to multiplying the texture by an exponential) the phase $\frac{2\pi}{n}k_1$. This will transform the texture to one of its disjoint matrices. We can repeat this process for the $m$ different $k$'s. This will lead to a set of $m+1$ disjoint matrices belonging to the same class. This defines a chain. The order of the smallest Abelian group that forms this chain has to be $m+1$, i.e. the number of distinct $k's$ plus the identity.

Next we present  an example of how to construct the possible chains for a given class, vector charge, as well as the Abelian groups that can be used to implement them.

\subsection{Building chains and associated vector charges}
We shall now present the general method for finding the vector charges. We use as an example the class $\mathbf{(2,2)}$. 
This class can be divided into two cases: with $\mathcal{P}^LA_2\mathcal{P}^R$; without $\mathcal{P}^LA_2\mathcal{P}^R$.

In the first case, i.e. with a texture $\mathcal{P}^LA_2\mathcal{P}^R$, the symmetry implementation is given by
\begin{equation}
\left\{
\begin{array}{l}
\mathcal{S}_R=\mathcal{P}^R\text{diag}\left(1,\,1,\,\omega^{k_{L2}}_n\right)\mathcal{P}^R\,,\\
\mathcal{S}_L=\mathcal{P}^L\text{diag}\left(1,\,1,\,\omega^{k_{L2}}_n\right)\mathcal{P}^L\,,
\end{array}\right.
\end{equation} 
leading to the phase transformation matrix $\Theta_{\mathcal{P}^LA_2\mathcal{P}^R}$
\begin{equation}
\frac{2\pi}{n}.\mathcal{P}^L
\begin{pmatrix}
0&0&k_{L2}\\
0&0&k_{L2}\\
-k_{L2}&-k_{L2}&0
\end{pmatrix}\mathcal{P}^R\,.
\end{equation}
In this case we have two possibilities:
\begin{itemize}
\item[(i)] $k_{L2}\neq -k_{L2}$

This implies $k_{L2}\neq n/2$. The order of the group has to be $n\geq 3$, leading to the chain
\begin{equation}
Z_{n\geq3}:\, \mathcal{P}^L\left\{A_2\oplus A_8\oplus A_{11}\right\}\mathcal{P}^R\,.
\end{equation}
The associated charge vector is
\begin{equation}
\left(1,\omega_n^{-k_{L2}},\omega_n^{k_{L2}}\right)\,.
\end{equation}

\item[(ii)] $k_{L2}= -k_{L2}$

This implies $k_{L2}=n/2$. The order of the group has to be $n\in 2\, \mathbb{Z}$, leading to the chain
\begin{equation}
Z_{2n}:\,\mathcal{P}^L\left\{ A_2\oplus A_3\right\}\mathcal{P}^R\,.
\end{equation}
We have made the redefinition $n\rightarrow 2n$. The associated charge vector is
\begin{equation}
\left(1,\omega_{2n}^{n}\right)\,,\quad k_{L2}=n\,.
\end{equation}
\end{itemize}

We now turn to the second case, i.e. without the texture $\mathcal{P}^LA_2\mathcal{P}^R$. The symmetry implementation is given by
\begin{equation}
\left\{
\begin{array}{l}
\mathcal{S}_R=\mathcal{P}^R\text{diag}\left(1,\,1,\,\omega^{k_{3}}_n\right)\mathcal{P}^R\,,\\
\mathcal{S}_L=\mathcal{P}^L\text{diag}\left(1,\,1,\,\omega^{k_{L2}}_n\right)\mathcal{P}^L\,,
\end{array}\right.
\end{equation} 
leading to the phase transformation matrix $\Theta_{\mathcal{P}^LA_7\mathcal{P}^R}$
\begin{equation}
\frac{2\pi}{n}.\mathcal{P}^L
\begin{pmatrix}
0&0&k_3\\
0&0&k_3\\
-k_{L2}&-k_{L2}&k_3-k_{L2}
\end{pmatrix}\mathcal{P}^R\, .
\end{equation}
In this case we have the following possibilities:
\begin{itemize}
\item[(i)]$k_3=-k_{L2}$

This implies $k_{L2}\neq n/2$. The group order has to be $n\geq 3$, leading to the chain
\begin{equation}
Z_{n\geq 3}:\, \mathcal{P}^L\left\{A_7\oplus A_3\oplus A_{12}\right\}\mathcal{P}^R\,,
\end{equation}
with the associated charge vector
\begin{equation}
\left(1,\omega_n^{k_{L2}},\omega_n^{2k_{L2}}\right)\,.
\end{equation}

\item[(ii)]$k_3\neq-k_{L2}$

The group order has to be $n\geq 4$, leading to the chain
\begin{equation}
Z_{n\geq 4}:\, \mathcal{P}^L\left\{A_7\oplus A_8\oplus A_{11}\oplus A_{12}\right\}\mathcal{P}^R\,.
\end{equation}
The associated charge vector is 
\begin{equation}
\left(1,\omega_{n}^{-k_3},\omega_{n}^{k_{L2}},\omega_{n}^{k_{L2}-k_3}\right)\,.
\end{equation}
\end{itemize}

The same procedure should be done for the nine classes. Details on this construction are given in Appendix.~\ref{chainsimple}. The Tables~\ref{C1i} and \ref{C2i} summarize the set of chains and their associated vector charge possible for the classes $\mathbf{(1,i)}$ and $\mathbf{(2,i)}$, respectively. Table~\ref{C3i} presents the chains for classes $\mathbf{(3,i)}$, while the associated vector charges are presented in Table~\ref{CV3ifull},
relegated to Appendix~\ref{Tables} due to its size. Also, in Appendix~\ref{Tables} the Table~\ref{SymTab} presents the symmetry groups that can be used to implement each chain.

\begin{table}[H]
\begin{center}
\begin{tabular}{r|l|l}
\hline\hline
$C_1^{\mathbf{(1,1)}}$&$A_1$&$1$\\
&&\\
$C_1^{\mathbf{(1,2)}}$&$\left\{A_6\oplus A_{10}\right\}\mathcal{P}^R$&$(1,\omega_n^{-k_3})$\\
&&\\
$C_1^{\mathbf{(1,3)}}$&$A_{10}\oplus A_{10}\mathcal{P}_{23}\oplus A_{10}\mathcal{P}_{13}$&$(1,\omega_n^{-k_2},\omega_n^{-k_1})$\\
\hline\hline
\end{tabular}
\end{center}
\caption{\label{C1i} Chains and associated charge vector for the classes $\mathbf{(1,i)}$.}
\end{table}

\begin{widetext}
\begin{table}[H]
\begin{center}
\begin{tabular}{r|l|l}
\hline\hline
$C_1^{\mathbf{(2,1)}}$&$\mathcal{P}^L\left\{A_5\oplus A_9\right\}$&$(1,\omega_n^{k_{L2}})$\\
&&\\
$C_1^{\mathbf{(2,2)}}$&$\mathcal{P}^L\left\{A_2\oplus A_3\right\}\mathcal{P}^{R}$&$(1,\omega_{2n}^n)\,,\,k_{L2}=n$\\
$C_2^{\mathbf{(2,2)}}$&$\mathcal{P}^L\left\{A_2\oplus A_8\oplus A_{11}
\right\}\mathcal{P}^{R}$&$(1,\omega_n^{-k_{L2}},\omega_n^{k_{L2}})$\\
$C_3^{\mathbf{(2,2)}}$&$\mathcal{P}^L\left\{A_7\oplus A_3\oplus A_{12}
\right\}\mathcal{P}^{R}$&$(1,\omega_n^{k_{L2}},\omega_n^{2k_{L2}})$\\
$C_4^{\mathbf{(2,2)}}$&$\mathcal{P}^L\left\{A_7\oplus A_8\oplus A_{11}\oplus A_{12}
\right\}\mathcal{P}^{R}$&$(1,\omega_n^{-k_3},\omega_n^{k_{L2}},\omega_n^{k_{L2}-k_3})$\\
&&\\
$C_1^\mathbf{(2,3)}$&$\mathcal{P}^L\left\{A_{4}\oplus A_{4}\mathcal{P}_{321}
\oplus A_{4}\mathcal{P}_{123}\right\}$&$(1,\omega_{3n}^{2n},\omega_{3n}^n)\,,\,k_{L2}=2n$\\
$C_2^\mathbf{(2,3)}$&$\mathcal{P}^L\left\{A_{4}\oplus A_{4}\mathcal{P}_{23}\oplus A_{8}\mathcal{P}_{13}
\oplus A_{12}\mathcal{P}_{13}\right\}$&$(1,\omega_{2(n+1)}^{n+1},\omega_{2(n+1)}^{-k_1},\omega_{2(n+1)}^{-k_1+n+1}),\, k_{L2}=n+1$\\
$C_3^\mathbf{(2,3)}$&$\mathcal{P}^L\left\{A_{4}\oplus A_{4}\mathcal{P}_{123}
\oplus A_{8}\mathcal{P}_{13}\oplus A_{12}\right\}$&$(1,\omega_n^{-k_{L2}},\omega_n^{-2k_{L2}},\omega_n^{k_{L2}})$\\
$C_4^\mathbf{(2,3)}$&$\mathcal{P}^L\left\{A_{4}\oplus A_{4}\mathcal{P}_{321}\oplus A_{12}\mathcal{P}_{13}\oplus A_{8}\mathcal{P}_{23}
\right\}$&$(1,\omega_n^{k_{L2}},\omega_n^{2k_{L2}},\omega_n^{-k_{L2}})$\\
$C_5^\mathbf{(2,3)}$&$\mathcal{P}^L\left\{A_{4}\oplus A_{8}\mathcal{P}_{23}\oplus A_{8}\mathcal{P}_{13}
\oplus A_{12}\oplus A_{12}\mathcal{P}_{13}\right\}$&$(1,\omega_n^{-k_{L2}},\omega_n^{-k_1},\omega_n^{k_{L2}},\omega_n^{k_{L2}-k_1})$\\
$C_6^\mathbf{(2,3)}$&$\mathcal{P}^L\left\{ A_{8}\oplus A_{8}\mathcal{P}_{23}
\oplus A_{8}\mathcal{P}_{13}\oplus A_{12}
\oplus A_{12}\mathcal{P}_{23}\oplus A_{12}\mathcal{P}_{13}\right\}$&$\left(1,\omega_{n}^{-k_2},\omega_n^{-k_{1}},\omega_{n}^{k_{L2}},\omega_{n}^{k_{L2}-k_2},\omega_{n}^{k_{L2}-k_{1}}\right)$\\
\hline\hline
\end{tabular}
\end{center}
\caption{\label{C2i} Chains and associated charge vector for the classes $\mathbf{(2,i)}$.}
\end{table}

\begin{table}[H]
\begin{center}
\begin{tabular}{r|l}
\hline\hline
$C_1^\mathbf{(3,1)}$&$A_9\oplus \mathcal{P}_{23}A_9\oplus \mathcal{P}_{13}A_9$\\
&\\
$C_1^\mathbf{(3,2)}$&$\left\{1,\mathcal{P}_{12}\right\}\left\{A_{14}\oplus\mathcal{P}_{123}A_{14}
\oplus\mathcal{P}_{321}A_{14}\right\}$\\
$C_2^\mathbf{(3,2)}$&$\left\{1,\mathcal{P}_{12},\mathcal{P}_{13}\right\}\left\{A_{14}\oplus\mathcal{P}_{23}A_{14}\oplus\mathcal{P}_{13}A_{11}
\oplus\mathcal{P}_{13}A_{12}\right\}$\\
$C_3^\mathbf{(3,2)}$&$\left\{\mathcal{P}\right\}\left\{A_{14}\oplus\mathcal{P}_{123}A_{14}
\oplus\mathcal{P}_{13}A_{11}\oplus\mathcal{P}_{23}A_{12}\right\}$\\
$C_4^\mathbf{(3,2)}$&$\left\{\mathcal{P}\right\}\left\{A_{14}\oplus A_{11}\oplus\mathcal{P}_{13}A_{11}
\oplus\mathcal{P}_{23}A_{12}\oplus\mathcal{P}_{13}A_{12}\right\}$\\
$C_5^\mathbf{(3,2)}$&$ A_{11}\oplus\mathcal{P}_{23}A_{11}
\oplus\mathcal{P}_{13}A_{11}\oplus A_{12}
\oplus\mathcal{P}_{23}A_{12}\oplus\mathcal{P}_{13}A_{12}$\\
&\\
$C_1^\mathbf{(3,3)}$&$A_{13}\oplus\mathcal{P}_{123}A_{13}\oplus\mathcal{P}_{321}A_{13}$\\
$C_2^\mathbf{(3,3)}$&$\left\{1,\mathcal{P}_{12},\mathcal{P}_{23}\right\}\left\{A_{13}\oplus\mathcal{P}_{321}A_{15}
\oplus A_{15}\mathcal{P}_{123}\oplus\mathcal{P}_{123}A_{15}\mathcal{P}_{12}\right\}$\\
$C_3^\mathbf{(3,3)}$&$\left\{1,\mathcal{P}_{12},\mathcal{P}_{23}\right\}\left\{A_{13}\oplus\mathcal{P}_{321}A_{15}\oplus
A_{15}\mathcal{P}_{123}\oplus\mathcal{P}_{13}A_{12}\oplus
A_{12}\mathcal{P}_{13}\right\}$\\
$C_4^\mathbf{(3,3)}$&$\left\{1,\mathcal{P}_{13},\mathcal{P}_{23}\right\}\left\{A_{13}\oplus\mathcal{P}_{321}A_{15}\mathcal{P}_{13}
\oplus A_{12}\mathcal{P}_{23}\oplus\mathcal{P}_{23}A_{12}
\oplus A_{12}\mathcal{P}_{13}\oplus\mathcal{P}_{13}A_{12}\right\}$\\
$C_{5}^\mathbf{(3,3)}$&$A_{13}\oplus A_{12}\mathcal{P}_{23}
\oplus\mathcal{P}_{23}A_{12}\oplus A_{12}\mathcal{P}_{13}
\oplus\mathcal{P}_{13}A_{12}\oplus\mathcal{P}_{13}A_{12}\mathcal{P}_{23}
\oplus\mathcal{P}_{23}A_{12}\mathcal{P}_{13}$\\
$C_{6}^\mathbf{(3,3)}$&$\left\{1,\mathcal{P}_{12},\mathcal{P}_{13}\right\}\left\{A_{15}\oplus\mathcal{P}_{23}A_{15}
\oplus\mathcal{P}_{321}A_{15}\mathcal{P}_{13}\oplus\mathcal{P}_{123}
A_{15}\mathcal{P}_{12}\oplus\mathcal{P}_{13}A_{12}\mathcal{P}_{13}\oplus A_0\right\}$\\
$C_{7}^\mathbf{(3,3)}$&$\left\{1,\mathcal{P}_{12},\mathcal{P}_{13}\right\}\left\{A_{15}\oplus A_{15}\mathcal{P}_{123}\oplus
\mathcal{P}_{321}A_{15}\mathcal{P}_{12}\oplus
\mathcal{P}_{123}A_{15}\mathcal{P}_{12}\oplus\mathcal{P}_{13}
A_{12}\mathcal{P}_{23}\right\}$\\
$C_{8}^\mathbf{(3,3)}$&$A_{15}\oplus\mathcal{P}_{321}A_{15}
\oplus\mathcal{P}_{123}A_{15}\oplus A_{12}\mathcal{P}_{13}\oplus\mathcal{P}_{23}A_{12}\mathcal{P}_{13}
\oplus\mathcal{P}_{13}A_{12}\mathcal{P}_{13}$\\
$C_{9}^\mathbf{(3,3)}$&$A_{15}\oplus\mathcal{P}_{12}A_{15}\mathcal{P}_{13}
\oplus\mathcal{P}_{13}A_{15}\mathcal{P}_{12}\oplus\mathcal{P}_{13}A_{12}\mathcal{P}_{13}
\oplus A_{12}\mathcal{P}_{23}\oplus\mathcal{P}_{23}A_{12}\oplus A_0
$\\
$C_{10}^\mathbf{(3,3)}$&$\left\{1,\mathcal{P}_{12},\mathcal{P}_{13}\right\}\left\{A_{15}\oplus A_{15}\mathcal{P}_{123}
\oplus A_{15}\mathcal{P}_{321}\oplus \mathcal{P}_{13}A_{12}\oplus\mathcal{P}_{13}A_{12}\mathcal{P}_{23}
\oplus\mathcal{P}_{13}A_{12}\mathcal{P}_{13}
\right\}$\\
$C_{11}^\mathbf{(3,3)}$&$\left\{\mathcal{P}\right\}\left\{A_{15}\oplus\mathcal{P}_{123}A_{15}\oplus
\mathcal{P}_{13}A_{15}\mathcal{P}_{123}\oplus\mathcal{P}_{13}A_{12}\mathcal{P}_{13}
\oplus\mathcal{P}_{23}A_{12}\oplus A_{12}\mathcal{P}_{13}\right\}$\\
$C_{12}^\mathbf{(3,3)}$&$\left\{\mathcal{P}\right\}\left\{A_{15}\oplus A_{15}\mathcal{P}_{321}\oplus
\mathcal{P}_{13}A_{15}\mathcal{P}_{123}\oplus A_{12}\mathcal{P}_{23}
\oplus\mathcal{P}_{13}A_{12}\oplus\mathcal{P}_{13}A_{12}\mathcal{P}_{13}\right\}$\\
$C_{13}^\mathbf{(3,3)}$&$\left\{\mathcal{P}\right\}\left\{A_{15}\oplus\mathcal{P}_{321}A_{15}\mathcal{P}_{13}
\oplus\mathcal{P}_{13}A_{12}\mathcal{P}_{13}\oplus\mathcal{P}_{23}A_{12}
\oplus\mathcal{P}_{13}A_{12}\oplus A_{12}\mathcal{P}_{23}\oplus
A_{12}\mathcal{P}_{13}\oplus A_0\right\}$\\
$C_{14}^\mathbf{(3,3)}$&$\left\{1,\mathcal{P}_{12},\mathcal{P}_{13},\mathcal{P}_{23}\right\}\left\{A_{15}\oplus\mathcal{P}_{321}A_{15}\oplus
\mathcal{P}_{13}A_{12}\mathcal{P}_{13}\oplus\mathcal{P}_{13}A_{12}
\oplus A_{12}\mathcal{P}_{23}\oplus A_{12}\mathcal{P}_{13}
\oplus\mathcal{P}_{23}A_{12}\mathcal{P}_{13}\right\}$\\
$C_{15}^\mathbf{(3,3)}$&$\left\{1,\mathcal{P}_{12},\mathcal{P}_{13},\mathcal{P}_{23}\right\}\left\{A_{15}\oplus A_{15}\mathcal{P}_{123}\oplus
A_{12}\mathcal{P}_{13}\oplus\mathcal{P}_{23}A_{12}\oplus
\mathcal{P}_{13}A_{12}\oplus\mathcal{P}_{13}A_{12}\mathcal{P}_{23}
\oplus\mathcal{P}_{13}A_{12}\mathcal{P}_{13}\right\}$\\
$C_{16}^\mathbf{(3,3)}$&$\left\{1,\mathcal{P}_{12},\mathcal{P}_{13}\right\}\left\{
A_{15}\oplus\mathcal{P}_{23}A_{15}\oplus\mathcal{P}_{13}A_{12}\mathcal{P}_{13}
\oplus\mathcal{P}_{13}A_{12}\oplus\mathcal{P}_{13}A_{12}\mathcal{P}_{23}
\oplus A_{12}\mathcal{P}_{13}\oplus\mathcal{P}_{23}A_{12}\mathcal{P}_{13}\oplus
A_0
\right\}$\\
$C_{17}^\mathbf{(3,3)}$&$\left\{1,\mathcal{P}_{12},\mathcal{P}_{13},\mathcal{P}_{23}\right\}\left\{
A_{15}\oplus A_{12}\mathcal{P}_{23}\oplus A_{12}\mathcal{P}_{13}\oplus
\mathcal{P}_{23}A_{12}\oplus\mathcal{P}_{23}A_{12}\mathcal{P}_{13}
\oplus \mathcal{P}_{13}A_{12}\oplus\mathcal{P}_{23}A_{12}\mathcal{P}_{23}
\oplus\mathcal{P}_{13}A_{12}\mathcal{P}_{13}
\right\}$\\
$C_{18}^\mathbf{(3,3)}$&$
A_{12}\oplus\mathcal{P}_{23}A_{12}\mathcal{P}_{23}\oplus A_{12}\mathcal{P}_{23}
\oplus A_{12}\mathcal{P}_{13}\oplus\mathcal{P}_{23}A_{12}
\oplus\mathcal{P}_{23}A_{12}\mathcal{P}_{13}
\oplus\mathcal{P}_{13}A_{12}\oplus\mathcal{P}_{13}A_{12}\mathcal{P}_{23}
\oplus\mathcal{P}_{13}A_{12}\mathcal{P}_{13}$\\
\hline\hline
\end{tabular}
\caption{\label{C3i} Chains for the classes $\mathbf{(3,i)}$.}
\end{center}
\end{table}
\end{widetext}

\section{Connecting up and down Yukawa sectors}\label{Cupdown}
Until now we have only studied textures and symmetries of matrices. No information of how an actual model would look like was given. In this section we shall clash the up and down sector to see what are the kind of textures, and minimal symmetries, that one can construct in a multi-Higgs model. The relation between the two sectors comes from the left-handed sector and the scalar fields. It is obvious that, even though the number of textures and chains is finite, the number of possible models with a large number of Higgs fields becomes impossible to deal with.

The steps to construct  the possible models from the vectors charge are the following:
\begin{itemize}
\item[(1)] Choose two chains, one for the down and another for the up sector, belonging to any of the three classes $\mathbf{(i,1)}$, $\mathbf{(i,2)}$, and $\mathbf{(i,3)}$. Each chain has its associated vector charge. To each charge  one defines a node. Draw a column of nodes for each sector (without lost of generality we choose the column of the down sector to be on the left). Any $A_0$ texture that was not presented in the minimal chain implementation is denoted by a white node.
\bigskip

Example:
\begin{equation}
\begin{array}{l}
\text{Down/Up sector: }\\
^0C_1^{\mathbf{(2,1)}}=
\left\{
\begin{array}{l}
\left\{A_5\,,A_9\,,A_0\right\}\\
(1\,,\omega_n^{k_{L2}}\,,\omega_n^k)
\end{array}\right.
\end{array}
\begin{array}{c}
\longrightarrow
\end{array}
\begin{array}{l}
\begin{tikzpicture}
[scale=.6,auto=left]
 \draw[fill]  (1,10) circle [radius=0.1cm];
 \node [left] at (0.75,10) {$\Gamma_{5}$};
 \draw[fill] (1,9) circle [radius=0.1cm];
  \node [left] at (0.75,9) {$\Gamma_{9}$};
   \draw (1,8) circle [radius=0.1cm];
  \node [left] at (0.75,8) {$\Gamma_{0}$};
   \draw[fill]  (1.5,10) circle [radius=0.1cm];
   \node [right] at (1.75,10) {$\Delta_{5}$};
   \draw[fill]  (1.5,9) circle [radius=0.1cm];
  \node [right] at (1.75,9) {$\Delta_{9}$};
    \draw  (1.5,8) circle [radius=0.1cm];
  \node [right] at (1.75,8) {$\Delta_{0}$}; 
\end{tikzpicture}
\end{array}
\end{equation}

\item[(2)] Write on each node of the left column the associated charges and on the right columns the conjugated ones.
\bigskip

Example:
\begin{equation}
\begin{array}{l}
\begin{tikzpicture}
[scale=.6,auto=left]
 \draw[fill]  (1,10) circle [radius=0.1cm];
 \node [left] at (0.75,10) {$0$};
 \draw[fill] (1,9) circle [radius=0.1cm];
  \node [left] at (0.75,9) {$k_{L2}$};
   \draw (1,8) circle [radius=0.1cm];
  \node [left] at (0.75,8) {$k$};
   \draw[fill]  (1.5,10) circle [radius=0.1cm];
   \node [right] at (1.75,10) {$0$};
   \draw[fill]  (1.5,9) circle [radius=0.1cm];
  \node [right] at (1.75,9) {$-k_{L2}$};
    \draw  (1.5,8) circle [radius=0.1cm];
  \node [right] at (1.75,8) {$-k$}; 
\end{tikzpicture}
\end{array}
\end{equation}

\item[(3)] Connect the nodes of the two columns. Each nontrivial connection gives a constraint on some of the symmetry phases. Constraints that impose two nodes, for the same column, with the same charge are not allowed.
\bigskip

Example (cases with no massless quark):
\begin{equation}
\begin{array}{l}
\left\{
\begin{array}{l}
\begin{tikzpicture}
[scale=.6,auto=left,every node/.style={draw,shape=circle,minimum size=0.1cm,inner sep=0}]
  \node[fill] (n1) at (1,10) {};
  \node[fill] (n2) at (1,9) {};
  \node[fill] (n3) at (1.5,10) {};
  \node[fill] (n4) at (1.5,9) {};
   \node (n5) at (1,8) {};
    \node (n6) at (1.5,8) {};
   \foreach \from/\to in {n1/n3,n2/n4}
    \draw[thick] (\from) -- (\to);
 \draw (1.25,9.2) node [draw,shape=circle,minimum size=0.1cm,inner sep=0,above] {2};   
\end{tikzpicture}
\end{array}\quad
\begin{array}{l}
\text{constraints:}\\
k_{L2}=-k_{L2}\, \text{mod}(n)\Rightarrow k_{L2}=\frac{n}{2}
\end{array}\right.
\\\\
\left\{
\begin{array}{l}
\begin{tikzpicture}
[scale=.6,auto=left,every node/.style={draw,shape=circle,minimum size=0.1cm,inner sep=0}]
  \node[fill] (n1) at (1,10) {};
  \node[fill] (n2) at (1,9) {};
  \node[fill] (n3) at (1.5,10) {};
  \node[fill] (n4) at (1.5,9) {};
   \node (n5) at (1,8) {};
    \node (n6) at (1.5,8) {};
   \foreach \from/\to in {n1/n3,n2/n6,n5/n4}
    \draw[thick] (\from) -- (\to);
\end{tikzpicture}
\end{array}
\quad
\begin{array}{l}
\text{constraints:}\\
k=-k_{L2}
\end{array}\right.
\end{array}
\end{equation}
The symbol $
\begin{tikzpicture}
[scale=.6,auto=left,every node/.style={draw,shape=circle,minimum size=0.1cm,inner sep=0}]
\draw (2,8.9) node [draw,shape=circle,minimum size=0.1cm,inner sep=0,above] {n};
\end{tikzpicture}$ states that the order of the diagram is $n\mathbb{Z}$.

\item[(4)] Use the freedom of a global phase transformation to change the position of the $0$ phase. With no loss of generality we can do it to the right column.
\bigskip

Example: we get two more cases
\begin{equation}
\begin{array}{l}
\begin{tikzpicture}
[scale=.6,auto=left]
 \draw[fill]  (1,10) circle [radius=0.1cm];
 \node [left] at (0.75,10) {$0$};
 \draw[fill] (1,9) circle [radius=0.1cm];
  \node [left] at (0.75,9) {$k_{L2}$};
   \draw (1,8) circle [radius=0.1cm];
  \node [left] at (0.75,8) {$k$};
   \draw[fill]  (1.5,10) circle [radius=0.1cm];
   \node [right] at (1.75,10) {$k_{L2}$};
   \draw[fill]  (1.5,9) circle [radius=0.1cm];
  \node [right] at (1.75,9) {$0$};
    \draw  (1.5,8) circle [radius=0.1cm];
  \node [right] at (1.75,8) {$-k+k_{L2}$}; 
\end{tikzpicture}
\end{array}\,,\,
\begin{array}{l}
\begin{tikzpicture}
[scale=.6,auto=left]
 \draw[fill]  (1,10) circle [radius=0.1cm];
 \node [left] at (0.75,10) {$0$};
 \draw[fill] (1,9) circle [radius=0.1cm];
  \node [left] at (0.75,9) {$k_{L2}$};
   \draw (1,8) circle [radius=0.1cm];
  \node [left] at (0.75,8) {$k$};
   \draw[fill]  (1.5,10) circle [radius=0.1cm];
   \node [right] at (1.75,10) {$k$};
   \draw[fill]  (1.5,9) circle [radius=0.1cm];
  \node [right] at (1.75,9) {$-k_{L2}+k$};
    \draw  (1.5,8) circle [radius=0.1cm];
  \node [right] at (1.75,8) {$0$}; 
\end{tikzpicture}
\end{array}\,.
\end{equation}

\item[(5)] Repeat step $(3)$.

\end{itemize}

\textbf{Statement:}
\textit{Given a diagram, the minimal order of the symmetry group is given by the number of nodes in the largest column (nnlc). If in that column the white node is not connected, the size of the group is reduced by one unit. 
If the condition $n/k\in \mathbb{Z}$ is imposed by one line, then the group order will be the smaller number bigger/equal than nnlc that is divisible by $k$.}

\subsection{Models up to $N=3$}

In this section we shall present some model implementations for up to three Higgs doublets. We start with models belonging to class $\mathbf{(1,i)}$. In Table~\ref{tab1i} , we present the combinations of textures that have three mixing angles. In order for a model to be phenomenologically viable, at least one of these combinations has to be present in one sector. 

\begin{table}[H]
\begin{center}
\begin{tabular}{r||c}
&Classes\\
\hline\hline
$A_1\oplus A_{0}\oplus A_{0}$&$^0C_1^{\mathbf{(1,1)}}$\\
$A_6\oplus A_{10}\oplus A_{0}$&$^0C^{\mathbf{(1,2)}}_1$\\
$A_{10}\oplus A_{10}\mathcal{P}_{23}\oplus A_{10}\mathcal{P}_{13}$&$C^{\mathbf{(1,3)}}_1$\\
\hline\hline
\end{tabular}
\caption{\label{tab1i} Combinations with $N=3$ which lead to three mixing angles for classes $\mathbf{(1,i)}$.}
\end{center}
\end{table}
The cases with $N=1,2$, which has three mixing angles, can be easily extracted from Table~\ref{tab1i}. In order to exemplify some properties we shall study in detail the implementation of models with $C_1^{\mathbf{(1,3)}}$ in both sectors [step $(1)$]. We start by drawing the diagram and associated charges [step $(2)$]

\begin{equation}
\begin{array}{l}
\begin{tikzpicture}
[scale=.6,auto=left]
 \draw[fill]  (1,10) circle [radius=0.1cm];
 \node [left] at (0.75,10) {$0$};
 \draw[fill] (1,9) circle [radius=0.1cm];
  \node [left] at (0.75,9) {$-k_2^d$};
   \draw[fill] (1,8) circle [radius=0.1cm];
  \node [left] at (0.75,8) {$-k_1^d$};
   \draw[fill]  (1.5,10) circle [radius=0.1cm];
   \node [right] at (1.75,10) {$0$};
   \draw[fill]  (1.5,9) circle [radius=0.1cm];
  \node [right] at (1.75,9) {$k_2^u$};
    \draw[fill]  (1.5,8) circle [radius=0.1cm];
  \node [right] at (1.75,8) {$k_1^u$}; 
\end{tikzpicture}
\end{array}
\end{equation}
Since we have the same chain in both sectors we should have on the right column the conjugated charges (or antisymmetric phases). However, the phases coming from the right-handed fields are different for the up and down quarks. Therefore, the only phases that we truly need to conjugate are the ones coming from the left-handed fields and the ones associated to extra null textures.

The next step is to join the nodes [step $(3)$]. However, in order to guarantee that the model can explain the six quark masses we need Table~\ref{detup3}, where the combinations of textures in a chain that gives $\text{det}(\mathbf{M}_{u,d})\neq 0$ up to three Higgs fields is presented. From Table~\ref{detup3}, we see that only models with three Higgs bosons are allowed and, therefore, all nodes of the diagrams must be connected. For the cases with the zero in the first position we get
\begin{equation}
\begin{array}{l}
\begin{tikzpicture}
[scale=.6,auto=left]
 \draw[fill]  (1,10) circle [radius=0.1cm];
 \node [left] at (0.75,10) {$0$};
 \draw[fill] (1,9) circle [radius=0.1cm];
  \node [left] at (0.75,9) {$-k_2^d$};
   \draw[fill] (1,8) circle [radius=0.1cm];
  \node [left] at (0.75,8) {$-k_1^d$};
   \draw[fill]  (1.5,10) circle [radius=0.1cm];
   \node [right] at (1.75,10) {$0$};
   \draw[fill]  (1.5,9) circle [radius=0.1cm];
  \node [right] at (1.75,9) {$k_2^u$};
    \draw[fill]  (1.5,8) circle [radius=0.1cm];
  \node [right] at (1.75,8) {$k_1^u$}; 
\end{tikzpicture}
\end{array}
\begin{array}{l}
=
\end{array}
\begin{array}{l}
\begin{tikzpicture}
[scale=.6,auto=left,every node/.style={draw,shape=circle,minimum size=0.1cm,inner sep=0}]
  \node[fill] (n1) at (1,10) {};
  \node[fill] (n2) at (1,9) {};
  \node[fill] (n3) at (1.5,10) {};
  \node[fill] (n4) at (1.5,9) {};
   \node[fill] (n5) at (1,8) {};
    \node[fill] (n6) at (1.5,8) {}; 
   \foreach \from/\to in {n1/n3,n2/n4,n5/n6}
    \draw[thick] (\from) -- (\to);
\end{tikzpicture}\quad
\begin{tikzpicture}
[scale=.6,auto=left,every node/.style={draw,shape=circle,minimum size=0.1cm,inner sep=0}]
  \node[fill] (n1) at (1,10) {};
  \node[fill] (n2) at (1,9) {};
  \node[fill] (n3) at (1.5,10) {};
  \node[fill] (n4) at (1.5,9) {};
   \node[fill] (n5) at (1,8) {};
    \node[fill] (n6) at (1.5,8) {}; 
   \foreach \from/\to in {n1/n3,n2/n6,n5/n4}
    \draw[thick] (\from) -- (\to);
\end{tikzpicture}
\end{array}
\end{equation}
The first diagram implies $k_2^u=-k_2^d$ and $k_1^u=-k_1^d$, while the second diagram implies $k_1^u=-k_2^d$ and $k_2^u=-k_1^d$. All these relations are true up to $\text{mod}(n)$. Next we use the freedom of a global phase transformation to change the position of the zero on the right column [step $(4)$[. Subtracting $k_2^u$ on the right columns and following step $(3)$, we get
\begin{equation}
\begin{array}{l}
\begin{tikzpicture}
[scale=.6,auto=left]
 \draw[fill]  (1,10) circle [radius=0.1cm];
 \node [left] at (0.75,10) {$0$};
 \draw[fill] (1,9) circle [radius=0.1cm];
  \node [left] at (0.75,9) {$-k_2^d$};
   \draw[fill] (1,8) circle [radius=0.1cm];
  \node [left] at (0.75,8) {$-k_1^d$};
   \draw[fill]  (1.5,10) circle [radius=0.1cm];
   \node [right] at (1.75,10) {$-k_2^u$};
   \draw[fill]  (1.5,9) circle [radius=0.1cm];
  \node [right] at (1.75,9) {$0$};
    \draw[fill]  (1.5,8) circle [radius=0.1cm];
  \node [right] at (1.75,8) {$k_1^u-k_2^u$}; 
\end{tikzpicture}
\end{array}
\begin{array}{l}
=
\end{array}
\begin{array}{l}
\begin{tikzpicture}
[scale=.6,auto=left,every node/.style={draw,shape=circle,minimum size=0.1cm,inner sep=0}]
  \node[fill] (n1) at (1,10) {};
  \node[fill] (n2) at (1,9) {};
  \node[fill] (n3) at (1.5,10) {};
  \node[fill] (n4) at (1.5,9) {};
   \node[fill] (n5) at (1,8) {};
    \node[fill] (n6) at (1.5,8) {}; 
   \foreach \from/\to in {n1/n4,n2/n3,n5/n6}
    \draw[thick] (\from) -- (\to);
\end{tikzpicture}\quad
\begin{tikzpicture}
[scale=.6,auto=left,every node/.style={draw,shape=circle,minimum size=0.1cm,inner sep=0}]
  \node[fill] (n1) at (1,10) {};
  \node[fill] (n2) at (1,9) {};
  \node[fill] (n3) at (1.5,10) {};
  \node[fill] (n4) at (1.5,9) {};
   \node[fill] (n5) at (1,8) {};
    \node[fill] (n6) at (1.5,8) {}; 
   \foreach \from/\to in {n1/n4,n2/n6,n5/n3}
    \draw[thick] (\from) -- (\to);
\end{tikzpicture}
\end{array}
\end{equation}
The first diagram implies $k_2^d=k_2^u$ and $k_1^d=k_2^u-k_1^u$, while the second diagram implies $k_1^d=k_2^u$ and $k_2^d=k_2^u-k_1^u$. We repeat the last steps, but now subtracting the phase $k_1^u$. We get the diagrams
\begin{equation}
\begin{array}{l}
\begin{tikzpicture}
[scale=.6,auto=left]
 \draw[fill]  (1,10) circle [radius=0.1cm];
 \node [left] at (0.75,10) {$0$};
 \draw[fill] (1,9) circle [radius=0.1cm];
  \node [left] at (0.75,9) {$-k_2^d$};
   \draw[fill] (1,8) circle [radius=0.1cm];
  \node [left] at (0.75,8) {$-k_1^d$};
   \draw[fill]  (1.5,10) circle [radius=0.1cm];
   \node [right] at (1.75,10) {$-k_1^u$};
   \draw[fill]  (1.5,9) circle [radius=0.1cm];
  \node [right] at (1.75,9) {$k_2^u-k_1^u$};
    \draw[fill]  (1.5,8) circle [radius=0.1cm];
  \node [right] at (1.75,8) {$0$}; 
\end{tikzpicture}
\end{array}
\begin{array}{l}
=
\end{array}
\begin{array}{l}
\begin{tikzpicture}
[scale=.6,auto=left,every node/.style={draw,shape=circle,minimum size=0.1cm,inner sep=0}]
  \node[fill] (n1) at (1,10) {};
  \node[fill] (n2) at (1,9) {};
  \node[fill] (n3) at (1.5,10) {};
  \node[fill] (n4) at (1.5,9) {};
   \node[fill] (n5) at (1,8) {};
    \node[fill] (n6) at (1.5,8) {}; 
   \foreach \from/\to in {n1/n6,n2/n4,n5/n3}
    \draw[thick] (\from) -- (\to);
\end{tikzpicture}\quad
\begin{tikzpicture}
[scale=.6,auto=left,every node/.style={draw,shape=circle,minimum size=0.1cm,inner sep=0}]
  \node[fill] (n1) at (1,10) {};
  \node[fill] (n2) at (1,9) {};
  \node[fill] (n3) at (1.5,10) {};
  \node[fill] (n4) at (1.5,9) {};
   \node[fill] (n5) at (1,8) {};
    \node[fill] (n6) at (1.5,8) {}; 
   \foreach \from/\to in {n1/n6,n2/n3,n5/n4}
    \draw[thick] (\from) -- (\to);
\end{tikzpicture}
\end{array}
\end{equation}
The first diagram implies $k_1^d=k_1^u$ and $k_2^d=k_1^u-k_2^u$, while the second one implies $k_2^d=k_1^u$ and $k_1^d=k_1^u-k_2^u$. This completes the identification of the models available for this case.

We shall use the notation $C_k^{\mathbf{(i,j)}}\otimes\, C_n^{\mathbf{(i,j^\prime)}}$ to represent the case where we have for the down-quark sector the chain $C_k^{\mathbf{(i,j)}}$ and for the up-quark sector the chain $C_n^{\mathbf{(i,j^\prime)}}$. Multiple lines connecting the same nodes represent  several Higgs fields with the same charge. In order to keep in mind that the down-quark sector is represented by the left column, we represent the textures $A_i$ by $\Gamma_i$. For the up-quark sector we do the same, but replacing $A_i$ by $\Delta_i$. We should regard $\Gamma_i$ and $\Delta_i$ as just matrices from Eq.~\eqref{Textures}. Their relation with the Yukawa couplings $\mathbf{\Gamma}_a$ and $\mathbf{\Delta}_a$ has to do with the labeling given for the Higgs fields. For example, the model with $\mathbf{\Gamma}_1=\Gamma_{10}$, $\mathbf{\Gamma}_2=\Gamma_{10}\mathcal{P}_{23}$, and $\mathbf{\Gamma}_3=\Gamma_{10}\mathcal{P}_{13}$ is telling us that $\Phi_1$ couples with $\Gamma_{10}$, $\Phi_2$ with $\Gamma_{10}\mathcal{P}_{23}$ and $\Phi_3$ with $\Gamma_{10}\mathcal{P}_{23}$.

Next we present the complete set of models for $N\geq 3$ for the classes $\mathbf{(1,i)}$:

\begin{subequations}\label{class1i1}
\begin{align}
&
\label{C113C113}
\end{equation}

We now turn to models belonging to classes $\mathbf{(2,i)}$. The corresponding diagrams are presented in Appendix~\ref{21N3}. Here we shall construct explicitly one particular example. We need Table~\ref{tab2i}, which is the equivalent of Table~\ref{tab1i} for the models of class $\mathbf{(2,i)}$.
\begin{table}[H]
\begin{center}
\begin{tabular}{r||c}
&Classes\\
\hline\hline
$A_5\oplus A_9\oplus A_0$&$^0C^{\mathbf{(2,1)}}_1$\\
$A_2\oplus A_3\oplus A_0$&$^0C^{\mathbf{(2,2)}}_1$\\
$A_2\oplus A_8\oplus A_0$&$^0C^{\mathbf{(2,2)}}_2$\\
$A_2\oplus A_{11}\oplus A_0$&$^0C^{\mathbf{(2,2)}}_2$\\
$A_2\oplus A_8\oplus A_{11}$&$C^{\mathbf{(2,2)}}_2$\\
$A_7\oplus A_3\oplus A_0$&$^0C^{\mathbf{(2,2)}}_3$\\
$A_7\oplus A_3\oplus A_{12}$&$C^{\mathbf{(2,2)}}_3$\\
\hline\hline
\end{tabular}
\caption{\label{tab2i} Combinations with $N=3$ which lead to three mixing angles for classes $\mathbf{(2,i)}$.}
\end{center}
\end{table}

We start by choosing the chain $^0C_1^{\mathbf{2,1}}$ for the down sector and the chain $^0C_3^{\mathbf{(2,2)}}$ for the up sector. The diagram is

\begin{equation}
\begin{array}{l}
\begin{tikzpicture}
[scale=.6,auto=left]
 \draw[fill]  (1,10) circle [radius=0.1cm];
 \node [left] at (0.75,10) {$0$};
 \draw[fill] (1,9) circle [radius=0.1cm];
  \node [left] at (0.75,9) {$k_{L2}$};
   \draw (1,8) circle [radius=0.1cm];
  \node [left] at (0.75,8) {$k$};
   \draw[fill]  (1.5,10) circle [radius=0.1cm];
   \node [right] at (1.75,10) {$0$};
   \draw[fill]  (1.5,9) circle [radius=0.1cm];
  \node [right] at (1.75,9) {$-k_{L2}$};
    \draw[fill]  (1.5,8) circle [radius=0.1cm];
  \node [right] at (1.75,8) {$-2k_{L2}$}; 
    \draw  (1.5,7) circle [radius=0.1cm];
  \node [right] at (1.75,7) {$-k^\prime$}; 
\end{tikzpicture}
\end{array}
\end{equation}
At this point we must go to Table~\ref{detup3} and check what combinations of one, two, and three textures we can make with nonzero determinant. For class $^0C_1^{\mathbf{2,1}}$ we always need to have $A_5$ and $A_9$, therefore, we can only implement models with at least two Higgs. For class $^0C_3^{\mathbf{2,2}}$ we always need to have the texture $A_7$ conjugated with at least another nonzero texture, i.e. $A_3$ or $A_{12}$. Knowing the cases with nonzero masses we just need to find on Table~\ref{tab2i} which combinations allow for three mixing angles. We see in its first line that the presence of $A_5$ and $A_9$ guarantees that the model has three mixing angles. Therefore, in the construction of these diagrams, we just need to take care of the non-zero masses, since the three mixing angles are guaranteed in that case. 

Drawing the first set of diagrams, we get
\begin{equation}
\begin{array}{l}
\begin{tikzpicture}
[scale=.6,auto=left]
 \draw[fill]  (1,10) circle [radius=0.1cm];
 \node [left] at (0.75,10) {$0$};
 \draw[fill] (1,9) circle [radius=0.1cm];
  \node [left] at (0.75,9) {$k_{L2}$};
   \draw (1,8) circle [radius=0.1cm];
  \node [left] at (0.75,8) {$k$};
   \draw[fill]  (1.5,10) circle [radius=0.1cm];
   \node [right] at (1.75,10) {$0$};
   \draw[fill]  (1.5,9) circle [radius=0.1cm];
  \node [right] at (1.75,9) {$-k_{L2}$};
    \draw[fill]  (1.5,8) circle [radius=0.1cm];
  \node [right] at (1.75,8) {$-2k_{L2}$}; 
    \draw  (1.5,7) circle [radius=0.1cm];
  \node [right] at (1.75,7) {$-k^\prime$}; 
\end{tikzpicture}
\end{array}
\begin{array}{l}
=
\end{array}
\begin{array}{l}
\begin{tikzpicture}
[scale=.6,auto=left,every node/.style={draw,shape=circle,minimum size=0.1cm,inner sep=0}]
  \node[fill] (n1) at (1,10) {};
  \node[fill] (n2) at (1,9) {};
  \node[fill] (n3) at (1.5,10) {};
  \node[fill] (n4) at (1.5,9) {};
   \node (n5) at (1,8) {};
    \node[fill] (n6) at (1.5,8) {};
    \node (n7) at (1.5,7) {};
  
   \foreach \from/\to in {n1/n3}
    \draw[style=double,thick] (\from) -- (\to);
    \foreach \from/\to in {n2/n6}
    \draw[thick] (\from) -- (\to);
    
\draw (1.25,9.2) node [draw,shape=circle,minimum size=0.1cm,inner sep=0,above] {3}; 
\end{tikzpicture}\quad
\begin{tikzpicture}
[scale=.6,auto=left,every node/.style={draw,shape=circle,minimum size=0.1cm,inner sep=0}]
  \node[fill] (n1) at (1,10) {};
  \node[fill] (n2) at (1,9) {};
  \node[fill] (n3) at (1.5,10) {};
  \node[fill] (n4) at (1.5,9) {};
   \node (n5) at (1,8) {};
    \node[fill] (n6) at (1.5,8) {};
    \node (n7) at (1.5,7) {};
  
   \foreach \from/\to in {n1/n3}
    \draw[thick] (\from) -- (\to);
    \foreach \from/\to in {n2/n6}
    \draw[style=double,thick] (\from) -- (\to);
    
\draw (1.25,9.2) node [draw,shape=circle,minimum size=0.1cm,inner sep=0,above] {3}; 
\end{tikzpicture}\quad
\begin{tikzpicture}
[scale=.6,auto=left,every node/.style={draw,shape=circle,minimum size=0.1cm,inner sep=0}]
  \node[fill] (n1) at (1,10) {};
  \node[fill] (n2) at (1,9) {};
  \node[fill] (n3) at (1.5,10) {};
  \node[fill] (n4) at (1.5,9) {};
   \node (n5) at (1,8) {};
    \node[fill] (n6) at (1.5,8) {};
    \node (n7) at (1.5,7) {};
  
   \foreach \from/\to in {n1/n3,n2/n6,n5/n4}
    \draw[thick] (\from) -- (\to);

 \draw (1.25,9.2) node [draw,shape=circle,minimum size=0.1cm,inner sep=0,above] {3};  
\end{tikzpicture}\quad
\begin{tikzpicture}
[scale=.6,auto=left,every node/.style={draw,shape=circle,minimum size=0.1cm,inner sep=0}]
  \node[fill] (n1) at (1,10) {};
  \node[fill] (n2) at (1,9) {};
  \node[fill] (n3) at (1.5,10) {};
  \node[fill] (n4) at (1.5,9) {};
   \node (n5) at (1,8) {};
    \node[fill] (n6) at (1.5,8) {};
    \node (n7) at (1.5,7) {};
  
   \foreach \from/\to in {n1/n3,n2/n6,n5/n7}
    \draw[thick] (\from) -- (\to);

 \draw (1.25,9.2) node [draw,shape=circle,minimum size=0.1cm,inner sep=0,above] {3};  
\end{tikzpicture}
\end{array}
\end{equation}
The order of the group has to be $3\mathbb{Z}$ due to the line connecting the second node on the left with the third one on the right. The second nodes of each column cannot be connected since it would imply $k_{L2}=n/2$ and, therefore, the first and third node of the right column equal. All models can be implemented with a $Z_3$ except the last one. This last model has the white node of the largest column connected, which increases the order of the group by one unit. However, since the order of the group has to be $3\mathbb{Z}$, the minimal symmetry group is $Z_6$. We continue by shifting the zero on the right column one position down; we get
\begin{equation}
\begin{array}{l}
\begin{tikzpicture}
[scale=.6,auto=left]
 \draw[fill]  (1,10) circle [radius=0.1cm];
 \node [left] at (0.75,10) {$0$};
 \draw[fill] (1,9) circle [radius=0.1cm];
  \node [left] at (0.75,9) {$k_{L2}$};
   \draw (1,8) circle [radius=0.1cm];
  \node [left] at (0.75,8) {$k$};
   \draw[fill]  (1.5,10) circle [radius=0.1cm];
   \node [right] at (1.75,10) {$k_{L2}$};
   \draw[fill]  (1.5,9) circle [radius=0.1cm];
  \node [right] at (1.75,9) {$0$};
    \draw[fill]  (1.5,8) circle [radius=0.1cm];
  \node [right] at (1.75,8) {$-k_{L2}$}; 
    \draw  (1.5,7) circle [radius=0.1cm];
  \node [right] at (1.75,7) {$-k^\prime+k_{L2}$}; 
\end{tikzpicture}
\end{array}
\begin{array}{l}
=
\end{array}
\begin{array}{l}
\begin{tikzpicture}
[scale=.6,auto=left,every node/.style={draw,shape=circle,minimum size=0.1cm,inner sep=0}]
  \node[fill] (n1) at (1,10) {};
  \node[fill] (n2) at (1,9) {};
  \node[fill] (n3) at (1.5,10) {};
  \node[fill] (n4) at (1.5,9) {};
   \node (n5) at (1,8) {};
    \node[fill] (n6) at (1.5,8) {};
    \node (n7) at (1.5,7) {};
  
   \foreach \from/\to in {n1/n4}
    \draw[style=double,thick] (\from) -- (\to);
    \foreach \from/\to in {n2/n3}
    \draw[thick] (\from) -- (\to);
    
\end{tikzpicture}\quad
\begin{tikzpicture}
[scale=.6,auto=left,every node/.style={draw,shape=circle,minimum size=0.1cm,inner sep=0}]
  \node[fill] (n1) at (1,10) {};
  \node[fill] (n2) at (1,9) {};
  \node[fill] (n3) at (1.5,10) {};
  \node[fill] (n4) at (1.5,9) {};
   \node (n5) at (1,8) {};
    \node[fill] (n6) at (1.5,8) {};
    \node (n7) at (1.5,7) {};
  
   \foreach \from/\to in {n1/n4}
    \draw[thick] (\from) -- (\to);
    \foreach \from/\to in {n2/n3}
    \draw[style=double,thick] (\from) -- (\to);
    
\end{tikzpicture}\quad
\begin{tikzpicture}
[scale=.6,auto=left,every node/.style={draw,shape=circle,minimum size=0.1cm,inner sep=0}]
  \node[fill] (n1) at (1,10) {};
  \node[fill] (n2) at (1,9) {};
  \node[fill] (n3) at (1.5,10) {};
  \node[fill] (n4) at (1.5,9) {};
   \node (n5) at (1,8) {};
    \node[fill] (n6) at (1.5,8) {};
    \node (n7) at (1.5,7) {};
  
   \foreach \from/\to in {n1/n4,n2/n3,n5/n6}
    \draw[thick] (\from) -- (\to);

\end{tikzpicture}\quad
\begin{tikzpicture}
[scale=.6,auto=left,every node/.style={draw,shape=circle,minimum size=0.1cm,inner sep=0}]
  \node[fill] (n1) at (1,10) {};
  \node[fill] (n2) at (1,9) {};
  \node[fill] (n3) at (1.5,10) {};
  \node[fill] (n4) at (1.5,9) {};
   \node (n5) at (1,8) {};
    \node[fill] (n6) at (1.5,8) {};
    \node (n7) at (1.5,7) {};
  
   \foreach \from/\to in {n1/n4,n2/n3,n5/n7}
    \draw[thick] (\from) -- (\to);

\end{tikzpicture}
\end{array}
\end{equation}
In contrast with the previous case, the last diagram can be implemented with a $Z_4$. The last two cases are

\begin{equation}
\begin{array}{l}
\begin{tikzpicture}
[scale=.6,auto=left]
 \draw[fill]  (1,10) circle [radius=0.1cm];
 \node [left] at (0.75,10) {$0$};
 \draw[fill] (1,9) circle [radius=0.1cm];
  \node [left] at (0.75,9) {$k_{L2}$};
   \draw (1,8) circle [radius=0.1cm];
  \node [left] at (0.75,8) {$k$};
   \draw[fill]  (1.5,10) circle [radius=0.1cm];
   \node [right] at (1.75,10) {$2k_{L2}$};
   \draw[fill]  (1.5,9) circle [radius=0.1cm];
  \node [right] at (1.75,9) {$k_{L2}$};
    \draw[fill]  (1.5,8) circle [radius=0.1cm];
  \node [right] at (1.75,8) {$0$}; 
    \draw  (1.5,7) circle [radius=0.1cm];
  \node [right] at (1.75,7) {$-k^\prime+2k_{L2}$}; 
\end{tikzpicture}
\end{array}
\begin{array}{l}
=
\end{array}
\begin{array}{l}
\begin{tikzpicture}
[scale=.6,auto=left,every node/.style={draw,shape=circle,minimum size=0.1cm,inner sep=0}]
  \node[fill] (n1) at (1,10) {};
  \node[fill] (n2) at (1,9) {};
  \node[fill] (n3) at (1.5,10) {};
  \node[fill] (n4) at (1.5,9) {};
   \node (n5) at (1,8) {};
    \node[fill] (n6) at (1.5,8) {};
    \node (n7) at (1.5,7) {};
  
   \foreach \from/\to in {n1/n6,n2/n4,n5/n3}
    \draw[thick] (\from) -- (\to);
\end{tikzpicture}
\end{array}
\end{equation}
and
\begin{equation}
\begin{array}{l}
\begin{tikzpicture}
[scale=.6,auto=left]
 \draw[fill]  (1,10) circle [radius=0.1cm];
 \node [left] at (0.75,10) {$0$};
 \draw[fill] (1,9) circle [radius=0.1cm];
  \node [left] at (0.75,9) {$k_{L2}$};
   \draw (1,8) circle [radius=0.1cm];
  \node [left] at (0.75,8) {$k$};
   \draw[fill]  (1.5,10) circle [radius=0.1cm];
   \node [right] at (1.75,10) {$k^\prime$};
   \draw[fill]  (1.5,9) circle [radius=0.1cm];
  \node [right] at (1.75,9) {$-k_{L2}+k^\prime$};
    \draw[fill]  (1.5,8) circle [radius=0.1cm];
  \node [right] at (1.75,8) {$-2k_{L2}+k^\prime$}; 
    \draw  (1.5,7) circle [radius=0.1cm];
  \node [right] at (1.75,7) {$0$}; 
\end{tikzpicture}
\end{array}
\begin{array}{l}
=
\end{array}
\begin{array}{l}
\begin{tikzpicture}
[scale=.6,auto=left,every node/.style={draw,shape=circle,minimum size=0.1cm,inner sep=0}]
  \node[fill] (n1) at (1,10) {};
  \node[fill] (n2) at (1,9) {};
  \node[fill] (n3) at (1.5,10) {};
  \node[fill] (n4) at (1.5,9) {};
   \node (n5) at (1,8) {};
    \node[fill] (n6) at (1.5,8) {};
    \node (n7) at (1.5,7) {};
  
   \foreach \from/\to in {n1/n7,n2/n6,n5/n3}
    \draw[thick] (\from) -- (\to);
\end{tikzpicture}
\end{array}
\end{equation}
These two implementations only allow models with three Higgs bosons (or more), contrarily to the previous cases, where models with two Higgs bosons are allowed. We summarize the possible model implementations:

\begin{equation}
\begin{array}{l}
\underline{^0C_1^{\mathbf{(2,1)}}\otimes ^0C_3^{\mathbf{(2,3)}}}\\
\begin{array}{r}
\begin{array}{l}
\begin{tikzpicture}
[scale=.6,auto=left]
 \draw[fill]  (1,10) circle [radius=0.1cm];
 \node [left] at (0.75,10) {$\Gamma_{5}$};
 \draw[fill] (1,9) circle [radius=0.1cm];
  \node [left] at (0.75,9) {$\Gamma_{9}$};
   \draw (1,8) circle [radius=0.1cm];
  \node [left] at (0.75,8) {$\Gamma_{0}$};
   \draw[fill]  (1.5,10) circle [radius=0.1cm];
   \node [right] at (1.75,10) {$\Delta_{7}$};
   \draw[fill]  (1.5,9) circle [radius=0.1cm];
  \node [right] at (1.75,9) {$\Delta_{3}$};
    \draw[fill]  (1.5,8) circle [radius=0.1cm];
  \node [right] at (1.75,8) {$\Delta_{12}$}; 
    \draw  (1.5,7) circle [radius=0.1cm];
  \node [right] at (1.75,7) {$\Delta_{0}$}; 
\end{tikzpicture}
\end{array}
\begin{array}{l}
=
\end{array}
\begin{array}{l}
\begin{tikzpicture}
[scale=.6,auto=left,every node/.style={draw,shape=circle,minimum size=0.1cm,inner sep=0}]
  \node[fill] (n1) at (1,10) {};
  \node[fill] (n2) at (1,9) {};
  \node[fill] (n3) at (1.5,10) {};
  \node[fill] (n4) at (1.5,9) {};
   \node (n5) at (1,8) {};
    \node[fill] (n6) at (1.5,8) {};
    \node (n7) at (1.5,7) {};
  
   \foreach \from/\to in {n1/n3}
    \draw[style=double,thick] (\from) -- (\to);
    \foreach \from/\to in {n2/n6}
    \draw[thick] (\from) -- (\to);
    
\draw (1.25,9.2) node [draw,shape=circle,minimum size=0.1cm,inner sep=0,above] {3}; 
\end{tikzpicture}\quad
\begin{tikzpicture}
[scale=.6,auto=left,every node/.style={draw,shape=circle,minimum size=0.1cm,inner sep=0}]
  \node[fill] (n1) at (1,10) {};
  \node[fill] (n2) at (1,9) {};
  \node[fill] (n3) at (1.5,10) {};
  \node[fill] (n4) at (1.5,9) {};
   \node (n5) at (1,8) {};
    \node[fill] (n6) at (1.5,8) {};
    \node (n7) at (1.5,7) {};
  
   \foreach \from/\to in {n1/n3}
    \draw[thick] (\from) -- (\to);
    \foreach \from/\to in {n2/n6}
    \draw[style=double,thick] (\from) -- (\to);
    
\draw (1.25,9.2) node [draw,shape=circle,minimum size=0.1cm,inner sep=0,above] {3}; 
\end{tikzpicture}\quad
\begin{tikzpicture}
[scale=.6,auto=left,every node/.style={draw,shape=circle,minimum size=0.1cm,inner sep=0}]
  \node[fill] (n1) at (1,10) {};
  \node[fill] (n2) at (1,9) {};
  \node[fill] (n3) at (1.5,10) {};
  \node[fill] (n4) at (1.5,9) {};
   \node (n5) at (1,8) {};
    \node[fill] (n6) at (1.5,8) {};
    \node (n7) at (1.5,7) {};
  
   \foreach \from/\to in {n1/n3,n2/n6,n5/n4}
    \draw[thick] (\from) -- (\to);

 \draw (1.25,9.2) node [draw,shape=circle,minimum size=0.1cm,inner sep=0,above] {3};  
\end{tikzpicture}\quad
\begin{tikzpicture}
[scale=.6,auto=left,every node/.style={draw,shape=circle,minimum size=0.1cm,inner sep=0}]
  \node[fill] (n1) at (1,10) {};
  \node[fill] (n2) at (1,9) {};
  \node[fill] (n3) at (1.5,10) {};
  \node[fill] (n4) at (1.5,9) {};
   \node (n5) at (1,8) {};
    \node[fill] (n6) at (1.5,8) {};
    \node (n7) at (1.5,7) {};
  
   \foreach \from/\to in {n1/n3,n2/n6,n5/n7}
    \draw[thick] (\from) -- (\to);

 \draw (1.25,9.2) node [draw,shape=circle,minimum size=0.1cm,inner sep=0,above] {3};  
\end{tikzpicture}\quad
\begin{tikzpicture}
[scale=.6,auto=left,every node/.style={draw,shape=circle,minimum size=0.1cm,inner sep=0}]
  \node[fill] (n1) at (1,10) {};
  \node[fill] (n2) at (1,9) {};
  \node[fill] (n3) at (1.5,10) {};
  \node[fill] (n4) at (1.5,9) {};
   \node (n5) at (1,8) {};
    \node[fill] (n6) at (1.5,8) {};
    \node (n7) at (1.5,7) {};
  
   \foreach \from/\to in {n1/n4}
    \draw[style=double,thick] (\from) -- (\to);
   \foreach \from/\to in {n2/n3}
    \draw[thick] (\from) -- (\to);
\end{tikzpicture}
\end{array}\\\\
\begin{array}{l}
\begin{tikzpicture}
[scale=.6,auto=left,every node/.style={draw,shape=circle,minimum size=0.1cm,inner sep=0}]
  \node[fill] (n1) at (1,10) {};
  \node[fill] (n2) at (1,9) {};
  \node[fill] (n3) at (1.5,10) {};
  \node[fill] (n4) at (1.5,9) {};
   \node (n5) at (1,8) {};
    \node[fill] (n6) at (1.5,8) {};
    \node (n7) at (1.5,7) {};
  
   \foreach \from/\to in {n1/n4}
    \draw[thick] (\from) -- (\to);
   \foreach \from/\to in {n2/n3}
    \draw[style=double,thick] (\from) -- (\to);
\end{tikzpicture}\quad
\begin{tikzpicture}
[scale=.6,auto=left,every node/.style={draw,shape=circle,minimum size=0.1cm,inner sep=0}]
  \node[fill] (n1) at (1,10) {};
  \node[fill] (n2) at (1,9) {};
  \node[fill] (n3) at (1.5,10) {};
  \node[fill] (n4) at (1.5,9) {};
   \node (n5) at (1,8) {};
    \node[fill] (n6) at (1.5,8) {};
    \node (n7) at (1.5,7) {};
  
   \foreach \from/\to in {n1/n4,n2/n3,n5/n6}
    \draw[thick] (\from) -- (\to);
\end{tikzpicture}\quad
\begin{tikzpicture}
[scale=.6,auto=left,every node/.style={draw,shape=circle,minimum size=0.1cm,inner sep=0}]
  \node[fill] (n1) at (1,10) {};
  \node[fill] (n2) at (1,9) {};
  \node[fill] (n3) at (1.5,10) {};
  \node[fill] (n4) at (1.5,9) {};
   \node (n5) at (1,8) {};
    \node[fill] (n6) at (1.5,8) {};
    \node (n7) at (1.5,7) {};
  
   \foreach \from/\to in {n1/n4,n2/n3,n5/n7}
    \draw[thick] (\from) -- (\to);
\end{tikzpicture}\quad
\begin{tikzpicture}
[scale=.6,auto=left,every node/.style={draw,shape=circle,minimum size=0.1cm,inner sep=0}]
  \node[fill] (n1) at (1,10) {};
  \node[fill] (n2) at (1,9) {};
  \node[fill] (n3) at (1.5,10) {};
  \node[fill] (n4) at (1.5,9) {};
   \node (n5) at (1,8) {};
    \node[fill] (n6) at (1.5,8) {};
    \node (n7) at (1.5,7) {};
  
   \foreach \from/\to in {n1/n6,n2/n4,n5/n3}
    \draw[thick] (\from) -- (\to);
\end{tikzpicture}\quad
\begin{tikzpicture}
[scale=.6,auto=left,every node/.style={draw,shape=circle,minimum size=0.1cm,inner sep=0}]
  \node[fill] (n1) at (1,10) {};
  \node[fill] (n2) at (1,9) {};
  \node[fill] (n3) at (1.5,10) {};
  \node[fill] (n4) at (1.5,9) {};
   \node (n5) at (1,8) {};
    \node[fill] (n6) at (1.5,8) {};
    \node (n7) at (1.5,7) {};
  
   \foreach \from/\to in {n1/n7,n2/n6,n5/n3}
    \draw[thick] (\from) -- (\to);
\end{tikzpicture}
\end{array}
\end{array}
\end{array}
\end{equation}

We turn now to the last classes $\mathbf{(3,i)}$. The number chain combinations that we can build is around a few hundred, with in most cases a large number of diagrams for each combination. In what follows we shall only present the cases up to $N=2$, the case with three Higgs bosons can be extracted from Tables~\ref{tab3iI}, \ref{tab3iII} and \ref{detup3}. The table with the combinations that allows three mixing angles up to two Higgs bosons is presented in Table~\ref{tab3iN2}.

\begin{table}[H]
\begin{center}
\begin{tabular}{r||c}
&Classes\\
\hline\hline
$\left\{1,\mathcal{P}_{12}\right\}\left\{A_{14}\oplus\mathcal{P}_{123} A_{14}\right\}$&$C^{\mathbf{(3,2)}}_{1,3}$\\
$\left\{1,\mathcal{P}_{12}\right\}\left\{A_{14}\oplus\mathcal{P}_{321} A_{14}\right\}$&$C^{\mathbf{(3,2)}}_1$\\
$\left\{1,\mathcal{P}_{12}\right\}\left\{\mathcal{P}_{123}A_{14}\oplus\mathcal{P}_{321} A_{14}\right\}$&$C^{\mathbf{(3,2)}}_1$\\
$\left\{\mathcal{P}_{13},\mathcal{P}_{23},\mathcal{P}_{123},\mathcal{P}_{321}\right\}\left\{A_{14}\oplus\mathcal{P}_{123} A_{14}\right\}$&$C^{\mathbf{(3,2)}}_{3}$\\
$A_{13}\oplus\mathcal{P}_{321} A_{13}$&$C^{\mathbf{(3,3)}}_{1}$\\
$\left\{1,\mathcal{P}_{12},\mathcal{P}_{23}\right\}\left\{A_{13}\oplus\mathcal{P}_{321} A_{15}\right\}$&$C^{\mathbf{(3,3)}}_{2,3}$\\
$\left\{1,\mathcal{P}_{12},\mathcal{P}_{23}\right\}\left\{A_{13}\oplus A_{15}\mathcal{P}_{123}\right\}$&$C^{\mathbf{(3,3)}}_{2,3}$\\
\hline\hline
\end{tabular}
\caption{\label{tab3iN2} Combinations with $N=2$ which lead to three mixing angles for classes $\mathbf{(3,i)}$.}
\end{center}
\end{table}

In order to present some properties of models in classes $\mathbf{(3,i)}$ we shall, once more, study a particular case. We choose the case with the chain $C_3^{\mathbf{(3,2)}}$ for the down sector and $C_3^{\mathbf{(3,3)}}$ for the up sector. Before drawing the nodes and associated charges we should notice that, contrarily to what happens for other classes, the possible left permutations of chains belong to the same class and should be taken as different chains. Therefore, when choosing $C_3^{\mathbf{(3,2)}}$ we actually have to study the six cases $\mathcal{P}C_3^{\mathbf{(3,2)}}$ and for $C_3^{\mathbf{(3,3)}}$ the three cases $\left\{1\,,\mathcal{P}_{12}\,,\mathcal{P}_{23}\right\}C_3^{\mathbf{(3,3)}}$. The extended table for the vector charges of class $\mathbf{(3,i)}$, Table~\ref{CV3ifull}, gives us the information for each individual case. These vector charges have a correlation between $k_{L1}$ and $k_{L2}$, which are shown in Table~\ref{constraints}.
\begin{table}[H]
\begin{center}
\begin{tabular}{l|l}
$C_3^{\mathbf{(3,2)}}$&constraint\\
\hline
$1\,,\mathcal{P}_{12}$&$k_{L1}=2k_{L2}$\\
$\mathcal{P}_{13}\,,\mathcal{P}_{123}$&$k_{L_1}=-k_{L2}$\\
$\mathcal{P}_{23}\,,\mathcal{P}_{321}$&$k_{L2}=2k_{L1}$
\end{tabular}\quad
\begin{tabular}{l|l}
$C_3^{\mathbf{(3,3)}}$&constraint\\
\hline
$1$&$k_{L2}=2k_{L1}$\\
$\mathcal{P}_{12}$&$k_{L_2}=-k_{L1}$\\
$\mathcal{P}_{23}$&$k_{L1}=2k_{L2}$
\end{tabular}
\caption{\label{constraints} Constraints impose by each chain.}
\end{center}
\end{table}
From the constraints in Table~\ref{constraints}, we see that the only possible matchings are
\begin{itemize}
\item[(a)]$\left\{1\,,\mathcal{P}_{12}\right\}C_3^{\mathbf{(2,3)}}\otimes \mathcal{P}_{23}C_{3}^{\mathbf{(3,3)}}$,

\item[(b)]$\left\{\mathcal{P}_{13}\,,\mathcal{P}_{123}\right\}C_3^{\mathbf{(2,3)}}\otimes \mathcal{P}_{12}C_{3}^{\mathbf{(3,3)}}$,

\item[(c)]$\left\{\mathcal{P}_{23}\,,\mathcal{P}_{321}\right\}C_3^{\mathbf{(2,3)}}\otimes C_{3}^{\mathbf{(3,3)}}$,
\end{itemize}
Any other case will imply that $k_{L1}$ or $k_{L2}$ equal $n/3$, which always leads to at least two equal charges in the vector charge of each sector. Let us then start with the first case of $(a)$, i.e. $C_3^{\mathbf{(2,3)}}\otimes \mathcal{P}_{23}C_{3}^{\mathbf{(3,3)}}$. The diagrams with the $0$ in the first node are
\begin{equation}
\begin{array}{l}
\begin{tikzpicture}
[scale=.6,auto=left]
 \draw[fill]  (1,10) circle [radius=0.1cm];
 \node [left] at (0.75,10) {$0$};
 \draw[fill]  (1,9) circle [radius=0.1cm];
  \node [left] at (0.75,9) {$-k_{L2}$};
     \draw[fill]  (1,8) circle [radius=0.1cm];
  \node [left] at (0.75,8) {$-2k_{L2}$};
      \draw[fill]  (1,7) circle [radius=0.1cm];
  \node [left] at (0.75,7) {$k_{L2}$};
  \draw[fill]  (1.5,10) circle [radius=0.1cm];
   \node [right] at (1.75,10) {$0$};
   \draw[fill]  (1.5,9) circle [radius=0.1cm];
  \node [right] at (1.75,9) {$k_{L2}$};
     \draw[fill]  (1.5,8) circle [radius=0.1cm];
  \node [right] at (1.75,8) {$-k_{L2}$};
   \draw[fill]  (1.5,7) circle [radius=0.1cm];
  \node [right] at (1.75,7) {$2k_{L2}$};
   \draw[fill]  (1.5,6) circle [radius=0.1cm];
  \node [right] at (1.75,6) {$-2k_{L2}$};   
\end{tikzpicture}
\end{array}
\begin{array}{l}
=\\\\
\end{array}
\begin{array}{l}
\begin{tikzpicture}
[scale=.6,auto=left,every node/.style={draw,shape=circle,minimum size=0.1cm,inner sep=0}]
  \node[fill] (n1) at (1,10) {};
  \node[fill] (n2) at (1,9) {};
  \node[fill] (n3) at (1.5,10) {};
  \node[fill] (n4) at (1.5,9) {};
  \node[fill] (n5) at (1,8) {};
  \node[fill] (n6) at (1.5,8) {};
  \node[fill] (n7) at (1,7) {};
  \node[fill] (n8) at (1.5,7) {};
  \node[fill] (n9) at (1.5,6) {};

   \foreach \from/\to in {n1/n3,n2/n6}
    \draw[thick] (\from) -- (\to);
    
\end{tikzpicture}\quad
\begin{tikzpicture}
[scale=.6,auto=left,every node/.style={draw,shape=circle,minimum size=0.1cm,inner sep=0}]
  \node[fill] (n1) at (1,10) {};
  \node[fill] (n2) at (1,9) {};
  \node[fill] (n3) at (1.5,10) {};
  \node[fill] (n4) at (1.5,9) {};
  \node[fill] (n5) at (1,8) {};
  \node[fill] (n6) at (1.5,8) {};
  \node[fill] (n7) at (1,7) {};
  \node[fill] (n8) at (1.5,7) {};
  \node[fill] (n9) at (1.5,6) {};

   \foreach \from/\to in {n1/n3,n5/n9}
    \draw[thick] (\from) -- (\to);
    
\end{tikzpicture}\quad
\begin{tikzpicture}
[scale=.6,auto=left,every node/.style={draw,shape=circle,minimum size=0.1cm,inner sep=0}]
  \node[fill] (n1) at (1,10) {};
  \node[fill] (n2) at (1,9) {};
  \node[fill] (n3) at (1.5,10) {};
  \node[fill] (n4) at (1.5,9) {};
  \node[fill] (n5) at (1,8) {};
  \node[fill] (n6) at (1.5,8) {};
  \node[fill] (n7) at (1,7) {};
  \node[fill] (n8) at (1.5,7) {};
  \node[fill] (n9) at (1.5,6) {};

   \foreach \from/\to in {n1/n3,n7/n4}
    \draw[thick] (\from) -- (\to);
    
\end{tikzpicture}\quad
\begin{tikzpicture}
[scale=.6,auto=left,every node/.style={draw,shape=circle,minimum size=0.1cm,inner sep=0}]
  \node[fill] (n1) at (1,10) {};
  \node[fill] (n2) at (1,9) {};
  \node[fill] (n3) at (1.5,10) {};
  \node[fill] (n4) at (1.5,9) {};
  \node[fill] (n5) at (1,8) {};
  \node[fill] (n6) at (1.5,8) {};
  \node[fill] (n7) at (1,7) {};
  \node[fill] (n8) at (1.5,7) {};
  \node[fill] (n9) at (1.5,6) {};

   \foreach \from/\to in {n2/n6,n5/n9}
    \draw[thick] (\from) -- (\to);
    
\end{tikzpicture}\quad
\begin{tikzpicture}
[scale=.6,auto=left,every node/.style={draw,shape=circle,minimum size=0.1cm,inner sep=0}]
  \node[fill] (n1) at (1,10) {};
  \node[fill] (n2) at (1,9) {};
  \node[fill] (n3) at (1.5,10) {};
  \node[fill] (n4) at (1.5,9) {};
  \node[fill] (n5) at (1,8) {};
  \node[fill] (n6) at (1.5,8) {};
  \node[fill] (n7) at (1,7) {};
  \node[fill] (n8) at (1.5,7) {};
  \node[fill] (n9) at (1.5,6) {};

   \foreach \from/\to in {n2/n6,n7/n4}
    \draw[thick] (\from) -- (\to);
    
\end{tikzpicture}\quad
\begin{tikzpicture}
[scale=.6,auto=left,every node/.style={draw,shape=circle,minimum size=0.1cm,inner sep=0}]
  \node[fill] (n1) at (1,10) {};
  \node[fill] (n2) at (1,9) {};
  \node[fill] (n3) at (1.5,10) {};
  \node[fill] (n4) at (1.5,9) {};
  \node[fill] (n5) at (1,8) {};
  \node[fill] (n6) at (1.5,8) {};
  \node[fill] (n7) at (1,7) {};
  \node[fill] (n8) at (1.5,7) {};
  \node[fill] (n9) at (1.5,6) {};

   \foreach \from/\to in {n5/n9,n7/n4}
    \draw[thick] (\from) -- (\to);
    
\end{tikzpicture}
\end{array}
\end{equation}
The last four diagrams imply massless quarks. The second one does not have three mixing angles. Therefore only the first diagram survives. Shifting the $0$ to the node below we get
\begin{equation}
\begin{array}{l}
\begin{tikzpicture}
[scale=.6,auto=left]
 \draw[fill]  (1,10) circle [radius=0.1cm];
 \node [left] at (0.75,10) {$0$};
 \draw[fill]  (1,9) circle [radius=0.1cm];
  \node [left] at (0.75,9) {$-k_{L2}$};
     \draw[fill]  (1,8) circle [radius=0.1cm];
  \node [left] at (0.75,8) {$-2k_{L2}$};
      \draw[fill]  (1,7) circle [radius=0.1cm];
  \node [left] at (0.75,7) {$k_{L2}$};
  \draw[fill]  (1.5,10) circle [radius=0.1cm];
   \node [right] at (1.75,10) {$-k_{L2}$};
   \draw[fill]  (1.5,9) circle [radius=0.1cm];
  \node [right] at (1.75,9) {$0$};
     \draw[fill]  (1.5,8) circle [radius=0.1cm];
  \node [right] at (1.75,8) {$-2k_{L2}$};
   \draw[fill]  (1.5,7) circle [radius=0.1cm];
  \node [right] at (1.75,7) {$k_{L2}$};
   \draw[fill]  (1.5,6) circle [radius=0.1cm];
  \node [right] at (1.75,6) {$-3k_{L2}$};   
\end{tikzpicture}
\end{array}
\begin{array}{l}
=\\\\
\end{array}
\begin{array}{l}
\begin{tikzpicture}
[scale=.6,auto=left,every node/.style={draw,shape=circle,minimum size=0.1cm,inner sep=0}]
  \node[fill] (n1) at (1,10) {};
  \node[fill] (n2) at (1,9) {};
  \node[fill] (n3) at (1.5,10) {};
  \node[fill] (n4) at (1.5,9) {};
  \node[fill] (n5) at (1,8) {};
  \node[fill] (n6) at (1.5,8) {};
  \node[fill] (n7) at (1,7) {};
  \node[fill] (n8) at (1.5,7) {};
  \node[fill] (n9) at (1.5,6) {};

   \foreach \from/\to in {n1/n4,n2/n3}
    \draw[thick] (\from) -- (\to);
    
\end{tikzpicture}\quad
\begin{tikzpicture}
[scale=.6,auto=left,every node/.style={draw,shape=circle,minimum size=0.1cm,inner sep=0}]
  \node[fill] (n1) at (1,10) {};
  \node[fill] (n2) at (1,9) {};
  \node[fill] (n3) at (1.5,10) {};
  \node[fill] (n4) at (1.5,9) {};
  \node[fill] (n5) at (1,8) {};
  \node[fill] (n6) at (1.5,8) {};
  \node[fill] (n7) at (1,7) {};
  \node[fill] (n8) at (1.5,7) {};
  \node[fill] (n9) at (1.5,6) {};

   \foreach \from/\to in {n1/n4,n5/n6}
    \draw[thick] (\from) -- (\to);
    
\end{tikzpicture}\quad
\begin{tikzpicture}
[scale=.6,auto=left,every node/.style={draw,shape=circle,minimum size=0.1cm,inner sep=0}]
  \node[fill] (n1) at (1,10) {};
  \node[fill] (n2) at (1,9) {};
  \node[fill] (n3) at (1.5,10) {};
  \node[fill] (n4) at (1.5,9) {};
  \node[fill] (n5) at (1,8) {};
  \node[fill] (n6) at (1.5,8) {};
  \node[fill] (n7) at (1,7) {};
  \node[fill] (n8) at (1.5,7) {};
  \node[fill] (n9) at (1.5,6) {};

   \foreach \from/\to in {n1/n4,n7/n8}
    \draw[thick] (\from) -- (\to);
    
\end{tikzpicture}\quad
\begin{tikzpicture}
[scale=.6,auto=left,every node/.style={draw,shape=circle,minimum size=0.1cm,inner sep=0}]
  \node[fill] (n1) at (1,10) {};
  \node[fill] (n2) at (1,9) {};
  \node[fill] (n3) at (1.5,10) {};
  \node[fill] (n4) at (1.5,9) {};
  \node[fill] (n5) at (1,8) {};
  \node[fill] (n6) at (1.5,8) {};
  \node[fill] (n7) at (1,7) {};
  \node[fill] (n8) at (1.5,7) {};
  \node[fill] (n9) at (1.5,6) {};

   \foreach \from/\to in {n2/n3,n5/n6}
    \draw[thick] (\from) -- (\to);
    
\end{tikzpicture}\quad
\begin{tikzpicture}
[scale=.6,auto=left,every node/.style={draw,shape=circle,minimum size=0.1cm,inner sep=0}]
  \node[fill] (n1) at (1,10) {};
  \node[fill] (n2) at (1,9) {};
  \node[fill] (n3) at (1.5,10) {};
  \node[fill] (n4) at (1.5,9) {};
  \node[fill] (n5) at (1,8) {};
  \node[fill] (n6) at (1.5,8) {};
  \node[fill] (n7) at (1,7) {};
  \node[fill] (n8) at (1.5,7) {};
  \node[fill] (n9) at (1.5,6) {};

   \foreach \from/\to in {n2/n3,n7/n8}
    \draw[thick] (\from) -- (\to);
    
\end{tikzpicture}\quad
\begin{tikzpicture}
[scale=.6,auto=left,every node/.style={draw,shape=circle,minimum size=0.1cm,inner sep=0}]
  \node[fill] (n1) at (1,10) {};
  \node[fill] (n2) at (1,9) {};
  \node[fill] (n3) at (1.5,10) {};
  \node[fill] (n4) at (1.5,9) {};
  \node[fill] (n5) at (1,8) {};
  \node[fill] (n6) at (1.5,8) {};
  \node[fill] (n7) at (1,7) {};
  \node[fill] (n8) at (1.5,7) {};
  \node[fill] (n9) at (1.5,6) {};

   \foreach \from/\to in {n5/n6,n7/n8}
    \draw[thick] (\from) -- (\to);
    
\end{tikzpicture}
\end{array}
\end{equation}
Only the first diagram has no massless quarks and 3 mixing angles. Shifting the $0$ one node down we get
\begin{equation}
\begin{array}{l}
\begin{tikzpicture}
[scale=.6,auto=left]
 \draw[fill]  (1,10) circle [radius=0.1cm];
 \node [left] at (0.75,10) {$0$};
 \draw[fill]  (1,9) circle [radius=0.1cm];
  \node [left] at (0.75,9) {$-k_{L2}$};
     \draw[fill]  (1,8) circle [radius=0.1cm];
  \node [left] at (0.75,8) {$-2k_{L2}$};
      \draw[fill]  (1,7) circle [radius=0.1cm];
  \node [left] at (0.75,7) {$k_{L2}$};
  \draw[fill]  (1.5,10) circle [radius=0.1cm];
   \node [right] at (1.75,10) {$k_{L2}$};
   \draw[fill]  (1.5,9) circle [radius=0.1cm];
  \node [right] at (1.75,9) {$2k_{L2}$};
     \draw[fill]  (1.5,8) circle [radius=0.1cm];
  \node [right] at (1.75,8) {$0$};
   \draw[fill]  (1.5,7) circle [radius=0.1cm];
  \node [right] at (1.75,7) {$3k_{L2}$};
   \draw[fill]  (1.5,6) circle [radius=0.1cm];
  \node [right] at (1.75,6) {$-k_{L2}$};   
\end{tikzpicture}
\end{array}
\begin{array}{l}
=\\\\
\end{array}
\begin{array}{l}
\begin{tikzpicture}
[scale=.6,auto=left,every node/.style={draw,shape=circle,minimum size=0.1cm,inner sep=0}]
  \node[fill] (n1) at (1,10) {};
  \node[fill] (n2) at (1,9) {};
  \node[fill] (n3) at (1.5,10) {};
  \node[fill] (n4) at (1.5,9) {};
  \node[fill] (n5) at (1,8) {};
  \node[fill] (n6) at (1.5,8) {};
  \node[fill] (n7) at (1,7) {};
  \node[fill] (n8) at (1.5,7) {};
  \node[fill] (n9) at (1.5,6) {};

   \foreach \from/\to in {n1/n6,n2/n9}
    \draw[thick] (\from) -- (\to);
    
\end{tikzpicture}\quad
\begin{tikzpicture}
[scale=.6,auto=left,every node/.style={draw,shape=circle,minimum size=0.1cm,inner sep=0}]
  \node[fill] (n1) at (1,10) {};
  \node[fill] (n2) at (1,9) {};
  \node[fill] (n3) at (1.5,10) {};
  \node[fill] (n4) at (1.5,9) {};
  \node[fill] (n5) at (1,8) {};
  \node[fill] (n6) at (1.5,8) {};
  \node[fill] (n7) at (1,7) {};
  \node[fill] (n8) at (1.5,7) {};
  \node[fill] (n9) at (1.5,6) {};

   \foreach \from/\to in {n1/n6,n5/n8}
    \draw[thick] (\from) -- (\to);
      \draw (1.25,6.2) node [draw,shape=circle,minimum size=0.1cm,inner sep=0,above] {5};    
\end{tikzpicture}\quad
\begin{tikzpicture}
[scale=.6,auto=left,every node/.style={draw,shape=circle,minimum size=0.1cm,inner sep=0}]
  \node[fill] (n1) at (1,10) {};
  \node[fill] (n2) at (1,9) {};
  \node[fill] (n3) at (1.5,10) {};
  \node[fill] (n4) at (1.5,9) {};
  \node[fill] (n5) at (1,8) {};
  \node[fill] (n6) at (1.5,8) {};
  \node[fill] (n7) at (1,7) {};
  \node[fill] (n8) at (1.5,7) {};
  \node[fill] (n9) at (1.5,6) {};

   \foreach \from/\to in {n1/n6,n7/n3}
    \draw[thick] (\from) -- (\to);
    
\end{tikzpicture}\quad
\begin{tikzpicture}
[scale=.6,auto=left,every node/.style={draw,shape=circle,minimum size=0.1cm,inner sep=0}]
  \node[fill] (n1) at (1,10) {};
  \node[fill] (n2) at (1,9) {};
  \node[fill] (n3) at (1.5,10) {};
  \node[fill] (n4) at (1.5,9) {};
  \node[fill] (n5) at (1,8) {};
  \node[fill] (n6) at (1.5,8) {};
  \node[fill] (n7) at (1,7) {};
  \node[fill] (n8) at (1.5,7) {};
  \node[fill] (n9) at (1.5,6) {};

   \foreach \from/\to in {n2/n9,n5/n8}
    \draw[thick] (\from) -- (\to);
      \draw (1.25,9.2) node [draw,shape=circle,minimum size=0.1cm,inner sep=0,above] {5};    
\end{tikzpicture}\quad
\begin{tikzpicture}
[scale=.6,auto=left,every node/.style={draw,shape=circle,minimum size=0.1cm,inner sep=0}]
  \node[fill] (n1) at (1,10) {};
  \node[fill] (n2) at (1,9) {};
  \node[fill] (n3) at (1.5,10) {};
  \node[fill] (n4) at (1.5,9) {};
  \node[fill] (n5) at (1,8) {};
  \node[fill] (n6) at (1.5,8) {};
  \node[fill] (n7) at (1,7) {};
  \node[fill] (n8) at (1.5,7) {};
  \node[fill] (n9) at (1.5,6) {};

   \foreach \from/\to in {n2/n9,n7/n3}
    \draw[thick] (\from) -- (\to);
    
\end{tikzpicture}\quad
\begin{tikzpicture}
[scale=.6,auto=left,every node/.style={draw,shape=circle,minimum size=0.1cm,inner sep=0}]
  \node[fill] (n1) at (1,10) {};
  \node[fill] (n2) at (1,9) {};
  \node[fill] (n3) at (1.5,10) {};
  \node[fill] (n4) at (1.5,9) {};
  \node[fill] (n5) at (1,8) {};
  \node[fill] (n6) at (1.5,8) {};
  \node[fill] (n7) at (1,7) {};
  \node[fill] (n8) at (1.5,7) {};
  \node[fill] (n9) at (1.5,6) {};

   \foreach \from/\to in {n5/n8,n7/n3}
    \draw[thick] (\from) -- (\to);
      \draw (1.25,6.2) node [draw,shape=circle,minimum size=0.1cm,inner sep=0,above] {5};    
\end{tikzpicture}
\end{array}
\end{equation}
Only the second diagram has no massless quarks. However, from Table~\ref{tab3iN2} we see that it does not have three mixing angles. Thus no diagrams survive in this case. We can continue the same procedure but no new diagram is found. Therefore we summarize our result as
\begin{equation}\label{case1}
\begin{array}{l}
\underline{C_3^{\mathbf{(3,2)}}\otimes \mathcal{P}_{23}C_3^{\mathbf{(3,3)}}}\\
\begin{array}{l}
\begin{tikzpicture}
[scale=.6,auto=left]
 \draw[fill]  (1,10) circle [radius=0.1cm];
 \node [left] at (0.75,10) {$\Gamma_{14}$};
 \draw[fill]  (1,9) circle [radius=0.1cm];
  \node [left] at (0.75,9) {$\mathcal{P}_{123}\Gamma_{14}$};
   \draw[fill]  (1,8) circle [radius=0.1cm];
  \node [left] at (0.75,8) {$\mathcal{P}_{13}\Gamma_{11}$};
    \draw[fill]  (1,7) circle [radius=0.1cm];
  \node [left] at (0.75,7) {$\mathcal{P}_{23}\Gamma_{12}$};
   \draw[fill]  (1.5,10) circle [radius=0.1cm];
   \node [right] at (1.75,10) {$\mathcal{P}_{23}\Delta_{13}$};
   \draw[fill]  (1.5,9) circle [radius=0.1cm];
  \node [right] at (1.75,9) {$\mathcal{P}_{12}\Delta_{15}$};
     \draw[fill]  (1.5,8) circle [radius=0.1cm];
  \node [right] at (1.75,8) {$\mathcal{P}_{23}\Delta_{15}\mathcal{P}_{123}$};
   \draw[fill]  (1.5,7) circle [radius=0.1cm];
  \node [right] at (1.75,7) {$\mathcal{P}_{13}\Delta_{12}$};
   \draw[fill]  (1.5,6) circle [radius=0.1cm];
  \node [right] at (1.75,6) {$\mathcal{P}_{23}\Delta_{12}\mathcal{P}_{13}$};
   
\end{tikzpicture}
\end{array}
\begin{array}{l}
=\\\\
\end{array}
\begin{array}{l}
\begin{tikzpicture}
[scale=.6,auto=left,every node/.style={draw,shape=circle,minimum size=0.1cm,inner sep=0}]
  \node[fill] (n1) at (1,10) {};
  \node[fill] (n2) at (1,9) {};
  \node[fill] (n3) at (1.5,10) {};
  \node[fill] (n4) at (1.5,9) {};
  \node[fill] (n5) at (1,8) {};
  \node[fill] (n6) at (1.5,8) {};
  \node[fill] (n7) at (1,7) {};
  \node[fill] (n8) at (1.5,7) {};
  \node[fill] (n9) at (1.5,6) {};

   \foreach \from/\to in {n1/n3,n2/n6}
    \draw[thick] (\from) -- (\to);
    
\end{tikzpicture}\quad
\begin{tikzpicture}
[scale=.6,auto=left,every node/.style={draw,shape=circle,minimum size=0.1cm,inner sep=0}]
  \node[fill] (n1) at (1,10) {};
  \node[fill] (n2) at (1,9) {};
  \node[fill] (n3) at (1.5,10) {};
  \node[fill] (n4) at (1.5,9) {};
  \node[fill] (n5) at (1,8) {};
  \node[fill] (n6) at (1.5,8) {};
  \node[fill] (n7) at (1,7) {};
  \node[fill] (n8) at (1.5,7) {};
  \node[fill] (n9) at (1.5,6) {};
   
   \foreach \from/\to in {n1/n4,n2/n3}
    \draw[thick] (\from) -- (\to);
    
\end{tikzpicture}
\end{array}
\end{array}
\end{equation}
From Table~\ref{detup3} we get that the class $C_3^{\mathbf{(3,3)}}$ can be implemented in models with just the first texture. So in principle we should have added a null texture and studied the class $^0C_3^{\mathbf{(3,3)}}$ instead. However, a quick examination shows us that no coupling with the null texture is possible. 

The second case of $(a)$, i.e. the down sector with the chain $\mathcal{P}_{12}C_3^{\mathbf{(2,3)}}\otimes \mathcal{P}_{23}C_{3}^{\mathbf{(3,3)}}$, is found in a similar way. The final diagrams are 

\begin{equation}\label{case2}
\begin{array}{l}
\underline{\mathcal{P}_{12}C_3^{\mathbf{(2,3)}}\otimes \mathcal{P}_{23}C_{3}^{\mathbf{(3,3)}}}\\
\begin{array}{l}
\begin{tikzpicture}
[scale=.6,auto=left]
 \draw[fill]  (1,10) circle [radius=0.1cm];
 \node [left] at (0.75,10) {$\mathcal{P}_{12}\Gamma_{14}$};
 \draw[fill]  (1,9) circle [radius=0.1cm];
  \node [left] at (0.75,9) {$\mathcal{P}_{23}\Gamma_{14}$};   
   \draw[fill]  (1,8) circle [radius=0.1cm];
  \node [left] at (0.75,8) {$\mathcal{P}_{321}\Gamma_{11}$};
    \draw[fill]  (1,7) circle [radius=0.1cm];
  \node [left] at (0.75,7) {$\mathcal{P}_{123}\Gamma_{12}$};  
   \draw[fill]  (1.5,10) circle [radius=0.1cm];
   \node [right] at (1.75,10) {$\mathcal{P}_{23}\Delta_{13}$};
   \draw[fill]  (1.5,9) circle [radius=0.1cm];
  \node [right] at (1.75,9) {$\mathcal{P}_{12}\Delta_{15}$};
     \draw[fill]  (1.5,8) circle [radius=0.1cm];
  \node [right] at (1.75,8) {$\mathcal{P}_{23}\Delta_{15}\mathcal{P}_{123}$};
   \draw[fill]  (1.5,7) circle [radius=0.1cm];
  \node [right] at (1.75,7) {$\mathcal{P}_{13}\Delta_{12}$};
   \draw[fill]  (1.5,6) circle [radius=0.1cm];
  \node [right] at (1.75,6) {$\mathcal{P}_{23}\Delta_{12}\mathcal{P}_{13}$};
\end{tikzpicture}
\end{array}
\begin{array}{l}
=\\\\
\end{array}
\begin{array}{l}
\begin{tikzpicture}
[scale=.6,auto=left,every node/.style={draw,shape=circle,minimum size=0.1cm,inner sep=0}]
  \node[fill] (n1) at (1,10) {};
  \node[fill] (n2) at (1,9) {};
  \node[fill] (n3) at (1.5,10) {};
  \node[fill] (n4) at (1.5,9) {};
  \node[fill] (n5) at (1,8) {};
  \node[fill] (n6) at (1.5,8) {};
  \node[fill] (n7) at (1,7) {};
  \node[fill] (n8) at (1.5,7) {};
  \node[fill] (n9) at (1.5,6) {};

   \foreach \from/\to in {n1/n3,n2/n4}
    \draw[thick] (\from) -- (\to);
    
\end{tikzpicture}\quad
\begin{tikzpicture}
[scale=.6,auto=left,every node/.style={draw,shape=circle,minimum size=0.1cm,inner sep=0}]
  \node[fill] (n1) at (1,10) {};
  \node[fill] (n2) at (1,9) {};
  \node[fill] (n3) at (1.5,10) {};
  \node[fill] (n4) at (1.5,9) {};
  \node[fill] (n5) at (1,8) {};
  \node[fill] (n6) at (1.5,8) {};
  \node[fill] (n7) at (1,7) {};
  \node[fill] (n8) at (1.5,7) {};
  \node[fill] (n9) at (1.5,6) {};
   
   \foreach \from/\to in {n1/n6,n2/n3}
    \draw[thick] (\from) -- (\to);
    
\end{tikzpicture}
\end{array}
\end{array}\,.
\end{equation}
The other cases are correlated with these two last results. Actually we can easily check that multiplying the chain $C_3^{\mathbf{(3,2)}}\otimes \mathcal{P}_{23}C_3^{\mathbf{(3,3)}}$ by $\mathcal{P}_{23}$ on the left we get the first chain of $(3)$, and by $\mathcal{P}_{123}$ we get the second case of $(2)$. The same can be applied to the other three cases. We can them summarize this as: $\left\{1,\mathcal{P}_{23},\mathcal{P}_{123}\right\}\times C_3^{\mathbf{(3,2)}}\otimes \mathcal{P}_{23}C_3^{\mathbf{(3,3)}}$ corresponds to Eq.~\eqref{case1} and $\left\{1,\mathcal{P}_{23},\mathcal{P}_{123}\right\}\times \mathcal{P}_{12}C_3^{\mathbf{(3,2)}}\otimes \mathcal{P}_{23}C_3^{\mathbf{(3,3)}}$ corresponds to Eq.~\eqref{case2}.

The full set of diagrams for $N=2$ are presented below.

\begin{subequations}
\begin{align}
&

\end{array}
\end{array}
\end{equation}

\section{Direct product of Abelian Groups}\label{DPAG}
Until now, all the results found are associated with cyclic groups. However, the fundamental theorem of finite Abelian groups states that any finite Abelian group $G$ is isomorphic to a direct product of cyclic groups of prime-power order. This allows the group $G$ to be written as a direct product of cyclic groups in either of the following ways:
\begin{itemize}
\item[(i)] $G\cong Z_{q_1}\times\cdots\times Z_{q_n}$, where each $q_j$ is a power of a prime;
\item[(ii)] $G\cong Z_{r_1}\times \cdots\times Z_{r_m}$, where $r_j$ divides $r_{j+1}$ for $1\leq j\leq m-1$.
\end{itemize}
Any group satisfying $(i)$ or $(ii)$ is not isomorphic to a cyclic group. 

The main idea is to have the fields transforming under a set of $n$ diagonal generators, leading to the symmetry equation
\begin{equation}
\left(\prod_{i=1}^n \mathcal{S}_L^{i\dagger}\right)\mathcal{A}_a\left(\prod_{i=1}^n \mathcal{S}_R^{i}\right)\left(\mathcal{S}_H\right)_{aa}=\mathcal{A}_a
\end{equation}
Since the product of $n$ generators can be reproduced by a single generator where the entries are the product of the $n$ phases, the textures previously found for $\mathcal{A}_a$ are not altered. However, this successive product of generators can add new chains. In order to find these new chains we shall introduce a, less common, matrix product:

\textit{Definition (Hadamard product):} Let $A$ and $B$ be two matrices with the same dimension $m\times n$. The Hadamard product $A\circ B$ is given by
\begin{equation}
(A\circ B)_{ij}=(A)_{ij}(B)_{ij}
\end{equation}

The Hadamard product is associative, distributive, and commutative (unlike the usual matrix product).

We can now state the necessary and sufficient steps in order to find the chains obtained by the product of cyclic groups. The three steps are as follows:
\begin{itemize}
\item[(1)] Find the chains for each individual cyclic group,
\item[(2)] Pick one texture from each of these chains and multiply them using the Hadamard product. The resulting matrix is one texture of the final chain,
\item[(3)] Repeat step $(2)$ for all possible combination.
\end{itemize}
In order to make the procedure clear, we shall present an example. Let us suppose that we have $Z_2\times Z_2$, where one $Z_2$ generates the chain $A_5\oplus A_9$ and the other the chain $\mathcal{P}_{23}\left\{A_5\oplus A_9\right\}$. The Hadamard products of these textures gives
\begin{eqnarray}
\begin{split}
&A_5\circ \mathcal{P}_{23}A_5=\mathcal{P}_{13}A_9\,,\quad A_5\circ \mathcal{P}_{23}A_9=\mathcal{P}_{23}A_9\,,\\
&A_9\circ \mathcal{P}_{23}A_5=A_9\,,\quad A_9\circ \mathcal{P}_{23}A_9=A_0\,.
\end{split}
\end{eqnarray}
The final chain is then given by
\begin{equation}\label{chainZ2Z2}
A_9\oplus\mathcal{P}_{23}A_9\oplus\mathcal{P}_{13}A_9\oplus A_0\,,
\end{equation}
which, in this case, can also be implemented from a $Z_4$ corresponding to the chain $^0C_1^{\mathbf{(3,1)}}$. However, there are solutions for model implementations that can be implemented with the $Z_2\times Z_2$ solution and not with the $Z_4$.  In order to understand this issue, it is convenient to write the vector charge associated with the chain in Eq.~\eqref{chainZ2Z2}, when it is a result of the action of two generators. The vector charge in this case is
\begin{equation}\label{veccharg}
\left((\omega_n^{k_{L2}},1),(1,\omega_n^{k^\prime_{L2}}),(1,1),(\omega_n^{k_{L2}},\omega_n^{k^\prime_{L2}})\right)\,.
\end{equation}
Contrarily to the cyclic groups, in this case each element of the group is specified by two phases, one from each cyclic group. The possible models constructed from the chain $^0C_{1}^{\mathbf{(3,1)}}$ in both sectors and three Higgs bosons are
\begin{equation}
\begin{array}{l}
\begin{tikzpicture}
[scale=.6,auto=left]
 \draw[fill]  (1,10) circle [radius=0.1cm];
 \node [left] at (0.75,10) {$\Gamma_{9}$};
 \draw[fill] (1,9) circle [radius=0.1cm];
  \node [left] at (0.75,9) {$\mathcal{P}_{23}\Gamma_{9}$};
   \draw[fill] (1,8) circle [radius=0.1cm];
  \node [left] at (0.75,8) {$\mathcal{P}_{13}\Gamma_{9}$};
    \draw (1,7) circle [radius=0.1cm];
  \node [left] at (0.75,7) {$\Gamma_{0}$};
   \draw[fill]  (1.5,10) circle [radius=0.1cm];
   \node [right] at (1.75,10) {$\Delta_{9}$};
   \draw[fill]  (1.5,9) circle [radius=0.1cm];
  \node [right] at (1.75,9) {$\mathcal{P}_{23}\Delta_{9}$};
    \draw[fill]  (1.5,8) circle [radius=0.1cm];
  \node [right] at (1.75,8) {$\mathcal{P}_{13}\Delta_{9}$}; 
      \draw (1.5,7) circle [radius=0.1cm];
  \node [right] at (1.75,7) {$\Delta_{0}$};
\end{tikzpicture}
\end{array}
\begin{array}{l}
=
\end{array}
\begin{array}{l}
\begin{tikzpicture}
[scale=.6,auto=left,every node/.style={draw,shape=circle,minimum size=0.1cm,inner sep=0}]
  \node[fill] (n1) at (1,10) {};
  \node[fill] (n2) at (1,9) {};
  \node[fill] (n3) at (1.5,10) {};
  \node[fill] (n4) at (1.5,9) {};
   \node[fill] (n5) at (1,8) {};
    \node[fill] (n6) at (1.5,8) {};
    \node (n7) at (1,7) {};
     \node (n8) at (1.5,7) {};
  
    \foreach \from/\to in {n1/n3,n2/n6,n5/n4}
    \draw[thick] (\from) -- (\to);
    
\end{tikzpicture}\quad
\begin{tikzpicture}
[scale=.6,auto=left,every node/.style={draw,shape=circle,minimum size=0.1cm,inner sep=0}]
  \node[fill] (n1) at (1,10) {};
  \node[fill] (n2) at (1,9) {};
  \node[fill] (n3) at (1.5,10) {};
  \node[fill] (n4) at (1.5,9) {};
   \node[fill] (n5) at (1,8) {};
    \node[fill] (n6) at (1.5,8) {};
    \node (n7) at (1,7) {};
     \node (n8) at (1.5,7) {};
  
    \foreach \from/\to in {n1/n4,n2/n3,n5/n6}
    \draw[thick] (\from) -- (\to);
    
\end{tikzpicture}\quad
\begin{tikzpicture}
[scale=.6,auto=left,every node/.style={draw,shape=circle,minimum size=0.1cm,inner sep=0}]
  \node[fill] (n1) at (1,10) {};
  \node[fill] (n2) at (1,9) {};
  \node[fill] (n3) at (1.5,10) {};
  \node[fill] (n4) at (1.5,9) {};
   \node[fill] (n5) at (1,8) {};
    \node[fill] (n6) at (1.5,8) {};
    \node (n7) at (1,7) {};
     \node (n8) at (1.5,7) {};
  
    \foreach \from/\to in {n1/n6,n5/n3,n2/n4}
    \draw[thick] (\from) -- (\to);
    
\end{tikzpicture}
\end{array}\,.
\end{equation}
Doing the same procedure, but now using the vector charge of Eq.~\eqref{veccharg}, we get a single diagram
\begin{equation}
\begin{array}{l}
\begin{tikzpicture}
[scale=.6,auto=left]
 \draw[fill]  (1,10) circle [radius=0.1cm];
 \node [left] at (0.75,10) {$\Gamma_{9}$};
 \draw[fill] (1,9) circle [radius=0.1cm];
  \node [left] at (0.75,9) {$\mathcal{P}_{23}\Gamma_{9}$};
   \draw[fill] (1,8) circle [radius=0.1cm];
  \node [left] at (0.75,8) {$\mathcal{P}_{13}\Gamma_{9}$};
    \draw (1,7) circle [radius=0.1cm];
  \node [left] at (0.75,7) {$\Gamma_{0}$};
   \draw[fill]  (1.5,10) circle [radius=0.1cm];
   \node [right] at (1.75,10) {$\Delta_{9}$};
   \draw[fill]  (1.5,9) circle [radius=0.1cm];
  \node [right] at (1.75,9) {$\mathcal{P}_{23}\Delta_{9}$};
    \draw[fill]  (1.5,8) circle [radius=0.1cm];
  \node [right] at (1.75,8) {$\mathcal{P}_{13}\Delta_{9}$}; 
      \draw (1.5,7) circle [radius=0.1cm];
  \node [right] at (1.75,7) {$\Delta_{0}$};
\end{tikzpicture}
\end{array}
\begin{array}{l}
=
\end{array}
\begin{array}{l}
\begin{tikzpicture}
[scale=.6,auto=left,every node/.style={draw,shape=circle,minimum size=0.1cm,inner sep=0}]
  \node[fill] (n1) at (1,10) {};
  \node[fill] (n2) at (1,9) {};
  \node[fill] (n3) at (1.5,10) {};
  \node[fill] (n4) at (1.5,9) {};
   \node[fill] (n5) at (1,8) {};
    \node[fill] (n6) at (1.5,8) {};
    \node (n7) at (1,7) {};
     \node (n8) at (1.5,7) {};
  
    \foreach \from/\to in {n1/n3,n2/n4,n5/n6}
    \draw[thick] (\from) -- (\to);
  \draw (1.25,9.2) node [draw,shape=circle,minimum size=0.1cm,inner sep=0,above] {2};  
\end{tikzpicture}
\end{array}\,,
\end{equation}
which is not one of the possible models implemented with a $Z_4$. Therefore, even if the chains found by direct products are already present for the cyclic groups, the model implementation may differ. We shall not pursue the determination of all possible model implementations for the chains that are shared by cyclic groups and direct products of cyclic groups. 

The cases we are most interested in are chains that can only be implemented through a direct product of cyclic groups. When the chains for the cyclic groups were found, in Appendix~\ref{chainsimple}, there were some combinations of textures not allowed. These cases are the ones we are interested in. From a simple inspection, we find out that the cases not allowed by cyclic symmetries are
\begin{equation}\label{chainDPAG}
\begin{array}{rl}
(1):&A_{13}\oplus \mathcal{P}_{23}A_{15}\oplus \mathcal{P}_{123}A_{15}\mathcal{P}_{12}\oplus
\mathcal{P}_{321}A_{15}\mathcal{P}_{13}\\
(2):&\left\{1\,,\mathcal{P}_{13}\,,\mathcal{P}_{23}\right\}\left\{A_{13}\oplus \mathcal{P}_{23}A_{15}\oplus \mathcal{P}_{123}A_{15}\mathcal{P}_{12}\right.\\
&\hspace{2.2cm}\left.\oplus
\mathcal{P}_{13}A_{12}\mathcal{P}_{23}\oplus
\mathcal{P}_{23}A_{12}\mathcal{P}_{13}\right\}\\
(3):&\left\{1\,,\mathcal{P}_{12}\,,\mathcal{P}_{23}\right\}\left\{A_{13}\oplus \mathcal{P}_{23}A_{15}\oplus \mathcal{P}_{321}A_{15}\mathcal{P}_{13}\right.\\
&\hspace{2.2cm}\left.\oplus
\mathcal{P}_{13}A_{12}\oplus
A_{12}\mathcal{P}_{13}\right\}\\
(4):&\left\{1\,,\mathcal{P}_{12}\,,\mathcal{P}_{13}\right\}\left\{A_{13}\oplus \mathcal{P}_{123}A_{15}\mathcal{P}_{12}\oplus
\mathcal{P}_{321}A_{15}\mathcal{P}_{13}\right.\\
&\hspace{2.3cm}\left.\oplus
\mathcal{P}_{23}A_{12}\oplus
A_{12}\mathcal{P}_{23}\right\}
\end{array}
\end{equation} 
The chains $(2)$ to $(4)$ have dimension 5. Therefore, they cannot be implemented through direct products, unless we had null textures. The minimal order is $8$, so, if we are able to build these chains, at least three null textures would have to be present. There are only seven chains whose products may end up with $A_{13}$ or permutations. These chains are: $C_{1,2}^{\mathbf{(2,2)}}$ and $C_{1,2,3,4,5}^{\mathbf{(3,3)}}$. The idea is to look now for products that lead to at least two textures of
\begin{equation}
\left\{ \mathcal{P}_{23}A_{15}\,,\, \mathcal{P}_{123}A_{15}\mathcal{P}_{12}\,,\,
\mathcal{P}_{321}A_{15}\mathcal{P}_{13}\right\}\,.
\end{equation}
Since the chains from class $\mathbf{(3,3)}$ do not have two of these textures, they can never be used to obtain chains $(1)$ to $(4)$. We are left with only two chains from class $\mathbf{(2,2)}$. In order to obtain a texture of the type $A_{13}$, we must use $C_{1,2}^{\mathbf{(2,2)}}\circ \mathcal{P}C_{1}^{\mathbf{(2,2)}}\mathcal{P}$, with $\mathcal{P}=\left\{\mathcal{P}_{13},\,\mathcal{P}_{23}\right\}$. Any of these cases leads to chain $(1)$. Therefore, $(1)$ is the only chain that can be exclusively implemented by direct products. The smallest implementation is given by the Hadamard product $C_{1}^{\mathbf{(2,2)}}\circ \mathcal{P}_{23}C_{1}^{\mathbf{(2,2)}}\mathcal{P}_{23}$, leading to the $Z_2\times Z_2$ group. The chain has the associated vector charge
\begin{equation}
\left((1,1),\,(-1,-1),\,(-1,1),\,(1,-1)\right)\,.
\end{equation}

\section{Quark models: general features and some examples}\label{QuaksM}
In general,
when analyzing the Yukawa sector of a NHDM, the scalar fields are transformed nontrivially under the horizontal symmetry. Since these fields can acquire vacuum expectation values, it is very important to avoid a (pseudo)Goldstone boson in the scalar potential. It is well known that the breaking of a continuous symmetry will lead to these massless particles. If only the scalar sector presents this property, then loop corrections can induce a mass to these scalars. Nevertheless, light scalars that couple to SM fermions and gauge fields are not desirable in a realistic model. Continuous symmetries in the scalar potential can be present by explicit construction, or accidentally. One shall focus on the second case.

Ivanov, Keus and Vdovin \cite{Ivanov:2011ae} have developed a strategy to identify all the discrete Abelian symmetries that can be implemented in NHDM and do not lead to an accidental continuous symmetry. The major result is the upper bound on the order of the Abelian discrete group $|G|\leq 2^{N-1}$, with $N$ the number of Higgs fields in the model. We shall use this information when classifying models.

Until now, we have only used the following experimental facts: quarks have nonzero masses, and the $V_{CKM}$ mixing matrix mixes all the flavor sectors. However, phenomena such as flavor changing neutral currents (FCNC), which are very suppressed in nature and appear only at loop level for both gauge and Higgs sectors in the SM, have no natural suppression in the NHDM without additional constraints. These FCNC are a consequence of the misalignment between the Yukawa couplings and the mass matrices. In the SM the mass matrix is proportional to the Yukawa coupling. However, in models with more scalar doublets this is no longer true and, in the mass eigenbasis, there will appear fermion interactions mediated by scalars that violate flavor. A simple way to obtain natural relations is through the use of symmetries in the Lagrangian; when they preclude FCNC it is said that the model has natural flavor conservation (NFC). Glashow and Weinberg~\cite{Glashow:1976nt} and Paschos \cite{Paschos:1976ay} pointed out that sequential extensions of the SM have a GIM-like mechanism~\cite{Glashow:1970gm} suppressing all direct neutral currents effects. From their work, NFC in NHDM can be formulated as the situation where all Yukawa couplings are simultaneously diagonalizable
\begin{equation}
\mathbf{U}_L^{n\dagger}\mathbf{\Gamma}_a\mathbf{U}_R^{n}=\text{diag}\quad\text{and}\quad \mathbf{U}_L^{p\dagger}\mathbf{\Delta}_a\mathbf{U}_R^{p}=\text{diag}\,,\,\forall_a\,,
\end{equation}
with $\mathbf{U}^{n,p}_{L,R}$ defined in Eq.	~\eqref{URL}. An alternative, and equivalent, way of expressing these conditions is through the definition of the sets
\begin{equation}
\mathbf{\Gamma}_{LL}=\left\{\mathbf{\Gamma}_a\mathbf{\Gamma}_b^\dagger\right\}\,,\quad \mathbf{\Gamma}_{RR}=\left\{\mathbf{\Gamma}_a^\dagger\mathbf{\Gamma}_b\right\}\,,
\end{equation}
for the down sector and
\begin{equation}
\mathbf{\Delta}_{LL}=\left\{\mathbf{\Delta}_a\mathbf{\Delta}_b^\dagger\right\}\,, \quad \mathbf{\Delta}_{RR}=\left\{\mathbf{\Delta}_a^\dagger\mathbf{\Delta}_b\right\}
\end{equation}
for the up sector. Requiring that each set $\mathbf{\Gamma}_{LL}$, $\mathbf{\Gamma}_{RR}$, $\mathbf{\Delta}_{LL}$ and $\mathbf{\Delta}_{RR}$ are Abelian is equivalent to the statement of NFC~\cite{Gatto:1979mr}. We shall use this second way of implementing NFC to classify the models.

For simplicity, we shall use $\mathcal{A}_{XX}$ to represent $\mathbf{\Gamma}_{XX}$ or $\mathbf{\Delta}_{XX}$. This set can be split into two parts
\begin{equation}
\mathcal{A}_{XX}=\left\{\mathcal{H}_{XX},\,\mathcal{A}_{XX}^{\text{off}}\right\}\,,\,
\left\{
\begin{array}{rl}
\mathcal{H}_{XX}\equiv&\left\{\mathcal{H}_{X}^a\right\}\,,\\ \mathcal{A}_{LL}^{\text{off}}\equiv&\left\{ \mathcal{A}_a\mathcal{A}_b^\dagger\right\}\\ \mathcal{A}_{RR}^{\text{off}}\equiv&\left\{ \mathcal{A}_a^\dagger\mathcal{A}_b\right\}
  \end{array}\right.
\end{equation}
and $a\neq b$. The usefulness of this separation has to do with the fact that we already know a lot from the structure of $\mathcal{H}_{X}^a$ when Abelian symmetries are in action, due to Table~\ref{classesij}. We shall now use the NFC condition and deviations from it as a way to classify these Abelian models.

\subsection{Model with NFC in just one sector}

We start this section by presenting the following theorem:
\bigskip 

\textit{Theorem (one sector NFC):} There are only six classes of models, within Abelian symmetries, that can implement NFC in one sector and have no massless fermion. The classes are as follows:
\begin{itemize}
\item[(i)] $A_1\oplus (N-1)A_0$,
\item[(ii)] $\mathcal{P}_L\left\{A_2\oplus(N-1)A_0\right\}$,
\item[(iii)] $\mathcal{P}_L\left\{A_7\oplus nA_{12}\oplus(N-(n+1))A_0\right\}$,
\item[(iv)] $\mathcal{P}\left\{nA_{13}\oplus (N-n)A_{0}\right\}\mathcal{P}^\prime$,
\item[(v)] $\mathcal{P}\left\{nA_{15}\oplus m\mathcal{P}_{13}A_{12}\mathcal{P}_{13}\oplus (N-(n+m))A_{0}\right\}\mathcal{P}^\prime$,
\item[(vi)] $
\mathcal{P}\left\{nA_{12}\oplus m\mathcal{P}_{23}A_{12}\mathcal{P}_{23}\oplus
k\mathcal{P}_{13}A_{12}\mathcal{P}_{13}\right.$
$
\quad\left.\oplus (N-(n+m+k))A_{0}\right\}\mathcal{P}^\prime$,
\end{itemize}
\textit{Proof} (see Appendix~\ref{Proof}).
\bigskip

From this theorem a very simple result on natural flavor conserving models can be extracted:

\textit{Corollary: }The only NFC model phenomenologically viable is the one with $(i)$ in both sectors.  

\textit{Proof:} In order for the model to be phenomenologically viable it has to have three mixing angles in the unitary matrix that mix the left rotation of both sectors. Since any of the possible NFC implementations for a given sector, i.e. $(i)$ to $(vi)$, belongs to the classes $\mathbf{(i,i)}$ with the texture of the chain equal to the texture of $\mathcal{H}^a_X$ of that class; only cases belonging to class $\mathbf{(1,1)}$ lead to three mixing angles. It follows immediately that the only allowed case is $(i)$ in both sectors.

These models correspond to some of the ones presented in Eq.~\eqref{C111C111}. We can have direct models, where $\Gamma_1$ is connected with $\Delta_1$ and $\Gamma_0$ with $\Delta_0$, or cross models, where $\Gamma_1$ is connected with $\Delta_0$ and vice versa. For any number $N$ of Higgs fields the minimal symmetry group that can be used to implement these models is $Z_2$. Therefore we can always implement NFC without the introduction of accidental symmetries.

There are other ways of implementing NFC in NHDM; however, these cannot be implemented through a symmetry. One common example is the Yukawa alignment in 2HDM~\cite{Pich:2009sp}. In this case NFC is achieved by requiring that all the Yukawa couplings, for each sector, are proportional,
\begin{equation}
\mathbf{\Gamma}_i= c_i \mathbf{\Gamma}_j\quad\text{and}\quad \mathbf{\Delta}_i= c_i \mathbf{\Delta}_j\,,\quad \forall_{i,j}\,.
\end{equation}
As shown in~\cite{Ferreira:2010xe}, no symmetry implementation can be used to implement this requirement. In~\cite{Serodio:2011hg}, alignment was seen as a low-energy effect of NFC models, while in~\cite{Varzielas:2011jr} its origin was related with flavor symmetries.

Another consequence of the above theorem follows:
\bigskip

\textit{Corollary: }Without the null texture, i.e. $A_0$, there are at most three phenomenological classes of models with NFC in one sector: classes $(iii)$, $(v)$ and $(vi)$.  

\textit{Proof: } Case $(ii)$ is excluded since there is no matrix, apart from $A_2$, in the classes $\mathbf{(2,i)}$ that has a nonzero determinant, and an $A_2$ texture alone cannot accommodate three mixing angles. In case $(iii)$ it is possible to find combinations of two matrices with determinants different from zero and three angles. For case $(iv)$ the texture has to belong to classes $\mathbf{(3,i)}$. By construction, no matrix in these classes can bring three mixing angles. For cases $(v)$ and $(vi)$ we can find combinations of two or three textures, respectively, from classes $\mathbf{(3,i)}$ that would allow phenomenological models.
\bigskip

This is an upper bound on these types of phenomenological models because we did not prove that we could implement them. This can only be done with the help of the chains. However, case $(iii)$ is implemented
in Eq.~\eqref{C121C322}, case $(v)$ in Eq.~\eqref{C233C933}, and case $(vi)$ (which needs a minimum of three Higgs bosons) can be implemented, for example, as
\begin{equation}\label{BGLM3}
\begin{array}{r}
\begin{array}{l}
\begin{tikzpicture}
[scale=.6,auto=left]
 \draw[fill]  (1,10) circle [radius=0.1cm];
 \node [left] at (0.75,10) {$\Gamma_{9}$};
 \draw[fill] (1,9) circle [radius=0.1cm];
  \node [left] at (0.75,9) {$\mathcal{P}_{23}\Gamma_{9}$};
   \draw[fill] (1,8) circle [radius=0.1cm];
  \node [left] at (0.75,8) {$\mathcal{P}_{13}\Gamma_{9}$};
  
   \draw[fill]  (1.5,10) circle [radius=0.1cm];
 \node [right] at (1.75,10) {$\Delta_{15}$};
 \draw[fill] (1.5,9) circle [radius=0.1cm];
  \node [right] at (1.75,9) {$\mathcal{P}_{12}\Delta_{15}\mathcal{P}_{13}$};
   \draw[fill] (1.5,8) circle [radius=0.1cm];
  \node [right] at (1.75,8) {$\mathcal{P}_{13}\Delta_{15}\mathcal{P}_{12}$};
     \draw[fill] (1.5,7) circle [radius=0.1cm];
  \node [right] at (1.75,7) {$\mathcal{P}_{13}\Delta_{12}\mathcal{P}_{13}$};
  \draw[fill] (1.5,6) circle [radius=0.1cm];
  \node [right] at (1.75,6) {$\Delta_{12}\mathcal{P}_{23}$};
     \draw[fill] (1.5,5) circle [radius=0.1cm];
  \node [right] at (1.75,5) {$\mathcal{P}_{23}\Delta_{12}$};
       \draw[fill] (1.5,4) circle [radius=0.1cm];
  \node [right] at (1.75,4) {$\Delta_{0}$};
\end{tikzpicture}
\end{array}
\begin{array}{l}
=\\\\
\end{array}
\begin{array}{l}
\begin{tikzpicture}
[scale=.6,auto=left,every node/.style={draw,shape=circle,minimum size=0.1cm,inner sep=0}]
  \node[fill] (n1) at (1,10) {};
  \node[fill] (n2) at (1,9) {};
  \node[fill] (n3) at (1.5,10) {};
  \node[fill] (n4) at (1.5,9) {};
   \node[fill] (n5) at (1,8) {};
    \node[fill] (n6) at (1.5,8) {};
     \node[fill] (n7) at (1.5,7) {};
     \node[fill] (n8) at (1.5,6) {};
      \node[fill] (n9) at (1.5,5) {};
     \node[fill] (n10) at (1.5,4) {};

   \foreach \from/\to in {n1/n8,n2/n9,n5/n7}
    \draw[thick] (\from) to (\to);   
     \draw (1.35,9.2) node [draw,shape=circle,minimum size=0.1cm,inner sep=0,above] {3};                     
\end{tikzpicture}
\end{array}
\end{array}
\end{equation}
Therefore, we can conclude that there are really three classes of phenomenological models with NFC in one sector and no null textures.

\textit{Definition (BGL models~\cite{Branco:1996bq}):} Models with NFC in just one sector and FCNCs in the other sector depending only on quark masses and $V_{CKM}$ elements.

\textit{Alternative definition~\cite{Botella:2012ab} (BGL model):} Models with NFC in one sector (up or down) and satisfying the constraint
\begin{equation}\label{BGL2}
  \mathbf{\Gamma}^\dagger_i \mathbf{\Delta}_j=0
  \quad (i\neq j)\,.
\end{equation}

Let us assume, without loss of generality, that we have NFC in the upper sector. Then Eq.~\eqref{BGL2} implies
\begin{equation}
\left(\mathbf{U}_{dR}^\dagger\, \mathbf{\Gamma}_i^\dagger \,\mathbf{U}_{dL}\right)\, \left(\mathbf{V}_{CKM}\,\mathbf{d}_{\Delta_j}\right)=0
\end{equation}
Since the second term $\mathbf{V}_{CKM}\,\mathbf{d}_{\Delta_j}\neq 0$ we can write the first one as $\mathbf{B}^\dagger\mathbf{V}_{CKM}^\dagger$, leading to (up to permutations)
\begin{equation}
\mathbf{B}^\dagger \, \mathbf{d}_{\Delta_j}=0\Leftrightarrow
\left\{
\begin{array}{ll}
\mathbf{B}=A_9&\text{for }\mathbf{d}_{\Delta_j}=(\times,\times,0)\\\\
\mathbf{B}=A_5&\text{for }\mathbf{d}_{\Delta_j}=(0,0\times)
\end{array}
\right.\,,
\end{equation}
where the texture of $\mathbf{d}_{\Delta_j}$ with no zeros is not available for models belonging to $(iii)$, $(v)$ or $(vi)$. In these classes the mixing coming from the NFC sector, the up sector in our case, is block diagonal. We can then conclude that $\mathbf{\Gamma}_i$ has to be a matrix of the set
\begin{equation}
\mathcal{P}_L\left\{A_5,\, A_9\right\}\,.
\end{equation}
The texture $A_5$ belongs to class $\mathbf{(2,i)}$, which implies that only models of class $(iii)$ can be implemented. These models are the BGL implementation in 2HDM~\cite{Branco:1996bq}; see Eq.~\eqref{C121C322}. The other two classes of models, i.e. $(v)$ and $(vi)$, need at least three Higgs bosons in order to have $\text{det}(\mathbf{M}_d)\neq 0$. This would imply that models of class $(v)$ would have two Higgs bosons coupling to the same texture in the up sector but to different textures in the down sector, such a case is not possible by construction. We are left with models of class $vi)$, that as seen in Eq.~\eqref{BGLM3}, can be constructed. This last case corresponds to the 3HDM implementation of BGL presented in~\cite{Botella:2009pq}. We then summarize the possible BGL implementations, in models without the null texture and up to $\mathbf{\Delta}_i\leftrightarrow\mathbf{\Gamma}_i$ exchanges:
\begin{itemize}

\item[(i)] BGL in 2HDM
\begin{equation}\label{BGL2HDM}
\begin{array}{l}
\mathbf{\Delta}_1=\mathcal{P}_L
\begin{pmatrix}
\times&\times&0\\
\times&\times&0\\
0&0&0
\end{pmatrix}\mathcal{P}_R\,,\,
\mathbf{\Delta}_2=\mathcal{P}_L
\begin{pmatrix}
0&0&0\\
0&0&0\\
0&0&\times
\end{pmatrix}\mathcal{P}_R\\
\text{and}\\
\mathbf{\Gamma}_1=\mathcal{P}_L
\begin{pmatrix}
\times&\times&\times\\
\times&\times&\times\\
0&0&0
\end{pmatrix}\,,\,
\mathbf{\Gamma}_2=\mathcal{P}_L
\begin{pmatrix}
0&0&0\\
0&0&0\\
\times&\times&\times
\end{pmatrix}\,.
\end{array}
\end{equation}

\item[(ii)] BGL in 3HDM
\begin{equation}
\begin{array}{l}
\mathbf{\Delta}_1=
\begin{pmatrix}
\times&0&0\\
0&0&0\\
0&0&0
\end{pmatrix}\,,\,
\mathbf{\Delta}_2=
\begin{pmatrix}
0&0&0\\
0&\times&0\\
0&0&0
\end{pmatrix}\,,\\\\
\mathbf{\Delta}_3=
\begin{pmatrix}
0&0&0\\
0&0&0\\
0&0&\times
\end{pmatrix}\quad \text{and}
\\\\
\mathbf{\Gamma}_1=
\begin{pmatrix}
\times&\times&\times\\
0&0&0\\
0&0&0
\end{pmatrix}\,,\,
\mathbf{\Gamma}_2=
\begin{pmatrix}
0&0&0\\
\times&\times&\times\\
0&0&0
\end{pmatrix}\,,\\\\
\mathbf{\Gamma}_3=
\begin{pmatrix}
0&0&0\\
0&0&0\\
\times&\times&\times
\end{pmatrix}\,.
\end{array}
\end{equation}
\end{itemize}
Models with more Higgs bosons, or three bosons for Eq.~\eqref{BGL2HDM}, cannot be BGL since  we will need to repeat textures, and Eq.~\eqref{BGL2} will not be satisfied for the full set of textures. Both of these BGL implementations lead to accidental symmetries in the scalar sector. A possible way out is the addition of extra Higgs doublets having no coupling to quarks (inert-like). 

\subsection{Nearest-neighbour-interaction}

The nearest-neighbour-interaction assumes that the light quarks acquire their masses through an interaction with their nearest neighbors. The mass matrices take the form
\begin{equation}\label{NNItexture}
\mathbf{M}_{u,d}=\begin{pmatrix}
0&\times&0\\
\times&0&\times\\
0&\times&\times
\end{pmatrix}\,.
\end{equation}
There have been many studies on NNI models within the SM~\cite{Fritzsch:1997fw} and extensions~\cite{Ito:1997ke,Fukuyama:2007ri,EmmanuelCosta:2011jq}. In this section we shall look for the minimal implementations of NNI within NHDMs. By minimal we mean that all the $N$ Higgs have nontrivial Yukawa textures associated and different charges under the Abelian group. 

We start by splitting the texture of Eq.~\eqref{NNItexture} into the largest set of non-null textures. We get the following set of textures:
\begin{equation}\label{NNI5}
\left\{\mathcal{P}_{23}A_{12}\mathcal{P}_{13},\,
\mathcal{P}_{13}A_{12}\mathcal{P}_{23},\,
A_{12}\mathcal{P}_{23},\,\mathcal{P}_{23}A_{12}
,\, A_{12}\right\}
\end{equation}
The smallest chains where this set of textures are present are the $\mathcal{P}_{321}C_{13}^{\mathbf{(3,3)}}\mathcal{P}_{123}$ and $\mathcal{P}_{12}C_{17}^{\mathbf{(3,3)}}\mathcal{P}_{321}$, leading to the diagrams
\begin{equation}
\begin{array}{r}
\begin{array}{l}
\begin{tikzpicture}
[scale=.6,auto=left]
 \draw[fill]  (1,10) circle [radius=0.1cm];
 \node [left] at (0.75,10) {$\mathcal{P}_{321}\Gamma_{15}\mathcal{P}_{123}$};
 \draw[fill] (1,9) circle [radius=0.1cm];
  \node [left] at (0.75,9) {$\mathcal{P}_{123}\Gamma_{15}\mathcal{P}_{12}$};
   \draw[fill] (1,8) circle [radius=0.1cm];
  \node [left] at (0.75,8) {$\Gamma_{12}$};
     \draw[fill] (1,7) circle [radius=0.1cm];
  \node [left] at (0.75,7) {$\mathcal{P}_{13}\Gamma_{12}\mathcal{P}_{23}$};
  \draw[fill] (1,6) circle [radius=0.1cm];
  \node [left] at (0.75,6) {$\Gamma_{12}\mathcal{P}_{23}$};
     \draw[fill] (1,5) circle [radius=0.1cm];
  \node [left] at (0.75,5) {$\mathcal{P}_{23}\Gamma_{12}\mathcal{P}_{13}$};
       \draw[fill] (1,4) circle [radius=0.1cm];
  \node [left] at (0.75,4) {$\mathcal{P}_{23}\Gamma_{12}$};
       \draw[fill] (1,3) circle [radius=0.1cm];
  \node [left] at (0.75,3) {$\Gamma_{0}$};
  
   \draw[fill]  (1.5,10) circle [radius=0.1cm];
 \node [right] at (1.75,10) {$\mathcal{P}_{321}\Delta_{15}\mathcal{P}_{123}$};
 \draw[fill] (1.5,9) circle [radius=0.1cm];
  \node [right] at (1.75,9) {$\mathcal{P}_{123}\Delta_{15}\mathcal{P}_{12}$};
   \draw[fill] (1.5,8) circle [radius=0.1cm];
  \node [right] at (1.75,8) {$\Delta_{12}$};
     \draw[fill] (1.5,7) circle [radius=0.1cm];
  \node [right] at (1.75,7) {$\mathcal{P}_{13}\Delta_{12}\mathcal{P}_{23}$};
  \draw[fill] (1.5,6) circle [radius=0.1cm];
  \node [right] at (1.75,6) {$\Delta_{12}\mathcal{P}_{23}$};
     \draw[fill] (1.5,5) circle [radius=0.1cm];
  \node [right] at (1.75,5) {$\mathcal{P}_{23}\Delta_{12}\mathcal{P}_{13}$};
       \draw[fill] (1.5,4) circle [radius=0.1cm];
  \node [right] at (1.75,4) {$\mathcal{P}_{23}\Delta_{12}$};
       \draw[fill] (1.5,3) circle [radius=0.1cm];
  \node [right] at (1.75,3) {$\Delta_{0}$};
\end{tikzpicture}
\end{array}
\begin{array}{l}
=\\\\
\end{array}
\begin{array}{l}
\begin{tikzpicture}
[scale=.6,auto=left,every node/.style={draw,shape=circle,minimum size=0.1cm,inner sep=0}]
  \node[fill] (n1) at (1,10) {};
  \node[fill] (n2) at (1,9) {};
  \node[fill] (n3) at (1.5,10) {};
  \node[fill] (n4) at (1.5,9) {};
   \node[fill] (n5) at (1,8) {};
    \node[fill] (n6) at (1.5,8) {};
     \node[fill] (n7) at (1,7) {};
     \node[fill] (n8) at (1.5,7) {};
      \node[fill] (n9) at (1,6) {};
     \node[fill] (n10) at (1.5,6) {};
      \node[fill] (n11) at (1,5) {};
      \node[fill] (n12) at (1.5,5) {};        
      \node[fill] (n13) at (1,4) {};
      \node[fill] (n14) at (1.5,4) {};               
      \node[fill] (n15) at (1,3) {};
      \node[fill] (n16) at (1.5,3) {};                     
                  
   \foreach \from/\to in {n5/n10,n7/n12,n9/n6,n11/n8,n13/n14}
    \draw[thick] (\from) to (\to);   
     \draw (1.25,8.2) node [draw,shape=circle,minimum size=0.1cm,inner sep=0,above] {8};                     
\end{tikzpicture}\quad
\begin{tikzpicture}
[scale=.6,auto=left,every node/.style={draw,shape=circle,minimum size=0.1cm,inner sep=0}]
  \node[fill] (n1) at (1,10) {};
  \node[fill] (n2) at (1,9) {};
  \node[fill] (n3) at (1.5,10) {};
  \node[fill] (n4) at (1.5,9) {};
   \node[fill] (n5) at (1,8) {};
    \node[fill] (n6) at (1.5,8) {};
     \node[fill] (n7) at (1,7) {};
     \node[fill] (n8) at (1.5,7) {};
      \node[fill] (n9) at (1,6) {};
     \node[fill] (n10) at (1.5,6) {};
      \node[fill] (n11) at (1,5) {};
      \node[fill] (n12) at (1.5,5) {};        
      \node[fill] (n13) at (1,4) {};
      \node[fill] (n14) at (1.5,4) {};               
      \node[fill] (n15) at (1,3) {};
      \node[fill] (n16) at (1.5,3) {};                     
                  
   \foreach \from/\to in {n5/n14,n7/n12,n9/n10,n11/n8,n13/n6}
    \draw[thick] (\from) to (\to);   
     \draw (1.25,8.2) node [draw,shape=circle,minimum size=0.1cm,inner sep=0,above] {8};                     
\end{tikzpicture}
\end{array}
\end{array}
\end{equation}
and
\begin{equation}
\begin{array}{r}
\begin{array}{l}
\begin{tikzpicture}
[scale=.6,auto=left]
 \draw[fill]  (1,10) circle [radius=0.1cm];
 \node [left] at (0.75,10) {$\mathcal{P}_{12}\Gamma_{15}\mathcal{P}_{321}$};
 \draw[fill] (1,9) circle [radius=0.1cm];
  \node [left] at (0.75,9) {$\Gamma_{12}$};
   \draw[fill] (1,8) circle [radius=0.1cm];
  \node [left] at (0.75,8) {$\Gamma_{12}\mathcal{P}_{23}$};
     \draw[fill] (1,7) circle [radius=0.1cm];
  \node [left] at (0.75,7) {$\mathcal{P}_{13}\Gamma_{12}\mathcal{P}_{13}$};
  \draw[fill] (1,6) circle [radius=0.1cm];
  \node [left] at (0.75,6) {$\mathcal{P}_{13}\Gamma_{12}\mathcal{P}_{23}$};
     \draw[fill] (1,5) circle [radius=0.1cm];
  \node [left] at (0.75,5) {$\mathcal{P}_{23}\Gamma_{12}\mathcal{P}_{13}$};
       \draw[fill] (1,4) circle [radius=0.1cm];
  \node [left] at (0.75,4) {$\mathcal{P}_{23}\Gamma_{12}$};
       \draw[fill] (1,3) circle [radius=0.1cm];
  \node [left] at (0.75,3) {$\mathcal{P}_{23}\Gamma_{12}\mathcal{P}_{23}$};
  
   \draw[fill]  (1.5,10) circle [radius=0.1cm];
 \node [right] at (1.75,10) {$\mathcal{P}_{12}\Delta_{15}\mathcal{P}_{321}$};
 \draw[fill] (1.5,9) circle [radius=0.1cm];
  \node [right] at (1.75,9) {$\Delta_{12}$};
   \draw[fill] (1.5,8) circle [radius=0.1cm];
  \node [right] at (1.75,8) {$\Delta_{12}\mathcal{P}_{23}$};
     \draw[fill] (1.5,7) circle [radius=0.1cm];
  \node [right] at (1.75,7) {$\mathcal{P}_{13}\Delta_{12}\mathcal{P}_{13}$};
  \draw[fill] (1.5,6) circle [radius=0.1cm];
  \node [right] at (1.75,6) {$\mathcal{P}_{13}\Delta_{12}\mathcal{P}_{23}$};
     \draw[fill] (1.5,5) circle [radius=0.1cm];
  \node [right] at (1.75,5) {$\mathcal{P}_{23}\Delta_{12}\mathcal{P}_{13}$};
       \draw[fill] (1.5,4) circle [radius=0.1cm];
  \node [right] at (1.75,4) {$\mathcal{P}_{23}\Delta_{12}$};
       \draw[fill] (1.5,3) circle [radius=0.1cm];
  \node [right] at (1.75,3) {$\mathcal{P}_{23}\Delta_{12}\mathcal{P}_{23}$};
\end{tikzpicture}
\end{array}
\begin{array}{l}
=\\\\
\end{array}
\begin{array}{l}
\begin{tikzpicture}
[scale=.6,auto=left,every node/.style={draw,shape=circle,minimum size=0.1cm,inner sep=0}]
  \node[fill] (n1) at (1,10) {};
  \node[fill] (n2) at (1,9) {};
  \node[fill] (n3) at (1.5,10) {};
  \node[fill] (n4) at (1.5,9) {};
   \node[fill] (n5) at (1,8) {};
    \node[fill] (n6) at (1.5,8) {};
     \node[fill] (n7) at (1,7) {};
     \node[fill] (n8) at (1.5,7) {};
      \node[fill] (n9) at (1,6) {};
     \node[fill] (n10) at (1.5,6) {};
      \node[fill] (n11) at (1,5) {};
      \node[fill] (n12) at (1.5,5) {};        
      \node[fill] (n13) at (1,4) {};
      \node[fill] (n14) at (1.5,4) {};               
      \node[fill] (n15) at (1,3) {};
      \node[fill] (n16) at (1.5,3) {};                     
                  
   \foreach \from/\to in {n2/n10,n5/n12,n9/n4,n11/n6,n13/n14}
    \draw[thick] (\from) to (\to);   
     \draw (1.25,9.2) node [draw,shape=circle,minimum size=0.1cm,inner sep=0,above] {4};                     
\end{tikzpicture}
\end{array}
\begin{array}{l}
,\\\\
\end{array}
\end{array}
\end{equation}
respectively. Diagrams where the order of the group was larger than $8$ where discarded. This allows us to state that NNI textures in a five Higgs doublet model can only be implemented with at least a $Z_8$ group.

We now turn to the case of four Higgs doublets. We should join two of the five textures of the previous case, in all possible combinations, and study each case. However, since each texture in a given chain belongs to the same class, we can just look for combinations that belong to class $\mathbf{(3,3)}$, since we always have a texture of the type $A_{12}$. For example, a possible union between to textures of Eq.~\eqref{NNI5} is
\begin{equation}
\mathcal{P}_{23}A_{12}\mathcal{P}_{13}\cup  \mathcal{P}_{13}A_{12}\mathcal{P}_{23}=\mathcal{P}_{321}A_{15}\mathcal{P}_{13} \in \mathbf{(3,3)}\,.
\end{equation}
However, the union
\begin{equation}
\mathcal{P}_{23}A_{12}\mathcal{P}_{13}\cup  \mathcal{P}_{23}A_{12}=\mathcal{P}_{23}A_{11}\mathcal{P}_{12} \in \mathcal{P}_{23}\mathbf{(2,2)}\,\text{or}\,\mathbf{(3,2)}\,,
\end{equation}
does not belong to the same class as the other textures. Doing this procedure for all combinations one finds five distinct cases
\begin{equation}
\begin{array}{rl}
(1):&\left\{\mathcal{P}_{321}A_{15}\mathcal{P}_{13},\,
A_{12}\mathcal{P}_{23},\,\mathcal{P}_{23}A_{12}
,\, A_{12}\right\}\\
(2):&\left\{A_{15}\mathcal{P}_{12},\,\mathcal{P}_{13}A_{12}\mathcal{P}_{23},\,
A_{12}\mathcal{P}_{23},\,\mathcal{P}_{23}A_{12}
\right\}\\
(3):&\left\{\mathcal{P}_{321}A_{15},\,
A_{12}\mathcal{P}_{23},\,\mathcal{P}_{23}A_{12}\mathcal{P}_{13}
,\, A_{12}\right\}\\
(4):&\left\{\mathcal{P}_{12}A_{15},\,
\mathcal{P}_{23}A_{12}\mathcal{P}_{13},\,\mathcal{P}_{23}A_{12}
,\, A_{12}\mathcal{P}_{23}\right\}\\
(5):&\left\{A_{15}\mathcal{P}_{23},\,
\mathcal{P}_{13}A_{12}\mathcal{P}_{23},\,\mathcal{P}_{23}A_{12}\mathcal{P}_{13}
,\, A_{12}\right\}
\end{array}
\end{equation}
The smallest chain that can implement case $(2)$ is $C_{11}^{\mathbf{(3,3)}}\mathcal{P}_{12}$, for case $(3)$ the smallest chains are $C_4^{\mathbf{(3,3)}}\mathcal{P}_{13}$ and $\mathcal{P}_{321}C_{12}^{\mathbf{(3,3)}}\mathcal{P}_{123}$, and for case $(4)$ it is $\mathcal{P}_{12}C_{12}^{\mathbf{(3,3)}}$. The cases $(1)$ and $(5)$ are implemented with larger chains and, therefore, will not be considered. Therefore, the diagram for case (2) is
\begin{equation}
\begin{array}{r}
\begin{array}{l}
\begin{tikzpicture}
[scale=.6,auto=left]
 \draw[fill]  (1,10) circle [radius=0.1cm];
 \node [left] at (0.75,10) {$\Gamma_{15}\mathcal{P}_{12}$};
 \draw[fill] (1,9) circle [radius=0.1cm];
  \node [left] at (0.75,9) {$\mathcal{P}_{321}\Gamma_{15}\mathcal{P}_{12}$};
   \draw[fill] (1,8) circle [radius=0.1cm];
  \node [left] at (0.75,8) {$\mathcal{P}_{13}\Gamma_{15}\mathcal{P}_{13}$};
     \draw[fill] (1,7) circle [radius=0.1cm];
  \node [left] at (0.75,7) {$\mathcal{P}_{13}\Gamma_{12}\mathcal{P}_{23}$};
  \draw[fill] (1,6) circle [radius=0.1cm];
  \node [left] at (0.75,6) {$\mathcal{P}_{23}\Gamma_{12}$};
     \draw[fill] (1,5) circle [radius=0.1cm];
  \node [left] at (0.75,5) {$\Gamma_{12}\mathcal{P}_{23}$};

 \draw[fill]  (1.5,10) circle [radius=0.1cm];
 \node [right] at (1.75,10) {$\Delta_{15}\mathcal{P}_{12}$};
 \draw[fill] (1.5,9) circle [radius=0.1cm];
  \node [right] at (1.75,9) {$\mathcal{P}_{321}\Delta_{15}\mathcal{P}_{12}$};
   \draw[fill] (1.5,8) circle [radius=0.1cm];
  \node [right] at (1.75,8) {$\mathcal{P}_{13}\Delta_{15}\mathcal{P}_{13}$};
     \draw[fill] (1.5,7) circle [radius=0.1cm];
  \node [right] at (1.75,7) {$\mathcal{P}_{13}\Delta_{12}\mathcal{P}_{23}$};
  \draw[fill] (1.5,6) circle [radius=0.1cm];
  \node [right] at (1.75,6) {$\mathcal{P}_{23}\Delta_{12}$};
     \draw[fill] (1.5,5) circle [radius=0.1cm];
  \node [right] at (1.75,5) {$\Delta_{12}\mathcal{P}_{23}$};
\end{tikzpicture}
\end{array}
\begin{array}{l}
=\\\\
\end{array}
\begin{array}{l}
\begin{tikzpicture}
[scale=.6,auto=left,every node/.style={draw,shape=circle,minimum size=0.1cm,inner sep=0}]
  \node[fill] (n1) at (1,10) {};
  \node[fill] (n2) at (1,9) {};
  \node[fill] (n3) at (1.5,10) {};
  \node[fill] (n4) at (1.5,9) {};
   \node[fill] (n5) at (1,8) {};
    \node[fill] (n6) at (1.5,8) {};
     \node[fill] (n7) at (1,7) {};
     \node[fill] (n8) at (1.5,7) {};
      \node[fill] (n9) at (1,6) {};
     \node[fill] (n10) at (1.5,6) {};
      \node[fill] (n11) at (1,5) {};
      \node[fill] (n12) at (1.5,5) {};                   
                  
   \foreach \from/\to in {n1/n12,n11/n3,n7/n10,n9/n8}
    \draw[thick] (\from) to (\to);   
               
\end{tikzpicture}
\end{array}
\end{array}
\end{equation}
In case (3), the chain $\mathcal{P}_{321}C_{12}^{\mathbf{(3,3)}}\mathcal{P}_{123}$ can only be implemented with a group of order larger than the one of chain $C_4^{\mathbf{(3,3)}}\mathcal{P}_{13}$. Therefore, for this case we have  
\begin{equation}
\begin{array}{r}
\begin{array}{l}
\begin{tikzpicture}
[scale=.6,auto=left]
 \draw[fill]  (1,10) circle [radius=0.1cm];
 \node [left] at (0.75,10) {$\Gamma_{13}\mathcal{P}_{13}$};
 \draw[fill] (1,9) circle [radius=0.1cm];
  \node [left] at (0.75,9) {$\mathcal{P}_{321}\Gamma_{15}$};
   \draw[fill] (1,8) circle [radius=0.1cm];
  \node [left] at (0.75,8) {$\Gamma_{12}\mathcal{P}_{23}$};
     \draw[fill] (1,7) circle [radius=0.1cm];
  \node [left] at (0.75,7) {$\mathcal{P}_{23}\Gamma_{12}\mathcal{P}_{13}$};
  \draw[fill] (1,6) circle [radius=0.1cm];
  \node [left] at (0.75,6) {$\Gamma_{12}$};
     \draw[fill] (1,5) circle [radius=0.1cm];
  \node [left] at (0.75,5) {$\mathcal{P}_{13}\Gamma_{12}\mathcal{P}_{13}$};
      
\draw[fill]  (1.5,10) circle [radius=0.1cm];
 \node [right] at (1.75,10) {$\Delta_{13}\mathcal{P}_{13}$};
 \draw[fill] (1.5,9) circle [radius=0.1cm];
  \node [right] at (1.75,9) {$\mathcal{P}_{321}\Delta_{15}$};
   \draw[fill] (1.5,8) circle [radius=0.1cm];
  \node [right] at (1.75,8) {$\Delta_{12}\mathcal{P}_{23}$};
     \draw[fill] (1.5,7) circle [radius=0.1cm];
  \node [right] at (1.75,7) {$\mathcal{P}_{23}\Delta_{12}\mathcal{P}_{13}$};
  \draw[fill] (1.5,6) circle [radius=0.1cm];
  \node [right] at (1.75,6) {$\Delta_{12}$};
     \draw[fill] (1.5,5) circle [radius=0.1cm];
  \node [right] at (1.75,5) {$\mathcal{P}_{13}\Delta_{12}\mathcal{P}_{13}$};
  
\draw (1.25,9.2) node [draw,shape=circle,minimum size=0.1cm,inner sep=0,above] {2};
\end{tikzpicture}
\end{array}
\begin{array}{l}
=\\\\
\end{array}
\begin{array}{l}
\begin{tikzpicture}
[scale=.6,auto=left,every node/.style={draw,shape=circle,minimum size=0.1cm,inner sep=0}]
  \node[fill] (n1) at (1,10) {};
  \node[fill] (n2) at (1,9) {};
  \node[fill] (n3) at (1.5,10) {};
  \node[fill] (n4) at (1.5,9) {};
   \node[fill] (n5) at (1,8) {};
    \node[fill] (n6) at (1.5,8) {};
     \node[fill] (n7) at (1,7) {};
     \node[fill] (n8) at (1.5,7) {};
      \node[fill] (n9) at (1,6) {};
     \node[fill] (n10) at (1.5,6) {};
      \node[fill] (n11) at (1,5) {};
      \node[fill] (n12) at (1.5,5) {};                   
                  
   \foreach \from/\to in {n2/n8,n7/n4,n5/n6,n9/n10}
    \draw[thick] (\from) to (\to);   

     \draw (1.25,5.1) node [draw,shape=circle,minimum size=0.1cm,inner sep=0,above] {6};                                
\end{tikzpicture}\quad
\begin{tikzpicture}
[scale=.6,auto=left,every node/.style={draw,shape=circle,minimum size=0.1cm,inner sep=0}]
  \node[fill] (n1) at (1,10) {};
  \node[fill] (n2) at (1,9) {};
  \node[fill] (n3) at (1.5,10) {};
  \node[fill] (n4) at (1.5,9) {};
   \node[fill] (n5) at (1,8) {};
    \node[fill] (n6) at (1.5,8) {};
     \node[fill] (n7) at (1,7) {};
     \node[fill] (n8) at (1.5,7) {};
      \node[fill] (n9) at (1,6) {};
     \node[fill] (n10) at (1.5,6) {};
      \node[fill] (n11) at (1,5) {};
      \node[fill] (n12) at (1.5,5) {};                   
                  
   \foreach \from/\to in {n2/n8,n7/n4,n5/n10,n9/n6}
    \draw[thick] (\from) to (\to);   

     \draw (1.25,5.1) node [draw,shape=circle,minimum size=0.1cm,inner sep=0,above] {6};                                                 
     \end{tikzpicture}
\end{array}
\end{array}
\end{equation}
Finally, for case (4) we have

\begin{equation}
\begin{array}{r}
\begin{array}{l}
\begin{tikzpicture}
[scale=.6,auto=left]
 \draw[fill]  (1,10) circle [radius=0.1cm];
 \node [left] at (0.75,10) {$\mathcal{P}_{12}\Gamma_{15}$};
 \draw[fill] (1,9) circle [radius=0.1cm];
  \node [left] at (0.75,9) {$\mathcal{P}_{12}\Gamma_{15}\mathcal{P}_{321}$};
   \draw[fill] (1,8) circle [radius=0.1cm];
  \node [left] at (0.75,8) {$\mathcal{P}_{321}\Gamma_{15}\mathcal{P}_{123}$};
     \draw[fill] (1,7) circle [radius=0.1cm];
  \node [left] at (0.75,7) {$\Gamma_{12}\mathcal{P}_{23}$};
  \draw[fill] (1,6) circle [radius=0.1cm];
  \node [left] at (0.75,6) {$\mathcal{P}_{23}\Gamma_{12}$};
     \draw[fill] (1,5) circle [radius=0.1cm];
  \node [left] at (0.75,5) {$\mathcal{P}_{23}\Gamma_{12}\mathcal{P}_{13}$};
      
 \draw[fill]  (1.5,10) circle [radius=0.1cm];
 \node [right] at (1.75,10) {$\mathcal{P}_{12}\Delta_{15}$};
 \draw[fill] (1.5,9) circle [radius=0.1cm];
  \node [right] at (1.75,9) {$\mathcal{P}_{12}\Delta_{15}\mathcal{P}_{321}$};
   \draw[fill] (1.5,8) circle [radius=0.1cm];
  \node [right] at (1.75,8) {$\mathcal{P}_{321}\Delta_{15}\mathcal{P}_{123}$};
     \draw[fill] (1.5,7) circle [radius=0.1cm];
  \node [right] at (1.75,7) {$\Delta_{12}\mathcal{P}_{23}$};
  \draw[fill] (1.5,6) circle [radius=0.1cm];
  \node [right] at (1.75,6) {$\mathcal{P}_{23}\Delta_{12}$};
     \draw[fill] (1.5,5) circle [radius=0.1cm];
  \node [right] at (1.75,5) {$\mathcal{P}_{23}\Delta_{12}\mathcal{P}_{13}$};
\end{tikzpicture}
\end{array}
\begin{array}{l}
=\\\\
\end{array}
\begin{array}{l}
\begin{tikzpicture}
[scale=.6,auto=left,every node/.style={draw,shape=circle,minimum size=0.1cm,inner sep=0}]
  \node[fill] (n1) at (1,10) {};
  \node[fill] (n2) at (1,9) {};
  \node[fill] (n3) at (1.5,10) {};
  \node[fill] (n4) at (1.5,9) {};
   \node[fill] (n5) at (1,8) {};
    \node[fill] (n6) at (1.5,8) {};
     \node[fill] (n7) at (1,7) {};
     \node[fill] (n8) at (1.5,7) {};
      \node[fill] (n9) at (1,6) {};
     \node[fill] (n10) at (1.5,6) {};
      \node[fill] (n11) at (1,5) {};
      \node[fill] (n12) at (1.5,5) {};                   
                  
   \foreach \from/\to in {n1/n8,n7/n3,n9/n12,n11/n10}
    \draw[thick] (\from) to (\to);   
                            
\end{tikzpicture}
\end{array}
\end{array}
\end{equation}

We now study the case of three Higgs doublets. Following the usual procedure one gets seven cases
\begin{equation}
\begin{array}{rl}
(1):&\left\{A_{13}\mathcal{P}_{12},\, A_{12}\mathcal{P}_{23},\,\mathcal{P}_{23}A_{12}\right\}\\
(2):&\left\{\mathcal{P}_{321}A_{15}\mathcal{P}_{13},\, A_{15}\mathcal{P}_{23},\,A_{12}\right\}\\
(3):&\left\{A_{15}\mathcal{P}_{12},\,\mathcal{P}_{321} A_{15},\,A_{12}\mathcal{P}_{23}\right\}\\
(4):&\left\{A_{15}\mathcal{P}_{12},\,A_{15}\mathcal{P}_{23},\,\mathcal{P}_{13}A_{12}\mathcal{P}_{23}\right\}\\
(5):&\left\{\mathcal{P}_{321}A_{15},\,A_{15}\mathcal{P}_{123},\,A_{12}\right\}\\
(6):&\left\{\mathcal{P}_{12}A_{15},\,A_{15}\mathcal{P}_{123},\,\mathcal{P}_{23}A_{12}\right\}\\
(7):&\left\{\mathcal{P}_{12}A_{15},\,A_{15}\mathcal{P}_{23},\,\mathcal{P}_{23}A_{12}\mathcal{P}_{13}\right\}\\
\end{array}
\end{equation} 
In this case, the smallest chain is the one containing the textures of case $(2)$, i.e. $C_3^{\mathbf{3,3}}\mathcal{P}_{13}$. The diagram is

\begin{equation}
\begin{array}{r}
\begin{array}{l}
\begin{tikzpicture}
[scale=.6,auto=left]
 \draw[fill]  (1,10) circle [radius=0.1cm];
 \node [left] at (0.75,10) {$\Gamma_{13}\mathcal{P}_{13}$};
 \draw[fill] (1,9) circle [radius=0.1cm];
  \node [left] at (0.75,9) {$\mathcal{P}_{321}\Gamma_{15}\mathcal{P}_{13}$};
   \draw[fill] (1,8) circle [radius=0.1cm];
  \node [left] at (0.75,8) {$\Gamma_{15}\mathcal{P}_{23}$};
     \draw[fill] (1,7) circle [radius=0.1cm];
  \node [left] at (0.75,7) {$\mathcal{P}_{13}\Gamma_{12}\mathcal{P}_{13}$};
  \draw[fill] (1,6) circle [radius=0.1cm];
  \node [left] at (0.75,6) {$\Gamma_{12}$};
      
 \draw[fill]  (1.5,10) circle [radius=0.1cm];
 \node [right] at (1.75,10) {$\Delta_{13}\mathcal{P}_{13}$};
 \draw[fill] (1.5,9) circle [radius=0.1cm];
  \node [right] at (1.75,9) {$\mathcal{P}_{321}\Delta_{15}\mathcal{P}_{13}$};
   \draw[fill] (1.5,8) circle [radius=0.1cm];
  \node [right] at (1.75,8) {$\Delta_{15}\mathcal{P}_{23}$};
     \draw[fill] (1.5,7) circle [radius=0.1cm];
  \node [right] at (1.75,7) {$\mathcal{P}_{13}\Delta_{12}\mathcal{P}_{13}$};
  \draw[fill] (1.5,6) circle [radius=0.1cm];
  \node [right] at (1.75,6) {$\Delta_{12}$};
\end{tikzpicture}
\end{array}
\begin{array}{l}
=\\\\
\end{array}
\begin{array}{l}
\begin{tikzpicture}
[scale=.6,auto=left,every node/.style={draw,shape=circle,minimum size=0.1cm,inner sep=0}]
  \node[fill] (n1) at (1,10) {};
  \node[fill] (n2) at (1,9) {};
  \node[fill] (n3) at (1.5,10) {};
  \node[fill] (n4) at (1.5,9) {};
   \node[fill] (n5) at (1,8) {};
    \node[fill] (n6) at (1.5,8) {};
     \node[fill] (n7) at (1,7) {};
     \node[fill] (n8) at (1.5,7) {};
      \node[fill] (n9) at (1,6) {};
     \node[fill] (n10) at (1.5,6) {};                  
                  
   \foreach \from/\to in {n2/n4,n5/n10,n9/n6}
    \draw[thick] (\from) to (\to);   
     \draw (1.25,9.1) node [draw,shape=circle,minimum size=0.1cm,inner sep=0,above] {5};                
\end{tikzpicture}
\end{array}
\end{array}
\end{equation}

We now study the case of two Higgs doublets, where one gets a single case
\begin{equation}\label{NNI2}
\left\{A_{13}\mathcal{P}_{12},\, A_{15}\mathcal{P}_{23}\right\}\,.
\end{equation}
The smallest chain to which this case belongs is $\mathcal{P}_{23}C_2^{\mathbf{(3,3)}}\mathcal{P}_{321}$. The corresponding diagram is
\begin{equation}
\begin{array}{r}
\begin{array}{l}
\begin{tikzpicture}
[scale=.6,auto=left]
 \draw[fill]  (1,10) circle [radius=0.1cm];
 \node [left] at (0.75,10) {$\Gamma_{13}\mathcal{P}_{12}$};
 \draw[fill] (1,9) circle [radius=0.1cm];
  \node [left] at (0.75,9) {$\mathcal{P}_{123}\Gamma_{15}\mathcal{P}_{13}$};
   \draw[fill] (1,8) circle [radius=0.1cm];
  \node [left] at (0.75,8) {$\Gamma_{15}\mathcal{P}_{23}$};
     \draw[fill] (1,7) circle [radius=0.1cm];
  \node [left] at (0.75,7) {$\mathcal{P}_{13}\Gamma_{15}\mathcal{P}_{13}$};
      
 \draw[fill]  (1.5,10) circle [radius=0.1cm];
 \node [right] at (1.75,10) {$\Delta_{13}\mathcal{P}_{12}$};
 \draw[fill] (1.5,9) circle [radius=0.1cm];
  \node [right] at (1.75,9) {$\mathcal{P}_{123}\Delta_{15}\mathcal{P}_{13}$};
   \draw[fill] (1.5,8) circle [radius=0.1cm];
  \node [right] at (1.75,8) {$\Delta_{15}\mathcal{P}_{23}$};
     \draw[fill] (1.5,7) circle [radius=0.1cm];
  \node [right] at (1.75,7) {$\mathcal{P}_{13}\Delta_{15}\mathcal{P}_{13}$};
\end{tikzpicture}
\end{array}
\begin{array}{l}
=\\\\
\end{array}
\begin{array}{l}
\begin{tikzpicture}
[scale=.6,auto=left,every node/.style={draw,shape=circle,minimum size=0.1cm,inner sep=0}]
  \node[fill] (n1) at (1,10) {};
  \node[fill] (n2) at (1,9) {};
  \node[fill] (n3) at (1.5,10) {};
  \node[fill] (n4) at (1.5,9) {};
   \node[fill] (n5) at (1,8) {};
    \node[fill] (n6) at (1.5,8) {};
     \node[fill] (n7) at (1,7) {};
     \node[fill] (n8) at (1.5,7) {};
                  
   \foreach \from/\to in {n1/n6,n5/n3}
    \draw[thick] (\from) to (\to);   
     \draw (1.25,7.1) node [draw,shape=circle,minimum size=0.1cm,inner sep=0,above] {4};                
\end{tikzpicture}
\end{array}
\end{array}
\end{equation}
In agreement with~\cite{EmmanuelCosta:2011jq}. We shall now present the symmetry implementation for this last case. For classes with one texture $\mathcal{P}_L A_{13}\mathcal{P}_{R}$, the symmetry generators are -- see section (9) of Appendix~\ref{chainsimple}.
\begin{equation}
\left\{
\begin{array}{l}
\mathcal{S}_L=\mathcal{P}_L\text{diag}\left(1\,,\omega_n^{k_{L1}}\,,\omega_n^{k_{L2}}\right)\mathcal{P}_L^T\\
\mathcal{S}_R=\mathcal{P}_R^T\text{diag}\left(1\,,\omega_n^{k_{L1}}\,,\omega_n^{k_{L2}}\right)\mathcal{P}_R
\end{array}\right.\,.
\end{equation}
The vector charges tell us that the order of the group is $4n$ and $k_{L2}=2k_{L1}=2n$. Since we permute on the left by $\mathcal{P}_{23}$, we get the new identification $k_{L1}=2k_{L2}=2n$, leading to the generator
\begin{equation}
\mathcal{S}_L=\text{diag}\left(1\,,\omega_{4n}^{2n}\,,\omega_{4n}^{n}\right)\,.
\end{equation}
For the generator of the right sector, we just need to do the permutation $1\rightarrow 2\rightarrow 3\rightarrow 1$ in the diagonal elements, leading to
\begin{equation}
\mathcal{S}_R=\text{diag}\left(\omega_{4n}^{2n}\,,1\,,\omega_{4n}^{n}\right)\,.
\end{equation}
Until this point, all the information used was in the chain $\mathcal{P}_{23}C_2^{\mathbf{(3,3)}}\mathcal{P}_{321}$ and its associated vector charge. Now we shall look at the corresponding diagram and extract the last piece of information. The diagram is telling us that the down sector does not connect the first texture with the first texture of the up sector. Since all the vector charges start with the trivial phase, the fact that the first texture is connected with the third one implies an overall phase transformation in the right-handed up sector, in order to put a trivial phase in the third entry. Therefore, the last step is to transform the up-quark right-handed fields with an additional $\omega_{4n}^n$ phase. The model symmetry implementation finally reads
\begin{equation}
\left\{
\begin{array}{l}
\mathcal{S}_L=\text{diag}\left(1\,,\omega_{4n}^{2n}\,,\omega_{4n}^{n}\right)\\
\mathcal{S}^n_R=\text{diag}\left(\omega_{4n}^{2n}\,,1\,,\omega_{4n}^{n}\right)\\
\mathcal{S}^p_R=\text{diag}\left(\omega_{4n}^{3n}\,,\omega_{4n}^{n}\,,\omega_{4n}^{2n}\right)
\end{array}\right.\,\text{and}\quad \mathcal{S}_{H}=\text{diag}\left(1\,,\omega_{4n}^{n}\right)\,.
\end{equation}
with the associated Yukawa sector
\begin{equation}
\begin{array}{l}
\mathbf{\Gamma}_1=
\begin{pmatrix}
0&\times&0\\
\times&0&0\\
0&0&\times
\end{pmatrix}\,,\quad
\mathbf{\Gamma}_2=
\begin{pmatrix}
0&0&0\\
0&0&\times\\
0&\times&0
\end{pmatrix}\,,\\\\
\mathbf{\Delta}_1=
\begin{pmatrix}
0&0&0\\
0&0&\times\\
0&\times&0
\end{pmatrix}\,,\quad
\mathbf{\Delta}_2=
\begin{pmatrix}
0&\times&0\\
\times&0&0\\
0&0&\times
\end{pmatrix}\,.
\end{array}
\end{equation}
In this last example the order of the group had to belong to $4\mathbb{Z}$, which in the minimal case is $4$ and the next-to-minimal $8$. However, the next-to-minimal order is only $8$ when implemented within the chain $\mathcal{P}_{23}C_2^{\mathbf{(3,3)}}\mathcal{P}_{321}$. The textures in Eq.~\eqref{NNI2} are also presented in the chain  $\mathcal{P}_{23}C_3^{\mathbf{(3,3)}}\mathcal{P}_{321}$, which is one texture larger than the previous one, and therefore, not included before when finding the minimal symmetry group.  With this chain we have
\begin{equation}
\begin{array}{r}
\begin{array}{l}
\begin{tikzpicture}
[scale=.6,auto=left]
 \draw[fill]  (1,10) circle [radius=0.1cm];
 \node [left] at (0.75,10) {$\Gamma_{13}\mathcal{P}_{12}$};
 \draw[fill] (1,9) circle [radius=0.1cm];
  \node [left] at (0.75,9) {$\mathcal{P}_{123}\Gamma_{15}\mathcal{P}_{13}$};
   \draw[fill] (1,8) circle [radius=0.1cm];
  \node [left] at (0.75,8) {$\Gamma_{15}\mathcal{P}_{23}$};
     \draw[fill] (1,7) circle [radius=0.1cm];
  \node [left] at (0.75,7) {$\mathcal{P}_{13}\Gamma_{12}\mathcal{P}_{13}$};
    \draw[fill] (1,6) circle [radius=0.1cm];
  \node [left] at (0.75,6) {$\mathcal{P}_{23}\Gamma_{12}\mathcal{P}_{23}$};
      
 \draw[fill]  (1.5,10) circle [radius=0.1cm];
 \node [right] at (1.75,10) {$\Delta_{13}\mathcal{P}_{12}$};
 \draw[fill] (1.5,9) circle [radius=0.1cm];
  \node [right] at (1.75,9) {$\mathcal{P}_{123}\Delta_{15}\mathcal{P}_{13}$};
   \draw[fill] (1.5,8) circle [radius=0.1cm];
  \node [right] at (1.75,8) {$\Delta_{15}\mathcal{P}_{23}$};
     \draw[fill] (1.5,7) circle [radius=0.1cm];
  \node [right] at (1.75,7) {$\mathcal{P}_{13}\Delta_{12}\mathcal{P}_{13}$};
      \draw[fill] (1.5,6) circle [radius=0.1cm];
  \node [right] at (1.75,6) {$\mathcal{P}_{23}\Delta_{12}\mathcal{P}_{23}$};
\end{tikzpicture}
\end{array}
\begin{array}{l}
=\\\\
\end{array}
\begin{array}{l}
\begin{tikzpicture}
[scale=.6,auto=left,every node/.style={draw,shape=circle,minimum size=0.1cm,inner sep=0}]
  \node[fill] (n1) at (1,10) {};
  \node[fill] (n2) at (1,9) {};
  \node[fill] (n3) at (1.5,10) {};
  \node[fill] (n4) at (1.5,9) {};
   \node[fill] (n5) at (1,8) {};
    \node[fill] (n6) at (1.5,8) {};
     \node[fill] (n7) at (1,7) {};
     \node[fill] (n8) at (1.5,7) {};
      \node[fill] (n9) at (1,6) {};
      \node[fill] (n10) at (1.5,6) {};
                  
   \foreach \from/\to in {n1/n6,n5/n3}
    \draw[thick] (\from) to (\to);                
\end{tikzpicture}
\end{array}
\end{array}
\end{equation} 
which allows the NNI texture to be implemented within the THDM  not only with $Z_{4n}$, but also with $Z_{n\geq 5}$ [and in particular $U(1)$]. We summarize in Table~\ref{NNIresume} the minimal NNI implementation groups and the existence or absence of accidental continuous symmetries.

\begin{table}[H]
\begin{center}
\begin{tabular}{r||c|c}
NNI&Minimal group&Accidental symmetry\\
\hline\hline
N=2&$Z_4$&Yes\\
N=3&$Z_5$&Yes\\
N=4&$Z_6$&No\\
N=5&$Z_8$&No\\
\hline\hline
\end{tabular}
\caption{\label{NNIresume}Minimal symmetry implementation in NNI models and the existence of accidental symmetries. }
\end{center}
\end{table}

There is a NNI extension, known as four-zero parallel texture~\cite{Fukuyama:2007ri} and given by
\begin{equation}
\mathbf{M}_{u,d}=\begin{pmatrix}
0&\times&0\\
\times&\times&\times\\
0&\times&\times
\end{pmatrix}\,.
\end{equation}
Up to $N=2$ this texture is an ansatz. The minimal number of Higgs bosons needed to implement this texture through a symmetry is $N=3$. In this case we have three sets of textures allowed:
\begin{equation}
\begin{array}{rl}
(1):&\left\{A_{13}\mathcal{P}_{12},\,A_{15}\mathcal{P}_{23},\,\mathcal{P}_{23}A_{12}\mathcal{P}_{23}\right\}\\
(2):&\left\{\mathcal{P}_{321}A_{15}\mathcal{P}_{13},\, A_{15},\, A_{15}\mathcal{P}_{23}\right\}\\
(3):&\left\{A_{15}\mathcal{P}_{123},\, \mathcal{P}_{321}A_{15},\,A_{15}\right\}
\end{array}\,.
\end{equation}
Case (1) is the one that can be implemented with the smallest chain, which is $\mathcal{P}_{23}C_{3}^{\mathbf{(3,3)}}\mathcal{P}_{321}$. The diagram is given by
\begin{equation}
\begin{array}{r}
\begin{array}{l}
\begin{tikzpicture}
[scale=.6,auto=left]
 \draw[fill]  (1,10) circle [radius=0.1cm];
 \node [left] at (0.75,10) {$\Gamma_{13}\mathcal{P}_{12}$};
 \draw[fill] (1,9) circle [radius=0.1cm];
  \node [left] at (0.75,9) {$\mathcal{P}_{123}\Gamma_{15}\mathcal{P}_{12}$};
   \draw[fill] (1,8) circle [radius=0.1cm];
  \node [left] at (0.75,8) {$\Gamma_{15}\mathcal{P}_{23}$};
     \draw[fill] (1,7) circle [radius=0.1cm];
  \node [left] at (0.75,7) {$\mathcal{P}_{13}\Gamma_{12}\mathcal{P}_{13}$};
    \draw[fill] (1,6) circle [radius=0.1cm];
  \node [left] at (0.75,6) {$\mathcal{P}_{23}\Gamma_{12}\mathcal{P}_{23}$};
      
 \draw[fill]  (1.5,10) circle [radius=0.1cm];
 \node [right] at (1.75,10) {$\Delta_{13}\mathcal{P}_{12}$};
 \draw[fill] (1.5,9) circle [radius=0.1cm];
  \node [right] at (1.75,9) {$\mathcal{P}_{123}\Delta_{15}\mathcal{P}_{12}$};
   \draw[fill] (1.5,8) circle [radius=0.1cm];
  \node [right] at (1.75,8) {$\Delta_{15}\mathcal{P}_{23}$};
     \draw[fill] (1.5,7) circle [radius=0.1cm];
  \node [right] at (1.75,7) {$\mathcal{P}_{13}\Delta_{12}\mathcal{P}_{13}$};
      \draw[fill] (1.5,6) circle [radius=0.1cm];
  \node [right] at (1.75,6) {$\mathcal{P}_{23}\Delta_{12}\mathcal{P}_{23}$};
\end{tikzpicture}
\end{array}
\begin{array}{l}
=\\\\
\end{array}
\begin{array}{l}
\begin{tikzpicture}
[scale=.6,auto=left,every node/.style={draw,shape=circle,minimum size=0.1cm,inner sep=0}]
  \node[fill] (n1) at (1,10) {};
  \node[fill] (n2) at (1,9) {};
  \node[fill] (n3) at (1.5,10) {};
  \node[fill] (n4) at (1.5,9) {};
   \node[fill] (n5) at (1,8) {};
    \node[fill] (n6) at (1.5,8) {};
     \node[fill] (n7) at (1,7) {};
     \node[fill] (n8) at (1.5,7) {};
      \node[fill] (n9) at (1,6) {};
      \node[fill] (n10) at (1.5,6) {};
                  
   \foreach \from/\to in {n1/n10,n9/n3,n5/n6}
    \draw[thick] (\from) to (\to);                
\end{tikzpicture}
\end{array}
\end{array}
\end{equation} 
Therefore, this model can be implemented with any Abelian group of order $n\geq 5$. This model contains an accidental global symmetry in the scalar sector.

\subsection{The $Z_2\times Z_2$ model}
In Sec.~\ref{DPAG} we found that the only chain that can be implemented by direct products of cyclic groups is chain $(1)$ of Eq.~\eqref{chainDPAG}. In this section we shall present the model implementation in more detail. We start by noticing that from the four textures available in the chain we need at least three; otherwise we get at least one vanishing mixing angle. The model implementations are given by

\begin{equation}\label{Z2Z2M}
\begin{array}{r}
\begin{array}{l}
\begin{tikzpicture}
[scale=.6,auto=left]
 \draw[fill]  (1,10) circle [radius=0.1cm];
 \node [left] at (0.75,10) {$\Gamma_{13}$};
 \draw[fill]  (1,9) circle [radius=0.1cm];
  \node [left] at (0.75,9) {$\mathcal{P}_{23}\Gamma_{15}$};
 \draw[fill]  (1,8) circle [radius=0.1cm];
  \node [left] at (0.75,8) {$\mathcal{P}_{123}\Gamma_{15}\mathcal{P}_{12}$};  
   \draw[fill]  (1,7) circle [radius=0.1cm];
  \node [left] at (0.75,7) {$\mathcal{P}_{321}\Gamma_{15}\mathcal{P}_{13}$};
   \draw[fill]  (1.5,10) circle [radius=0.1cm];
   \node [right] at (1.75,10) {$\Delta_{13}$};
   \draw[fill]  (1.5,9) circle [radius=0.1cm];
  \node [right] at (1.75,9) {$\mathcal{P}_{23}\Delta_{15}$};
     \draw[fill]  (1.5,8) circle [radius=0.1cm];
  \node [right] at (1.75,8) {$\mathcal{P}_{123}\Delta_{15}\mathcal{P}_{12}$};
     \draw[fill]  (1.5,7) circle [radius=0.1cm];
  \node [right] at (1.75,7) {$\mathcal{P}_{321}\Delta_{15}\mathcal{P}_{13}$};
\end{tikzpicture}
\end{array}
\begin{array}{l}
=\\\\
\end{array}
\begin{array}{l}
\begin{tikzpicture}
[scale=.6,auto=left,every node/.style={draw,shape=circle,minimum size=0.1cm,inner sep=0}]
  \node[fill] (n1) at (1,10) {};
  \node[fill] (n2) at (1,9) {};
  \node[fill] (n3) at (1.5,10) {};
  \node[fill] (n4) at (1.5,9) {};
  \node[fill] (n5) at (1,8) {};
  \node[fill] (n6) at (1.5,8) {};
  \node[fill] (n7) at (1,7) {};
  \node[fill] (n8) at (1.5,7) {};
   
   \foreach \from/\to in {n1/n3,n2/n4,n5/n6}
    \draw[thick] (\from) -- (\to);
    
\end{tikzpicture}\quad
\begin{tikzpicture}
[scale=.6,auto=left,every node/.style={draw,shape=circle,minimum size=0.1cm,inner sep=0}]
  \node[fill] (n1) at (1,10) {};
  \node[fill] (n2) at (1,9) {};
  \node[fill] (n3) at (1.5,10) {};
  \node[fill] (n4) at (1.5,9) {};
  \node[fill] (n5) at (1,8) {};
  \node[fill] (n6) at (1.5,8) {};
  \node[fill] (n7) at (1,7) {};
  \node[fill] (n8) at (1.5,7) {};
   
   \foreach \from/\to in {n1/n3,n2/n4,n7/n8}
    \draw[thick] (\from) -- (\to);
    
\end{tikzpicture}
\end{array}\\\\
\begin{array}{l}
\begin{tikzpicture}
[scale=.6,auto=left,every node/.style={draw,shape=circle,minimum size=0.1cm,inner sep=0}]
  \node[fill] (n1) at (1,10) {};
  \node[fill] (n2) at (1,9) {};
  \node[fill] (n3) at (1.5,10) {};
  \node[fill] (n4) at (1.5,9) {};
  \node[fill] (n5) at (1,8) {};
  \node[fill] (n6) at (1.5,8) {};
  \node[fill] (n7) at (1,7) {};
  \node[fill] (n8) at (1.5,7) {};
   
   \foreach \from/\to in {n1/n3,n5/n6,n7/n8}
    \draw[thick] (\from) -- (\to);
    
\end{tikzpicture}\quad
\begin{tikzpicture}
[scale=.6,auto=left,every node/.style={draw,shape=circle,minimum size=0.1cm,inner sep=0}]
  \node[fill] (n1) at (1,10) {};
  \node[fill] (n2) at (1,9) {};
  \node[fill] (n3) at (1.5,10) {};
  \node[fill] (n4) at (1.5,9) {};
  \node[fill] (n5) at (1,8) {};
  \node[fill] (n6) at (1.5,8) {};
  \node[fill] (n7) at (1,7) {};
  \node[fill] (n8) at (1.5,7) {};
   
   \foreach \from/\to in {n2/n4,n5/n6,n7/n8}
    \draw[thick] (\from) -- (\to);
    
\end{tikzpicture}\quad
\begin{tikzpicture}
[scale=.6,auto=left,every node/.style={draw,shape=circle,minimum size=0.1cm,inner sep=0}]
  \node[fill] (n1) at (1,10) {};
  \node[fill] (n2) at (1,9) {};
  \node[fill] (n3) at (1.5,10) {};
  \node[fill] (n4) at (1.5,9) {};
  \node[fill] (n5) at (1,8) {};
  \node[fill] (n6) at (1.5,8) {};
  \node[fill] (n7) at (1,7) {};
  \node[fill] (n8) at (1.5,7) {};
   
   \foreach \from/\to in {n1/n4,n2/n3,n5/n8}
    \draw[thick] (\from) -- (\to);
    
\end{tikzpicture}\quad
\begin{tikzpicture}
[scale=.6,auto=left,every node/.style={draw,shape=circle,minimum size=0.1cm,inner sep=0}]
  \node[fill] (n1) at (1,10) {};
  \node[fill] (n2) at (1,9) {};
  \node[fill] (n3) at (1.5,10) {};
  \node[fill] (n4) at (1.5,9) {};
  \node[fill] (n5) at (1,8) {};
  \node[fill] (n6) at (1.5,8) {};
  \node[fill] (n7) at (1,7) {};
  \node[fill] (n8) at (1.5,7) {};
   
   \foreach \from/\to in {n1/n4,n2/n3,n7/n6}
    \draw[thick] (\from) -- (\to);
    
\end{tikzpicture}\quad
\begin{tikzpicture}
[scale=.6,auto=left,every node/.style={draw,shape=circle,minimum size=0.1cm,inner sep=0}]
  \node[fill] (n1) at (1,10) {};
  \node[fill] (n2) at (1,9) {};
  \node[fill] (n3) at (1.5,10) {};
  \node[fill] (n4) at (1.5,9) {};
  \node[fill] (n5) at (1,8) {};
  \node[fill] (n6) at (1.5,8) {};
  \node[fill] (n7) at (1,7) {};
  \node[fill] (n8) at (1.5,7) {};
   
   \foreach \from/\to in {n1/n4,n5/n8,n7/n6}
    \draw[thick] (\from) -- (\to);
    
\end{tikzpicture}\quad
\begin{tikzpicture}
[scale=.6,auto=left,every node/.style={draw,shape=circle,minimum size=0.1cm,inner sep=0}]
  \node[fill] (n1) at (1,10) {};
  \node[fill] (n2) at (1,9) {};
  \node[fill] (n3) at (1.5,10) {};
  \node[fill] (n4) at (1.5,9) {};
  \node[fill] (n5) at (1,8) {};
  \node[fill] (n6) at (1.5,8) {};
  \node[fill] (n7) at (1,7) {};
  \node[fill] (n8) at (1.5,7) {};
   
   \foreach \from/\to in {n2/n3,n5/n8,n7/n6}
    \draw[thick] (\from) -- (\to);
    
\end{tikzpicture}\quad
\begin{tikzpicture}
[scale=.6,auto=left,every node/.style={draw,shape=circle,minimum size=0.1cm,inner sep=0}]
  \node[fill] (n1) at (1,10) {};
  \node[fill] (n2) at (1,9) {};
  \node[fill] (n3) at (1.5,10) {};
  \node[fill] (n4) at (1.5,9) {};
  \node[fill] (n5) at (1,8) {};
  \node[fill] (n6) at (1.5,8) {};
  \node[fill] (n7) at (1,7) {};
  \node[fill] (n8) at (1.5,7) {};
   
   \foreach \from/\to in {n1/n6,n2/n8,n5/n3}
    \draw[thick] (\from) -- (\to);
    
\end{tikzpicture}\quad
\begin{tikzpicture}
[scale=.6,auto=left,every node/.style={draw,shape=circle,minimum size=0.1cm,inner sep=0}]
  \node[fill] (n1) at (1,10) {};
  \node[fill] (n2) at (1,9) {};
  \node[fill] (n3) at (1.5,10) {};
  \node[fill] (n4) at (1.5,9) {};
  \node[fill] (n5) at (1,8) {};
  \node[fill] (n6) at (1.5,8) {};
  \node[fill] (n7) at (1,7) {};
  \node[fill] (n8) at (1.5,7) {};
   
   \foreach \from/\to in {n1/n6,n2/n8,n7/n4}
    \draw[thick] (\from) -- (\to);
    
\end{tikzpicture}
\end{array}\\\\
\begin{array}{l}
\begin{tikzpicture}
[scale=.6,auto=left,every node/.style={draw,shape=circle,minimum size=0.1cm,inner sep=0}]
  \node[fill] (n1) at (1,10) {};
  \node[fill] (n2) at (1,9) {};
  \node[fill] (n3) at (1.5,10) {};
  \node[fill] (n4) at (1.5,9) {};
  \node[fill] (n5) at (1,8) {};
  \node[fill] (n6) at (1.5,8) {};
  \node[fill] (n7) at (1,7) {};
  \node[fill] (n8) at (1.5,7) {};
   
   \foreach \from/\to in {n1/n6,n5/n3,n7/n4}
    \draw[thick] (\from) -- (\to);
    
\end{tikzpicture}\quad
\begin{tikzpicture}
[scale=.6,auto=left,every node/.style={draw,shape=circle,minimum size=0.1cm,inner sep=0}]
  \node[fill] (n1) at (1,10) {};
  \node[fill] (n2) at (1,9) {};
  \node[fill] (n3) at (1.5,10) {};
  \node[fill] (n4) at (1.5,9) {};
  \node[fill] (n5) at (1,8) {};
  \node[fill] (n6) at (1.5,8) {};
  \node[fill] (n7) at (1,7) {};
  \node[fill] (n8) at (1.5,7) {};
   
   \foreach \from/\to in {n2/n8,n5/n3,n7/n4}
    \draw[thick] (\from) -- (\to);
    
\end{tikzpicture}\quad
\begin{tikzpicture}
[scale=.6,auto=left,every node/.style={draw,shape=circle,minimum size=0.1cm,inner sep=0}]
  \node[fill] (n1) at (1,10) {};
  \node[fill] (n2) at (1,9) {};
  \node[fill] (n3) at (1.5,10) {};
  \node[fill] (n4) at (1.5,9) {};
  \node[fill] (n5) at (1,8) {};
  \node[fill] (n6) at (1.5,8) {};
  \node[fill] (n7) at (1,7) {};
  \node[fill] (n8) at (1.5,7) {};
   
   \foreach \from/\to in {n1/n8,n2/n6,n5/n4}
    \draw[thick] (\from) -- (\to);
    
\end{tikzpicture}\quad
\begin{tikzpicture}
[scale=.6,auto=left,every node/.style={draw,shape=circle,minimum size=0.1cm,inner sep=0}]
  \node[fill] (n1) at (1,10) {};
  \node[fill] (n2) at (1,9) {};
  \node[fill] (n3) at (1.5,10) {};
  \node[fill] (n4) at (1.5,9) {};
  \node[fill] (n5) at (1,8) {};
  \node[fill] (n6) at (1.5,8) {};
  \node[fill] (n7) at (1,7) {};
  \node[fill] (n8) at (1.5,7) {};
   
   \foreach \from/\to in {n1/n8,n2/n6,n7/n3}
    \draw[thick] (\from) -- (\to);
    
\end{tikzpicture}\quad
\begin{tikzpicture}
[scale=.6,auto=left,every node/.style={draw,shape=circle,minimum size=0.1cm,inner sep=0}]
  \node[fill] (n1) at (1,10) {};
  \node[fill] (n2) at (1,9) {};
  \node[fill] (n3) at (1.5,10) {};
  \node[fill] (n4) at (1.5,9) {};
  \node[fill] (n5) at (1,8) {};
  \node[fill] (n6) at (1.5,8) {};
  \node[fill] (n7) at (1,7) {};
  \node[fill] (n8) at (1.5,7) {};
   
   \foreach \from/\to in {n1/n8,n5/n4,n7/n3}
    \draw[thick] (\from) -- (\to);
    
\end{tikzpicture}\quad
\begin{tikzpicture}
[scale=.6,auto=left,every node/.style={draw,shape=circle,minimum size=0.1cm,inner sep=0}]
  \node[fill] (n1) at (1,10) {};
  \node[fill] (n2) at (1,9) {};
  \node[fill] (n3) at (1.5,10) {};
  \node[fill] (n4) at (1.5,9) {};
  \node[fill] (n5) at (1,8) {};
  \node[fill] (n6) at (1.5,8) {};
  \node[fill] (n7) at (1,7) {};
  \node[fill] (n8) at (1.5,7) {};
   
   \foreach \from/\to in {n2/n6,n5/n4,n7/n3}
    \draw[thick] (\from) -- (\to);
    
\end{tikzpicture}
\end{array}
\end{array}
\end{equation}
We shall now specify the symmetry implementation of theses models. The chain used was built from the Hadamard product of $C_{1}^{\mathbf{(2,2)}}$ with $\mathcal{P}_{23}C_{1}^{\mathbf{(2,2)}}\mathcal{P}_{23}$. The left generator for each chain is
\begin{equation}\label{genL}
\text{diag}\left(1,1,-1\right)\quad\text{and}\quad\text{diag}\left(1,-1,1\right)\,,
\end{equation}
respectively. The generators of the right sector take the same form as Eq.~\eqref{genL}. Therefore, the final symmetry generators are given by
\begin{equation}
\mathcal{S}^u_R=\mathcal{S}^d_R=\mathcal{S}_L=\text{diag}\left((1,1),\,(1,-1),\,(-1,1)\right)\,.
\end{equation}
This set of generators allows us to build the first 4 diagrams of Eq.~\eqref{Z2Z2M}; the other 12 diagrams are found through global phase transformations in one right-handed sector. The mass matrix $\mathbf{M}_{u,d}$ can take one of the four textures
\begin{equation}
\left\{
\begin{pmatrix}
0&\times&\times\\
\times&0&\times\\
\times&\times&0
\end{pmatrix},
\begin{pmatrix}
\times&0&\times\\
0&\times&\times\\
\times&\times&\times
\end{pmatrix},
\begin{pmatrix}
\times&\times&0\\
\times&\times&\times\\
0&\times&\times
\end{pmatrix},
\begin{pmatrix}
\times&\times&\times\\
\times&\times&0\\
\times&0&\times
\end{pmatrix}
\right\}\,.
\end{equation}
These models are free from accidental symmetries. Other models could be constructed with $Z_2\times Z_2$, for example connecting the chain used here with the one of Eq.~\eqref{chainZ2Z2}.

\section{Conclusions}\label{conclusions}
The presence of Abelian symmetries may restrict considerably the Yukawa textures of NHDM. In this work a general method for determining these textures and their implementations was given. The method allows us to determine all possible model implementations for a given number of Higgs fields. We have mapped all possibilities and presented several specific examples for the case of $N=2$ and $N=3$. It was shown that the number of Higgs fields only dictates the possible model implementations it has no effect on the available textures. This means that all the textures found could be implemented in the SM. However, these would in general lead to unphysical mass matrices. The presence of additional Higgs fields allow us to choose several textures, phenomenologically forbidden in the SM, and keep having no massless quark and 3 mixing angles. Therefore, all possible textures in NHDM were found, turning the construction of flavor models into a straightforward combinatorial problem.

We have also found that, within Abelian symmetries and without inertlike couplings, there are only three types of models with NFC in one sector, where two out of these are BGL-like. All these implementations introduce accidental symmetries in the scalar potential. 

The method presented is not only helpful in order to give a systematic classification of possible NHDM with Abelian symmetries, but it can also be used to find the minimal symmetry implementations giving the mass textures. The example presented was the NNI case. We found all the minimal implementations up to $N=5$ starting from the mass matrix textures; additional Higgs fields will introduce inertlike couplings or textures repetition. We found that only for $N\geq 4$ we are able to avoid accidental symmetries in the scalar sector. 

We have also looked at the possibility of having a direct product of cyclic groups. We found that there was only one single chain, not present in the case of a single cyclic group: the chain generated by $Z_2\times Z_2$. The model implementation of this case was  also presented.

All the work done assumed that the Higgs fields had no inert vacuum. These extra cases can be easily extracted from results present in this work.

\begin{acknowledgments}
I am very grateful to Jo\~ao P. Silva for invaluable discussions and careful revision of the manuscript. This work is funded by the European FEDER, Spanish MINECO, under Grant No. FPA2011-23596.
\end{acknowledgments}

\appendix

\section{Proof of the theorem for NFC in one sector}\label{Proof}

In order to guarantee that the set $\mathcal{A}_{XX}$ is Abelian, we can first look to the $\mathcal{H}_{XX}$ part. We do this by looking at each class of Table~\ref{classesij} and see what combination of textures we are allowed to have. 

For class $\mathbf{(1,1)}$, there is only one texture $A_1$ that has the most general texture and, therefore, can only be present one time, leading to $\mathcal{H}_{XX}$ Abelian and $\mathcal{A}^{\text{off}}_{XX}=\emptyset$ [case $(i)$]. 

For classes $\mathbf{(1,2)},\, \mathbf{(1,3)}$ [and similarly $\mathbf{(2,1)},\,\mathbf{(3,1)}$] we always need more than one of its textures in order to have the determinant different from zero. The presence of more than one of these textures will imply, due to the general form of $\mathcal{H}^a_{L}$ (or $\mathcal{H}^a_{R}$), the noncommutativity of $\mathcal{H}_{LL}$ (or $\mathcal{H}_{RR}$). 

For the class $\mathbf{(2,2)}$, we can split the cases in either with $A_2$ or without $A_2$. In the first case only the textures $A_{12}$ and $A_0$ lead to $\mathcal{H}_{XX}$ Abelian. However, since a chain is a set of disjoint matrices, only $A_0$ survives $\mathcal{A}^{\text{off}}_{XX}=\emptyset$ [case $(ii)$]. For the second case, i.e. without $A_2$, the only combination of textures that leads to a nonzero determinant is $A_7$, $A_{12}$, and $A_0$. The presence of more than one $A_7$ texture would leave $\mathcal{H}_{XX}$ non-Abelian. Therefore, the only nontrivial texture that can be repeated is $A_{12}$ leading to $\mathcal{H}_{XX}$ Abelian and $\mathcal{A}^{\text{off}}_{XX}=\emptyset$ [case $(iii)$].

For the class $\mathbf{(2,3)}$ [or similarly $\mathbf{(3,2)}$], in order to have a nonzero determinant we need at least an $A_{4}$ and an $A_{8}$ texture (with some appropriate permutation on the right). Both of these textures have a $2\times2$ block form for the combination $\mathcal{H}_{L}^a$. Therefore, no Abelian set can be constructed.

In the last class, i.e. $\mathbf{(3,3)}$, $\mathcal{H}_{XX}$ is trivially Abelian. We then need to look at the set $\mathcal{A}_{XX}^{\text{off}}$.

In order to tackle this case, we introduce the $3\times 3$ matrices $P_{ij}$ (not the permutation matrices), where the element $(i,j)$ is one and all other are zero. These matrices satisfy the following relation:
\begin{equation}
P_{ij}\, P_{kl}=P_{il}\,\delta_{jk}\,.
\end{equation}  
Any of the textures in this class can be written as 
\begin{equation}
A=a\,P_{ij}+b\,P_{kl}+c\,P_{mn}\,,\quad (i\neq k\neq m;\, j\neq l\neq n)\,.
\end{equation}
We may calculate the commutator $\left[AA^{\prime\dagger},A^\prime A^\dagger\right]$ and evaluate it to zero;
we get
\begin{align}\label{cases}
\begin{split}
&(aa^\prime)^2(1-\delta_{ii^\prime})\delta_{jj^\prime}
+(bb^\prime)^2(1-\delta_{kk^\prime})\delta_{ll^\prime}\\
+&(cc^\prime)^2(1-\delta_{mm^\prime})\delta_{nn^\prime}
+(ab^\prime)^2(1-\delta_{ik^\prime})\delta_{jl^\prime}\\
+&(ac^\prime)^2(1-\delta_{im^\prime})\delta_{jn^\prime}
+(ba^\prime)^2(1-\delta_{ki^\prime})\delta_{lj^\prime}\\
+&(bc^\prime)^2(1-\delta_{km^\prime})\delta_{lk^\prime}
+(ca^\prime)^2(1-\delta_{mi^\prime})\delta_{nj^\prime}\\
+&(cb^\prime)^2(1-\delta_{mk^\prime})\delta_{nl^\prime}=0
\end{split}
\end{align}
If $a,\,b,\,c$ and $a^\prime,\,b^\prime,\,c^\prime$ are different from zero, then Eq.~\eqref{cases} implies $A^\prime$ with the same texture as $A$. This leads to $\mathcal{A}_{XX}^{\text{off}}=\emptyset$ [case $(iv)$]. If one or two parameters of $A^\prime$ were zero, it would imply a texture with some of the entries that were nonzero to be zero. However, the nonzero final entries would overlap with entries of $A$, which is not possible in a chain.

If one parameter of each texture is zero, for example $a$ and $a^\prime$, we get
\begin{equation}
\begin{split}
&(bb^\prime)^2(1-\delta_{kk^\prime})\delta_{ll^\prime}+(cc^\prime)^2(1-\delta_{mm^\prime})\delta_{nn^\prime}\\
+&(bc^\prime)^2(1-\delta_{km^\prime})\delta_{lk^\prime}+(cb^\prime)^2(1-\delta_{mk^\prime})\delta_{nl^\prime}=0\,.
\end{split}
\end{equation}
This leads to two cases: $A$ and $A^\prime$ with the same texture and $A^\prime$ with one overlapping element. In the first case the mass matrix will have determinant zero, while in the second case the determinant is nonzero. However, since the second case has an overlapping element the parameter associated with that texture must be zero, leaving $A^\prime$ with just one parameter and $\mathcal{A}_{XX}^{\text{off}}=\emptyset$ [case $(v)$].

If two parameters of both matrices are zero, for example $a,\,b,\,a^\prime$, and $b^\prime$, we get
\begin{equation}
(cc^\prime)^2(1-\delta_{mm^\prime})\delta_{nn^\prime}=0\,.
\end{equation}
The only solution that does not lead to determinant zero is $n\neq n^\prime$ and another texture $A^{\prime\prime}$ with $a^{\prime\prime}=b^{\prime\prime}=0$ and $n^{\prime\prime}\neq n\neq n^\prime$, leading to $\mathcal{A}_{XX}^{\text{off}}=\emptyset$ [case $(vi)$].

\section{Symmetry implementation, chains and vectors charge}\label{chainsimple}
\subsection{\textbf{Class} $\mathbf{(1,1)}$}
No symmetry is needed in order to impose the texture $A_1$.

\subsection{\textbf{Class} $\mathbf{(1,2)}$}
The symmetry implementation
\begin{equation}
\left\{
\begin{array}{l}
\mathcal{S}_R=\mathcal{P}^R\text{diag}\left(1,\,1,\,\omega_n^{k_{3}}\right)\mathcal{P}^R\,,\\
\mathcal{S}_L=\text{diag}\left(1,\,1,\,1\right)\,,
\end{array}\right.
\end{equation}
leads to the phase transformation matrix $\Theta_{A_6\mathcal{P}^R}$
\begin{equation}
\frac{2\pi}{n}
\begin{pmatrix}
0&0&k_{3}\\
0&0&k_{3}\\
0&0&k_{3}
\end{pmatrix}\mathcal{P}^R\,.
\end{equation}
Therefore, the chain and its symmetry is
\begin{equation}
\begin{array}{rl}
Z_{n\geq2}:&\left\{A_6\oplus A_{10}\right\}\mathcal{P}^{R}\,,
\end{array}
\end{equation}
with the associated charge vector
\begin{equation}
\left(1,\omega_n^{-k_{3}}\right)\,.
\end{equation}
The tilde is present to remind us that, up to a global rephasing, this $k$ appears only in the $\mathcal{S}_R$ generator.

\subsection{\textbf{Class} $\mathbf{(1,3)}$}
The symmetry implementation
\begin{equation}
\left\{
\begin{array}{l}
\mathcal{S}_R=\mathcal{P}^T\text{diag}\left(\omega^{k_{1}}_n,\,\omega^{k_{2}}_n,\,1\right)\mathcal{P}\,,\\
\mathcal{S}_L=\text{diag}\left(1,\,1,\,1\right)
\end{array}\right.
\end{equation}
leads to transformation matrix $\Theta_{A_{10}\mathcal{P}}$
\begin{equation}
\frac{2\pi}{n}
\begin{pmatrix}
k_{1}&k_{2}&0\\
k_{1}&k_{2}&0\\
k_{1}&k_{2}&0
\end{pmatrix}\,.
\end{equation}
The chain and symmetry is
\begin{equation}
\begin{array}{rl}
Z_{n\geq3}:&\left\{A_{10}\oplus A_{10}\mathcal{P}_{23}\oplus A_{10}\mathcal{P}_{13}\right\}\mathcal{P}
\end{array}\,,
\end{equation}
with the associated charge vector
\begin{equation}
\left(1,\omega_n^{-\tilde{k}_{2}},\omega_n^{-\tilde{k}_{1}}\right)\,.
\end{equation}

\subsection{\textbf{Class} $\mathbf{(2,1)}$}
The symmetry implementation
\begin{equation}
\left\{
\begin{array}{l}
\mathcal{S}_R=\text{diag}\left(1,\,1,\,1\right)\,,\\
\mathcal{S}_L=\mathcal{P}^{L}\text{diag}\left(1,\,1,\,\omega_n^{k_{L2}}\right)\mathcal{P}^{L}\,.
\end{array}\right.
\end{equation}
leads to the phase transformation matrix $\Theta_{\mathcal{P}^LA_5}$
\begin{equation}
\frac{2\pi}{n}.\mathcal{P}^L
\begin{pmatrix}
0&0&0\\
0&0&0\\
k_{L2}&k_{L2}&k_{L2}
\end{pmatrix}\,.
\end{equation}
Therefore, the chain and its symmetry is
\begin{equation}
\begin{array}{rl}
Z_{n\geq2}:&\mathcal{P}^L\left\{A_5\oplus A_9\right\}\,,
\end{array}
\end{equation}
with the associated charge vector
\begin{equation}
\left(1,\omega_n^{k_{L2}}\right)\,.
\end{equation}

\subsection{\textbf{Class} $\mathbf{(2,2)}$}

\subsubsection{With $\mathcal{P}^LA_2\mathcal{P}^R$}
The symmetry implementation is given by
\begin{equation}
\left\{
\begin{array}{l}
\mathcal{S}_R=\mathcal{P}^R\text{diag}\left(1,\,1,\,\omega^{k_{L2}}_n\right)\mathcal{P}^R\,,\\
\mathcal{S}_L=\mathcal{P}^L\text{diag}\left(1,\,1,\,\omega^{k_{L2}}_n\right)\mathcal{P}^L\,,
\end{array}\right.
\end{equation} 
leading to the phase transformation matrix $\Theta_{\mathcal{P}^LA_2\mathcal{P}^R}$
\begin{equation}
\frac{2\pi}{n}.\mathcal{P}^L
\begin{pmatrix}
0&0&k_{L2}\\
0&0&k_{L2}\\
-k_{L2}&-k_{L2}&0
\end{pmatrix}\mathcal{P}^R\,.
\end{equation}
In this case we have two possibilities:

$\bullet$\,\underline{$k_{L2}\neq -k_{L2}$}

This implies $k_{L2}\neq n/2$. The order of the group has to be $n\geq 3$, leading to the chain
\begin{equation}
Z_{n\geq3}:\, \mathcal{P}^L\left\{A_2\oplus A_8\oplus A_{11}\right\}\mathcal{P}^R\,.
\end{equation}
The associated charge vector is
\begin{equation}
\left(1,\omega_n^{-k_{L2}},\omega_n^{k_{L2}}\right)\,.
\end{equation}

$\bullet$\,\underline{$k_{L2}= -k_{L2}$}

This implies $k_{L2}=n/2$. The order of the group has to be $n\in 2\, \mathbb{Z}$, leading to the chain
\begin{equation}
Z_{2n}:\,\mathcal{P}^L\left\{ A_2\oplus A_3\right\}\mathcal{P}^R\,.
\end{equation}
We have made the redefinition $n\rightarrow 2n$. The associated charge vector is
\begin{equation}
\left\{
\begin{array}{l}
\left(1,\omega_{2n}^{n}\right)\,,\\
k_{L2}=n
\end{array}\right.
\end{equation}

\subsubsection{Without $\mathcal{P}^LA_2\mathcal{P}^R$}
The symmetry implementation is given by
\begin{equation}
\left\{
\begin{array}{l}
\mathcal{S}_R=\mathcal{P}^R\text{diag}\left(1,\,1,\,\omega^{k_{3}}_n\right)\mathcal{P}^R\,,\\
\mathcal{S}_L=\mathcal{P}^L\text{diag}\left(1,\,1,\,\omega^{k_{L2}}_n\right)\mathcal{P}^L\,,
\end{array}\right.
\end{equation} 
leading to the phase transformation matrix $\Theta_{\mathcal{P}^LA_7\mathcal{P}^R}$
\begin{equation}
\frac{2\pi}{n}.\mathcal{P}^L
\begin{pmatrix}
0&0&k_3\\
0&0&k_3\\
-k_{L2}&-k_{L2}&k_3-k_{L2}
\end{pmatrix}\mathcal{P}^R
\end{equation}
In this case we have the following possibilities:

$\bullet$\,\underline{$k_3=-k_{L2}$}

This implies $k_{L2}\neq n/2$. The group order has to be $n\geq 3$, leading to the chain
\begin{equation}
Z_{n\geq 3}:\, \mathcal{P}^L\left\{A_7\oplus A_3\oplus A_{12}\right\}\mathcal{P}^R\,,
\end{equation}
with the associated charge vector
\begin{equation}
\left(1,\omega_n^{k_{L2}},\omega_n^{2k_{L2}}\right)\,.
\end{equation}

$\bullet$\,\underline{$k_3\neq-k_{L2}$}

The group order has to be $n\geq 4$, leading to the chain
\begin{equation}
Z_{n\geq 4}:\, \mathcal{P}^L\left\{A_7\oplus A_8\oplus A_{11}\oplus A_{12}\right\}\mathcal{P}^R\,.
\end{equation}
The associated charge vector is 
\begin{equation}
\left(1,\omega_{n}^{-k_3},\omega_{n}^{k_{L2}},\omega_{n}^{k_{L2}-k_3}\right)\,.
\end{equation}

\subsection{\textbf{Class} $\mathbf{(2,3)}$}

\subsubsection{With at least one matrix of the form $\mathcal{P}^LA_{4}\mathcal{P}$}
The symmetry implementation is given by
\begin{equation}
\left\{
\begin{array}{l}
\mathcal{S}_R=\mathcal{P}^T\text{diag}\left(\omega^{k_{1}},\,\omega^{k_{L2}},\,1\right)\mathcal{P}\,,\\
\mathcal{S}_L=\mathcal{P}^L
\text{diag}\left(1,\,1,\,\omega^{k_{L2}}\right)\mathcal{P}^L\,,
\end{array}\right.
\end{equation} 
leading to the phase transformation matrix $\Theta_{\mathcal{P}^LA_{4}\mathcal{P}}$
\begin{equation}
\frac{2\pi}{n}.\mathcal{P}^L
\begin{pmatrix}
k_{1}&k_{L2}&0\\
k_{1}&k_{L2}&0\\
k_{1}-k_{L2}&0&-k_{L2}
\end{pmatrix}\mathcal{P}\,,
\end{equation}
We now have the following cases:

$\bullet$\, Another two matrices from $\mathcal{P}^LA_{4}\mathcal{P}$

This leads to $k_{1}=-k_{L2}$ and $k_{L2}=k_1-k_{L2}$. This implies $k_{L2}=n/3$. The order of the group has to be $n\in 3\, \mathbb{Z}$, leading to the chain
\begin{equation}
Z_{3n}:
\mathcal{P}^L\left\{A_{4}\oplus A_{4}\mathcal{P}_{321}
\oplus A_{4}\mathcal{P}_{123}\right\}\mathcal{P}\,.
\end{equation}
The associated charge vector is
\begin{equation}
\left\{
\begin{array}{l}
\left(1,\omega_{3n}^{2n},\omega_{3n}^{n}\right)\,,\\
k_{L2}=2n
\end{array}\right.
\end{equation}

$\bullet$\, Another matrix from $\mathcal{P}^LA_{4}\mathcal{P}$

We get two cases:
\begin{itemize}
\item[(1)] \underline{$k_{L2}=-k_{L2}$}

This implies $k_{L2}=n/2$. The order of the group has to be $n\in 2\, \mathbb{Z}$ with $n\geq 4$, leading to the chain
\begin{equation}
Z_{2(n+1)}:\,
\mathcal{P}^L\left\{A_{4}\oplus A_{4}\mathcal{P}_{23}\oplus
A_{8}\mathcal{P}_{13}\oplus A_{12}\mathcal{P}_{13}\right\}\mathcal{P},
\end{equation}
with the associated charge vector
\begin{equation}
\left\{
\begin{array}{l}
\left(1,\omega_{2(n+1)}^{n+1},\omega_{2(n+1)}^{-k_{1}},\omega_{2(n+1)}^{-k_{1}+n+1}\right)\,,\\
k_{L2}=n+1
\end{array}\right.
\end{equation}

\item[(2)] \underline{$k_{L2}=k_1-k_{L2}$}

This implies $k_{1}=2k_{L2}$ and $k_{L2}\neq n/2,\, n/3$. The order of the group has to be $n\geq 4$, leading to the chain
\begin{equation}
Z_{n\geq 4}:\, \mathcal{P}^L\left\{A_{4}\oplus A_{4}\mathcal{P}_{123}\oplus A_{8}\mathcal{P}_{13}
\oplus A_{12}\right\}\mathcal{P}\,.
\end{equation}
The associated charge vector is
\begin{equation}
\left(1,\omega_{n}^{-k_{L2}},\omega_{n}^{-2k_{L2}},\omega_{n}^{k_{L2}}\right)\,.
\end{equation}

\item[(3)] \underline{$k_{1}=-k_{L2}$}

This implies $k_{L2}\neq n/2,\, n/3$. The order of the group has to be $n\geq 4$, leading to the chain
\begin{equation}
Z_{n\geq 4}:\, \mathcal{P}^L\left\{A_{4}\oplus A_{4}\mathcal{P}_{321}\oplus A_{12}\mathcal{P}_{13}\oplus A_{8}\mathcal{P}_{23}
\right\}\mathcal{P}\,.
\end{equation}
The associated charge vector is
\begin{equation}
\left(1,\omega_{n}^{k_{L2}},\omega_{n}^{2k_{L2}},\omega_{n}^{-k_{L2}}\right)\,.
\end{equation}
 
\end{itemize}

$\bullet$\, No additional matrices from $A_{4}\mathcal{P}$

The only possibility is with all the $k's$ different. It implies $k_{L2}\neq n/2$ and $k_{1}\neq-k_{L2}$. The order of the group has to be $n\geq 5$, leading to the chain
\begin{equation}
Z_{n\geq 5}:\,\mathcal{P}^L\left\{A_{4}\oplus A_{8}\mathcal{P}_{23}\oplus A_{8}\mathcal{P}_{13}
\oplus A_{12}\oplus A_{12}\mathcal{P}_{13}\right\}\mathcal{P}\,.
\end{equation}
The associated charge vector is
\begin{equation}
\left(1,\omega_n^{-k_{L2}},\omega_n^{-k_{1}},\omega_n^{k_{L2}},\omega_{n}^{k_{L2}-k_{1}}\right)
\end{equation}

\subsubsection{Without a matrix of the form $\mathcal{P}^LA_{4}\mathcal{P}$ and at least one form $\mathcal{P}^LA_{8}\mathcal{P}$}
The symmetry implementation is given by
\begin{equation}
\left\{
\begin{array}{l}
\mathcal{S}_{R}=\mathcal{P}^T\text{diag}\left(\omega_n^{k_{1}},\,\omega_n^{k_2},\,1\right)\mathcal{P}\,,\\
\mathcal{S}_{L}=\mathcal{P}^L\text{diag}\left(1,\,1,\,\omega_n^{k_{L2}}\right)\mathcal{P}^L
\end{array}\right.
\end{equation}
leading to the phase transformation matrix $\Theta_{\mathcal{P}^LA_{8}\mathcal{P}}$ 
\begin{equation}
\frac{2\pi}{n}.\mathcal{P}^L
\begin{pmatrix}
k_{1}&k_{2}&0\\
k_{1}&k_{2}&0\\
k_1-k_{L2}&k_2-k_{L2}&-k_{L2}
\end{pmatrix}\mathcal{P}
\end{equation}
The only case possible is with all $k's$ different. The order of the group has to be $n\geq 6$, leading to the chain
\begin{equation}
\begin{array}{rl}
Z_{n\geq 6}:&\mathcal{P}^L \left\{A_{8}\oplus A_{8}\mathcal{P}_{23}
\oplus A_{8}\mathcal{P}_{13}\oplus A_{12}\right.\\
&\quad\left.\oplus A_{12}\mathcal{P}_{23}
\oplus A_{12}\mathcal{P}_{13}\right\}\mathcal{P}\,.
\end{array}
\end{equation}
The associated charge vector is
\begin{equation}
\left(1,\omega_{n}^{-k_2},\omega_n^{-k_{1}},\omega_{n}^{k_{L2}},\omega_{n}^{k_{L2}-k_2},\omega_{n}^{k_{L2}-k_{1}}\right)\,.
\end{equation}

\subsection{\textbf{Class} $\mathbf{(3,1)}$}
The symmetry implementation is
\begin{equation}
\left\{
\begin{array}{l}
\mathcal{S}_R=\omega_n^{k_{L2}}\text{diag}\left(1,\,1,\,1\right)\,,\\
\mathcal{S}_L=\text{diag}\left(1,\,\omega_n^{k_{L1}},\,\omega_n^{k_{L2}}\right)\,,
\end{array}\right.
\end{equation}
leading to the phase transformation matrix $\Theta_{A_9}$
\begin{equation}
\frac{2\pi}{n}.\mathcal{P}^\prime
\begin{pmatrix}
k_{L2}&k_{L2}&k_{L2}\\
k_{L2}-k_{L1}&k_{L2}-k_{L1}&k_{L2}-k_{L1}\\
0&0&0
\end{pmatrix}\,.
\end{equation}
Therefore, the chain and its symmetry is
\begin{equation}
\begin{array}{rl}
Z_{n\geq 3}:&A_9\oplus\mathcal{P}_{23}A_9
\oplus\mathcal{P}_{13}A_9\,,
\end{array}
\end{equation}
with the associated charge vector
\begin{equation}
\left(1,\omega_{n}^{k_{L1}-k_{L2}},\omega_n^{-k_{L2}}\right)\,.
\end{equation}

\subsection{\textbf{Class} $\mathbf{(3,2)}$}

\subsubsection{With at least one matrix of the form $\mathcal{P}^\prime A_{14}\mathcal{P}^R$}
The symmetry implementation in order to obtain $A_{14}\mathcal{P}^R$ is given by
\begin{equation}
\left\{
\begin{array}{l}
\mathcal{S}_R=\mathcal{P}^R\text{diag}\left(\omega_n^{k_{L1}},\,\omega_n^{k_{L1}},\,\omega_n^{k_{L2}}\right)\mathcal{P}^R\,,\\
\mathcal{S}_L=\text{diag}\left(1,\,\omega_n^{k_{L1}},\,\omega_n^{k_{L2}}\right)\,,
\end{array}\right.
\end{equation}
leading to the phase transformation matrix $\Theta_{A_{14}\mathcal{P}^R}$
\begin{equation}
\frac{2\pi}{n}
\begin{pmatrix}
k_{L1}&k_{L1}&k_{L2}\\
0&0&k_{L2}-k_{L1}\\
k_{L1}-k_{L2}&k_{L1}-k_{L2}&0
\end{pmatrix}\mathcal{P}^R\,.
\end{equation}
This allows for the following three cases:

$\bullet$\, Another two matrices from $\mathcal{P}A_{14}\mathcal{P}^R$

This leads to $k_{L2}=k_{L1}-k_{L2}$ and $k_{L1}=k_{L2}-k_{L1}$. This implies $k_{L2}=2k_{L1}$ and $k_{L1}=n/3$. The order of the group has to be $n\in 3\, \mathbb{Z}$, leading to the chain
\begin{equation}
Z_{3n}:
\left\{A_{14}\oplus\mathcal{P}_{123}A_{14}
\oplus\mathcal{P}_{321}A_{14}\right\}\mathcal{P}^R
\end{equation}
and the associated charge vector
\begin{equation}
\left\{
\begin{array}{l}
\left(1,\omega_{3n}^{2n},\omega_{3n}^{n}\right)\,,\\
k_{L1}=2k_{L2}=2n
\end{array}\right.
\end{equation}

$\bullet$\, Another matrix from $\mathcal{P}A_{14}\mathcal{P}^R$

We get two cases:
\begin{itemize}
\item[(1)] \underline{$k_{L1}-k_{L2}=k_{L2}-k_{L1}$}

This implies $2k_{L1}=2k_{L2}$, or $k_{L2}=k_{L1}+n/2$. The order of the group has to be $n\in 2\, \mathbb{Z}$ with $n\geq 4$, leading to the chain
\begin{equation}
Z_{2(n+1)}:
\left\{A_{14}\oplus\mathcal{P}_{23}A_{14}\oplus
\mathcal{P}_{13}A_{11}\oplus\mathcal{P}_{13}A_{12}\right\}\mathcal{P}^R
\end{equation}
The associated charge vector is 
\begin{equation}
\left\{
\begin{array}{l}
\left(1,\omega_{2(n+1)}^{n+1},\omega_{2(n+1)}^{-k_{L1}},\omega_{2(n+1)}^{-k_{L1}+n+1}\right)\,,\\
k_{L2}=k_{L1}+n+1
\end{array}\right.
\end{equation}

\item[(2)] \underline{$k_{L1}-k_{L2}=k_{L2}$}

This implies $k_{L1}=2k_{L2}$ and $k_{L2}\neq n/2,\, n/3$. The order of the group has to be $n\geq 4$, leading to the chain
\begin{equation}
Z_{n\geq 4}:\,\left\{ A_{14}\oplus \mathcal{P}_{123}A_{14}\oplus\mathcal{P}_{13}A_{11}
\oplus\mathcal{P}_{23}A_{12}\right\}\mathcal{P}^R\,.
\end{equation}
The associated charge vector is 
\begin{equation}
\left\{
\begin{array}{l}
\left(1,\omega_{n}^{-k_2},\omega_{n}^{-2k_2},\omega_{n}^{k_2}\right)\,,\\ k_{L1}=2k_{L2}
\end{array}\right.
\end{equation}

\item[(3)] \underline{$k_{L1}=k_{L2}-k_{L1}$}

This implies $k_{L2}=2k_{L1}$ and $k_{L1}\neq n/2,\, n/3$. The order of the group has to be $n\geq 4$, leading to the chain

\begin{equation}
Z_{n\geq 4}:\,\left\{A_{14}\oplus \mathcal{P}_{321}A_{14}
\oplus\mathcal{P}_{13}A_{12}\oplus A_{11}\right\}\mathcal{P}^R\,.
\end{equation}
The associated charge vector is 
\begin{equation}
\left\{
\begin{array}{l}
\left(1,\omega_{n}^{-k_1},\omega_{n}^{-2k_1},\omega_{n}^{k_1}\right)\,,\\ k_{L2}=2k_{L1}
\end{array}\right.
\end{equation}
\end{itemize}

$\bullet$\, No additional matrices from $\mathcal{P}A_{14}\mathcal{P}^R$

The only possibility is with all the entries different. The order of the group has to be $n\geq 5$, leading to the chain
\begin{equation}
\begin{array}{rl}
Z_{n\geq 5}:&\left\{A_{14}\oplus A_{11}\oplus\mathcal{P}_{13}A_{11}
\oplus\mathcal{P}_{23}A_{12}\right.\\
&\quad\left.\oplus\mathcal{P}_{13}A_{12}\right\}
\mathcal{P}^R
\end{array}
\end{equation}
and the associated charge vector
\begin{equation}
\left(1,\omega_n^{k_{L2}-k_{L1}},\omega_n^{-k_{L1}},\omega_{n}^{k_{L1}-k_{L2}},\omega_n^{-k_{L2}}\right)\,.
\end{equation}

\subsubsection{Without a matrix of the form $\mathcal{P}^\prime A_{14}\mathcal{P}^R$ and at least one form $\mathcal{P}^\prime A_{11}\mathcal{P}^R$}

The implementation is given by
\begin{equation}
\left\{
\begin{array}{l}
\mathcal{S}_{R}=\mathcal{P}^R\text{diag}\left(\omega^{k_{L2}}_n,\,\omega^{k_{L2}}_n,\,\omega_n^{k_3}\right)\mathcal{P}^R\,,\\
\mathcal{S}_{L}=\text{diag}\left(1,\,\omega^{k_{L1}}_n,\,\omega^{k_{L2}}_n\right)
\end{array}\right.
\end{equation}
leading to the phase transformation matrix $\Theta_{ A_{11}\mathcal{P}^R}$
\begin{equation}
\frac{2\pi}{n}
\begin{pmatrix}
k_{L2}&k_{L2}&k_3\\
k_{L2}-k_{L1}&k_{L2}-k_{L1}&k_{3}-k_{L1}\\
0&0&k_{3}-k_{L2}
\end{pmatrix}\mathcal{P}^R
\end{equation}
The only case possible is with all $k's$ different. The order of the group has to be $n\geq 6$, leading to the chain
\begin{equation}
\begin{array}{rl}
Z_{n\geq 6}:& \left\{A_{11}\oplus\mathcal{P}_{23}A_{11}
\oplus\mathcal{P}_{13}A_{11}\oplus A_{12}\right.\\
&\quad\left.
\oplus\mathcal{P}_{23}A_{12}\oplus\mathcal{P}_{13}A_{12}\right\}\mathcal{P}^R
\end{array}
\end{equation}
and the associated charge vector
\begin{equation}
\left(1,\omega_n^{k_{L1}-k_{L2}},\omega_n^{-k_{L2}},\omega_n^{k_{L2}-k_3},\omega_n^{k_{L1}-k_{3}},\omega_n^{-k_3}\right)
\end{equation}

\subsection{\textbf{Class} $\mathbf{(3,3)}$}\label{class33}


\subsubsection{With one matrix belonging to $\mathcal{P}^\prime A_{13}\mathcal{P}$}

This can be implemented through the symmetry generators
\begin{equation}
\left\{
\begin{array}{l}
\mathcal{S}_R=\mathcal{P}^T\text{diag}\left(1,\,\omega_n^{k_{L1}},\,\omega_n^{k_{L2}}\right)\mathcal{P}\\
\mathcal{S}_L=\text{diag}\left(1,\,\omega_n^{k_{L1}},\,\omega_n^{k_{L2}}\right)
\end{array}
\right.
\end{equation}
leading to the phase transformation matrix $\Theta_{A_{13}\mathcal{P}}$
\begin{equation}
\frac{2\pi}{n}%
\mathcal{P}
\end{equation}
We now have the following possibilities:

$\bullet$\, Another disjoint matrix from $A_{13}\mathcal{P}$.

This imposes the additional condition $k_{L1}=k_{L2}-k_{L1}=-k_{L2}$. This implies $k_{L1}=n/3$ (or $k_{L2}=n/3$) and, therefore, $n\in 3\,\mathbb{Z}$. The chain is given by
\begin{equation}
Z_{3n}:
\left\{A_{13}\oplus\mathcal{P}_{123}A_{13}
\oplus\mathcal{P}_{321}A_{13}\right\}\mathcal{P}
\end{equation}
with the associated charge vector
\begin{equation}
\left\{
\begin{array}{l}
\left(1,\omega_{3n}^{n},\omega_{3n}^{2n}\right)\,,\\
\pm k_{L1}=\mp k_{L2}=n
\end{array}\right.
\end{equation}

$\bullet$\, Three matrices from $\mathcal{P}^\prime A_{15}\mathcal{P}$

\begin{itemize}

\item[(1)] \underline{$k_{L1}=k_{L2}-k_{L1},\, k_{L2}=-k_{L2}$}

This implies $k_{L2}=n/2$ and $k_{L1}=n/4$. The order of the group has to be $n\in 4\,\mathbb{Z}$, leading to the chain
\begin{equation}
\begin{array}{rl}
Z_{4n}:&\left\{A_{13}\oplus\mathcal{P}_{321}A_{15}\oplus A_{15}\mathcal{P}_{123}\right.\\
&\quad\left.
\oplus\mathcal{P}_{123}A_{15}\mathcal{P}_{12}\right\}\mathcal{P}
\end{array}
\end{equation}
with the associated charge vector
\begin{equation}
\left\{
\begin{array}{l}
\left(1,\omega_{4n}^{3n},\omega_{4n}^{n},\omega_{4n}^{2n}\right)\,,\\
k_{L2}=2k_{L1}=2n
\end{array}\right.
\end{equation}

\item[(2)] \underline{$k_{L2}-k_{L1}=-k_{L2},\,k_{L1}=-k_{L1}$}

This implies $k_{L1}=n/2$ and $k_{L2}=n/4$. The order of the group has to be $n\in 4\,\mathbb{Z}$, leading to the chain
\begin{equation}
\begin{array}{rl}
Z_{4n}:&\left\{A_{13}\oplus\mathcal{P}_{123}A_{15}\oplus A_{15}\mathcal{P}_{321}\right.\\
&\quad\left.
\oplus\mathcal{P}_{321}A_{15}\mathcal{P}_{13}\right\}\mathcal{P}
\end{array}
\end{equation}
with the associated charge vector
\begin{equation}
\left\{
\begin{array}{l}
\left(1,\omega_{4n}^{3n},\omega_{4n}^{n},\omega_{4n}^{2n}\right)\,,\\
k_{L1}=2k_{L2}=2n
\end{array}\right.
\end{equation}
It can be obtained from case $(1)$ when $\mathcal{P}_L=\mathcal{P}=\mathcal{P}_{12}$.

\item[(3)] \underline{$k_{L1}=-k_{L2},\,k_{L2}-k_{L1}=k_{L1}-k_{L2}$}

This implies $k_{L1}=n/4$ (or $k_{L2}=n/4$) and $k_{L1}=-k_{L2}$. The order of the group has to be $n\in 4\,\mathbb{Z}$, leading to the chain
\begin{equation}
\begin{array}{rl}
Z_{4n}:&\left\{A_{13}\oplus\mathcal{P}_{13}A_{15}\mathcal{P}_{12}\oplus \mathcal{P}_{12}A_{15}\mathcal{P}_{13}\right.\\
&\quad\left.
\oplus\mathcal{P}_{23}A_{15}\right\}\mathcal{P}
\end{array}
\end{equation}
with the associated charge vector
\begin{equation}
\left\{
\begin{array}{l}
\left(1,\omega_{4n}^{3n},\omega_{4n}^{n},\omega_{4n}^{2n}\right)\,,\\
\pm k_{L1}=\mp k_{L2}=n
\end{array}\right.
\end{equation}
It can be obtained from case $(1)$ when $\mathcal{P}_L=\mathcal{P}=\mathcal{P}_{23}$.

\end{itemize}

$\bullet$\, Two matrices from $\mathcal{P}^\prime A_{15}\mathcal{P}$

\begin{itemize}

\item[(1)] \underline{$k_{L1}=k_{L2}-k_{L1}$}

This implies $k_{L2}=2k_{L1}$ and $k_{L1}\neq n/2,\,n/4$. The order of the group has to be $n\geq 5$, leading to the chain
\begin{equation}
\begin{array}{rl}
Z_{n\geq5}:&\left\{
A_{13}\oplus\mathcal{P}_{321}A_{15}\oplus A_{15}\mathcal{P}_{123}
\right.\\
&\quad\left.\oplus\mathcal{P}_{13}A_{12}\oplus A_{12}\mathcal{P}_{13}\right\}\mathcal{P}\,.
\end{array}
\end{equation}
The associated charge vector is
\begin{equation}
\left\{
\begin{array}{l}
\left(1,\omega_n^{-k_{L1}},\omega_n^{k_{L1}},\omega_n^{-2k_{L1}},\omega_n^{2k_{L1}}\right)\,,\\
k_{L2}=2k_{L1}
\end{array}\right.
\end{equation}

\item[(2)] \underline{$k_{L2}=k_{L1}-k_{L2}$}

This implies $k_{L1}=2k_{L2}$ and $k_{L2}\neq n/2,\,n/4$. The order of the group has to be $n\geq 5$, leading to the chain
\begin{equation}
\begin{array}{rl}
Z_{n\geq5}:&\left\{
A_{13}\oplus\mathcal{P}_{123}A_{15}\oplus A_{15}\mathcal{P}_{321}
\right.\\
&\quad\left.\oplus\mathcal{P}_{13}A_{12}\mathcal{P}_{23}\oplus \mathcal{P}_{23}A_{12}\mathcal{P}_{13}\right\}\mathcal{P}\,.
\end{array}
\end{equation}
The associated charge vector is
\begin{equation}
\left\{
\begin{array}{l}
\left(1,\omega_n^{-k_{L2}},\omega_n^{k_{L2}},\omega_n^{-2k_{L2}},\omega_n^{2k_{L2}}\right)\,,\\
k_{L1}=2k_{L2}
\end{array}\right.
\end{equation}
It can be obtained from case $(1)$ when $\mathcal{P}_L=\mathcal{P}=\mathcal{P}_{23}$.

\item[(3)] \underline{$k_{L2}=-k_{L1}$}

The order of the group has to be $n\geq 5$, leading to the chain
\begin{equation}
\begin{array}{rl}
Z_{n\geq5}:&\left\{
A_{13}\oplus\mathcal{P}_{12}A_{15}\mathcal{P}_{13}\oplus \mathcal{P}_{13}A_{15}\mathcal{P}_{12}
\right.\\
&\quad\left.\oplus A_{12}\mathcal{P}_{23}\oplus \mathcal{P}_{23}A_{12}\right\}\mathcal{P}\,.
\end{array}
\end{equation}
The associated charge vector is
\begin{equation}
\left\{
\begin{array}{l}
\left(1,\omega_n^{-k_{L1}},\omega_n^{k_{L1}},\omega_n^{-2k_{L1}},\omega_n^{2k_{L1}}\right)\,,\\
k_{L2}=-k_{L1}
\end{array}\right.
\end{equation}
It can be obtained from case $(1)$ when $\mathcal{P}_L=\mathcal{P}=\mathcal{P}_{12}$.

\end{itemize}

$\bullet$\, One matrix from $\mathcal{P}^\prime A_{15}\mathcal{P}$

\begin{itemize}

\item[(1)] \underline{$k_{L1}=-k_{L1}$}

This implies $k_{L1}=n/2$. The order of the group has to be $n\in 2\,\mathbb{Z}$, leading to the chain
\begin{equation}
\begin{array}{rl}
Z_{2(n+2)}:&\left\{ A_{13}\oplus\mathcal{P}_{321}A_{15}\mathcal{P}_{13}\oplus
A_{12}\mathcal{P}_{23}\right.\\
&\quad\left.\oplus\mathcal{P}_{23}A_{12}\oplus
A_{12}\mathcal{P}_{13}\oplus\mathcal{P}_{13}A_{12}\right\}\mathcal{P}
\end{array}\,.
\end{equation} 
The associated charge vector is
\begin{equation}
\left\{
\begin{array}{l}
\left(1,\omega_{2(n+2)}^{n+2},\omega_{2(n+2)}^{k_{L2}+n+2},\omega_{2(n+2)}^{-k_{L2}+n+2},\omega_{2(n+2)}^{k_{L2}},\omega_{2(n+2)}^{-k_{L2}}\right)\,,\\
k_{L1}=n+2
\end{array}\right.
\end{equation}

\item[(2)] \underline{$k_{L2}=-k_{L2}$}

This implies $k_{L2}=n/2$. The order of the group has to be $n\in 2\,\mathbb{Z}$, leading to the chain
\begin{equation}
\begin{array}{rl}
Z_{2(n+2)}:&\left\{ A_{13}\oplus\mathcal{P}_{123}A_{15}\mathcal{P}_{12}\oplus
\mathcal{P}_{23}A_{12}\right.\\
&\quad\left.\oplus A_{12}\mathcal{P}_{23}\oplus
\mathcal{P}_{23}A_{12}\mathcal{P}_{13}\oplus\mathcal{P}_{13}
A_{12}\mathcal{P}_{23}\right\}\mathcal{P}
\end{array}\,.
\end{equation} 
The associated charge vector is
\begin{equation}
\left\{
\begin{array}{l}
\left(1,\omega_{2(n+2)}^{n+2},\omega_{2(n+2)}^{k_{L1}+n+2},\omega_{2(n+2)}^{-k_{L1}+n+2},\omega_{2(n+2)}^{k_{L1}},\omega_{2(n+2)}^{-k_{L1}}\right)\,,\\
k_{L2}=n+2
\end{array}\right.
\end{equation}
It can be obtained from case $(1)$ when $\mathcal{P}_{L}=\mathcal{P}=\mathcal{P}_{23}$.

\item[(3)] \underline{$k_{L2}-k_{L1}=k_{L1}-k_{L2}$}

This implies $2k_{L2}=2k_{L1}$, or $k_{L2}=k_{L1}+n/2$. The order of the group has to be $n\in 2\,\mathbb{Z}$, leading to the chain
\begin{equation}
\begin{array}{rl}
Z_{2(n+2)}:&\left\{ A_{13}\oplus\mathcal{P}_{23}A_{15}\oplus
A_{12}\mathcal{P}_{13}\oplus\mathcal{P}_{13}A_{12}\right.\\
&\quad\left.\oplus
\mathcal{P}_{23}A_{12}\mathcal{P}_{13}\oplus\mathcal{P}_{13}
A_{12}\mathcal{P}_{23}\right\}\mathcal{P}
\end{array}\,.
\end{equation} 
The associated charge vector is
\begin{equation}
\left\{
\begin{array}{l}
\left(1,\omega_{2(n+2)}^{n+2},\omega_{2(n+2)}^{k_{L1}+n+2},\omega_{2(n+2)}^{-k_{L1}+n+2},\omega_{2(n+2)}^{k_{L1}},\omega_{2(n+2)}^{-k_{L1}}\right)\,,\\
k_{L2}=k_{L1}+n+2
\end{array}\right.
\end{equation}
It can be obtained from case $(1)$ when $\mathcal{P}_{L}=\mathcal{P}=\mathcal{P}_{13}$.

\end{itemize}

$\bullet$\, Only matrices from $\mathcal{P}^\prime A_{12}\mathcal{P}$

In this case we have six distinct $k$'s, which leads to the chain
\begin{equation}
\begin{array}{rl}
Z_{n\geq 7}:&\left\{ A_{13}\oplus A_{12}\mathcal{P}_{23}
\oplus\mathcal{P}_{23}A_{12}\oplus A_{12}\mathcal{P}_{13}
\right.\\
&\left.\oplus\mathcal{P}_{13}A_{12}\oplus\mathcal{P}_{13}A_{12}\mathcal{P}_{23}
\oplus\mathcal{P}_{23}A_{12}\mathcal{P}_{13}\right\}\mathcal{P}\,.
\end{array}
\end{equation}
The associated charge vector is
\begin{equation}
\begin{array}{l}
\left(1,\omega_{n}^{k_{L2}+k_{L1}},\omega_n^{-(k_{L2}+k_{L1})},\omega_n^{2k_{L1}+k_{L2}},\right.\\
\quad\left.\omega_n^{-(2k_{L1}+k_{L2})},\omega_n^{-k_{L1}},\omega_n^{k_{L1}}\right)
\end{array}
\end{equation}

\subsubsection{No textures from $\mathcal{P}^\prime A_{13}\mathcal{P}$ and at least one from $\mathcal{P}^\prime A_{15}\mathcal{P}$}

The symmetry implementation
\begin{equation}
\left\{
\begin{array}{l}
\mathcal{S}_R=\mathcal{P}^T\text{diag}\left(\omega_n^{k_1},\, \omega_n^{k_{L1}},\, \omega_n^{k_{L2}}\right)\mathcal{P}\\
\mathcal{S}_L=\text{diag}\left(1,\, \omega_n^{k_{L1}},\, \omega_n^{k_{L2}}\right)
\end{array}\right.\,,
\end{equation}
which leads to the phase transformation matrix $\Theta_{A_{15}\mathcal{P}}$ is
\begin{equation}
\frac{2\pi}{n}
%
\mathcal{P}\,,
\end{equation} 
We have the following possibilities:

$\bullet$\, Four matrices from $\mathcal{P}^\prime A_{15}\mathcal{P}$

In this case we may have the following:
\begin{itemize}

\item[(1)] \underline{$k_{L1}=k_1-k_{L1},\,k_{L2}-k_{L1}=k_{L1}-k_{L2},$} \underline{$ k_{L2}=k_{1}-k_{L2}$}

This implies $k_1=2k_{L1}=2k_{L2}$ and $k_{L1}=k_{L2}+n/2$. The order of the group has to be $n\in 2\, \mathbb{Z}$, leading to the chain
\begin{equation}
\begin{array}{rl}
Z_{2(n+2)}:&\left\{A_{15}\oplus \mathcal{P}_{23}A_{15}\oplus\mathcal{P}_{321}A_{15}\mathcal{P}_{13}\oplus\right.\\
&\quad\left.\mathcal{P}_{123}A_{15}\mathcal{P}_{12}\oplus\mathcal{P}_{13}A_{12}\mathcal{P}_{13}\oplus A_0\right\}\mathcal{P}
\end{array}
\end{equation}
The symmetry automatically gives the $A_0$ in the chain. The associated charge vector is
\begin{equation}
\left\{
\begin{array}{l}
\left(1,\omega_{2(n+2)}^{n+2},\omega_{2(n+2)}^{-k_{L2}+n+2},\omega_{2(n+2)}^{-k_{L2}},\omega_{2(n+2)}^{-2k_{L2}},\omega_{2(n+2)}^{\tilde{k}}\right)\,,\\
k_{L1}=k_{L2}+n+2\,.
\end{array}\right.
\end{equation}

\item[(2)] \underline{$k_{L1}=k_1-k_{L2},\, k_{L2}=k_1-k_{L1},$} \underline{$k_{L2}-k_{L1}=k_{L1}-k_{L2}$}

This implies $k_1=k_{L1}+k_{L2}$ and $k_{L1}=k_{L2}+n/2$ (or $2k_{L1}=2k_{L2}$). The order of the group has to be $n\in 2\, \mathbb{Z}$, leading to the chain
\begin{equation}
\begin{array}{rl}
Z_{2(n+2)}:&\left\{A_{15}\oplus \mathcal{P}_{23}A_{15}\oplus\mathcal{P}_{12}A_{15}\mathcal{P}_{13}\oplus\right.\\
&\quad\left.\mathcal{P}_{13}A_{15}\mathcal{P}_{12}\oplus\mathcal{P}_{13}A_{12}\mathcal{P}_{13}\oplus A_0\right\}\mathcal{P}
\end{array}
\end{equation}
The symmetry automatically gives the $A_0$ in the chain. The associated charge vector is
\begin{equation}
\left\{
\begin{array}{l}
\left(1,\omega_{2(n+2)}^{n+2},\omega_{2(n+2)}^{-k_{L2}+n+2},\omega_{2(n+2)}^{-k_{L2}},\omega_{2(n+2)}^{-2k_{L2}+n+2},\omega_{2(n+2)}^{\tilde{k}}\right)\,,\\
k_{L1}=k_{L2}+n+2\,.
\end{array}\right.
\end{equation} 
It is included in case $(1)$ when $\mathcal{P}=\mathcal{P}_{23}$.

\item[(3)] \underline{$k_1=k_{L1}-k_{L2},\, k_{L1}=k_1-k_{L2},\,k_{L2}=k_{1}-k_{L1}$}

This implies $k_1=k_{L1}+k_{L2}$ and $k_{L2}=n/2$. The order of the group has to be $n\in 2\, \mathbb{Z}$, leading to the chain
\begin{equation}
\begin{array}{rl}
Z_{2(n+2)}:&\mathcal{P}^\prime\left\{A_{15}\oplus \mathcal{P}_{13}A_{15}\mathcal{P}_{12}\oplus\mathcal{P}_{12}A_{15}\mathcal{P}_{123}\oplus\right.\\
&\quad\left.\mathcal{P}_{12}A_{15}\mathcal{P}_{13}\oplus\mathcal{P}_{23}A_{12}\oplus A_0\right\}\mathcal{P}
\end{array}
\end{equation}
The symmetry automatically gives the $A_0$ in the chain. The associated charge vector is
\begin{equation}
\left\{
\begin{array}{l}
\left(1,\omega_{2(n+2)}^{n+2},\omega_{2(n+2)}^{-k_{L1}+n+2},\omega_{2(n+2)}^{-k_{L1}},\omega_{2(n+2)}^{k_{L1}+n+2},\omega_{2(n+2)}^{\tilde{k}}\right)\,,\\
k_{L2}=n+2\,.
\end{array}\right.
\end{equation} 
It is included in case $(1)$ when $\mathcal{P}_L=\mathcal{P}_{12}$ and $\mathcal{P}=\mathcal{P}_{123}$.

\item[(4)] \underline{$k_1=k_{L2}-k_{L1},\, k_1-k_{L1}=k_{L1}-k_{L2}$} \underline{$k_{L2}=k_{1}-k_{L2}$}

This implies $k_1=-2k_{L1},\, k_{L2}=-k_{L1}$, and $k_{L1}=\pm n/5$. The order of the group has to be $n\in 5\, \mathbb{Z}$, leading to the chain
\begin{equation}
\begin{array}{rl}
Z_{5n}:&\left\{ A_{15}\oplus A_{15}\mathcal{P}_{123}
\oplus \mathcal{P}_{321}A_{15}\mathcal{P}_{12}
\right.\\
&\quad\left.\oplus \mathcal{P}_{123}A_{15}\mathcal{P}_{12}\oplus
\mathcal{P}_{13}A_{12}\mathcal{P}_{23}\right\}\mathcal{P}\,.
\end{array}
\end{equation}
The associated charge vector is
\begin{equation}
\left\{
\begin{array}{l}
\left(1,\omega_{5n}^{2n},\omega_{5n}^{3n},\omega_{5n}^{4n},\omega_{5n}^n\right)\,,\\
\pm k_{L1}=\mp k_{L2}=n
\end{array}\right.
\end{equation}

\item[(5)] \underline{$k_1=k_{L1}-k_{L2},\, k_{L1}=k_{1}-k_{L1}$}, \underline{$k_{L2}-k_{L1}=k_1-k_{L2}$}

This implies $k_1=2k_{L1},\, k_{L2}=-k_{L1}$, and $k_{L1}=\pm n/5$. The order of the group has to be $n\in 5\, \mathbb{Z}$, leading to the chain
\begin{equation}
\begin{array}{rl}
Z_{5n}:&\left\{ A_{15}\oplus \mathcal{P}_{12}A_{15}\mathcal{P}_{123}
\oplus A_{15}\mathcal{P}_{321}
\right.\\
&\quad\left.\oplus
\mathcal{P}_{13}A_{12}\oplus \mathcal{P}_{321}A_{15}\mathcal{P}_{13}\right\}\mathcal{P}\,.
\end{array}
\end{equation}
The associated charge vector is
\begin{equation}
\left\{
\begin{array}{l}
\left(1,\omega_{5n}^{2n},\omega_{5n}^{3n},\omega_{5n}^{4n},\omega_{5n}^n\right)\,,\\
\pm k_{L1}=\mp k_{L2}=n
\end{array}\right.
\end{equation}
It is included in case $(4)$ when $\mathcal{P}_L=\mathcal{P}_{12}$ and $\mathcal{P}=\mathcal{P}_{123}$.

\item[(6)] \underline{$k_1=k_{L1}-k_{L2},\, k_{L1}=k_{L2}-k_{L1},\,k_{L2}=k_1-k_{L2}$}

This implies $k_1=4k_{L1},\, k_{L2}=2k_{L1}$, and $k_{L1}=n/5$. The order of the group has to be $n\in 5\, \mathbb{Z}$, leading to the chain
\begin{equation}
\begin{array}{rl}
Z_{5n}:&\left\{ A_{15}\oplus
\mathcal{P}_{23}A_{12}\mathcal{P}_{13}\oplus \mathcal{P}_{123}A_{15}\mathcal{P}_{12}
\right.\\
&\quad\left.\oplus \mathcal{P}_{321}A_{15}\oplus
 \mathcal{P}_{12}A_{15}\mathcal{P}_{123}\right\}\mathcal{P}\,.
\end{array}
\end{equation}
The associated charge vector is
\begin{equation}
\left\{
\begin{array}{l}
\left(1,\omega_{5n}^{2n},\omega_{5n}^{3n},\omega_{5n}^{4n},\omega_{5n}^n\right)\,,\\
k_{L2}=2k_{L1}=2n
\end{array}\right.
\end{equation}
It is included in case $(4)$ when $\mathcal{P}_L=\mathcal{P}_{12}$ and $\mathcal{P}=\mathcal{P}_{123}$.

\item[(7)] \underline{$k_1=k_{L2}-k_{L1},\, k_{L1}=k_{1}-k_{L1},\,k_{L2}=k_{L1}-k_{L2}$}

This implies $k_1=2k_{L1},\, k_{L1}=2k_{L2}$, and $k_{L2}=n/5$. The order of the group has to be $n\in 5\, \mathbb{Z}$, leading to the chain
\begin{equation}
\begin{array}{rl}
Z_{5n}:&\left\{ A_{15}\oplus
A_{12}\mathcal{P}_{13}\oplus \mathcal{P}_{321}A_{15}\mathcal{P}_{13}
\right.\\
&\quad\left.\oplus \mathcal{P}_{123}A_{15}\oplus
 \mathcal{P}_{321}A_{15}\mathcal{P}_{12}\right\}\mathcal{P}\,.
\end{array}
\end{equation}
The associated charge vector is
\begin{equation}
\left\{
\begin{array}{l}
\left(1,\omega_{5n}^{2n},\omega_{5n}^{3n},\omega_{5n}^{4n},\omega_{5n}^n\right)\,,\\
k_{L1}=2k_{L2}=2n
\end{array}\right.
\end{equation}
It is included in case $(4)$ when $\mathcal{P}_L=\mathcal{P}_{13}$ and $\mathcal{P}=\mathcal{P}_{123}$.
\end{itemize}

$\bullet$\, Three matrices from $\mathcal{P}^\prime A_{15}\mathcal{P}$.

In this case we have the following possibilities:
\begin{itemize}
\item[(1)] \underline{$k_{L1}=k_{L2}-k_{L1},\, k_{L2}=k_{L1}-k_{L2}$}

This implies $k_{L1}=-k_{L2}$ and $k_{L1}=\pm n/3$. The order of the group has to be $n\in 3\, \mathbb{Z}$ with $n\geq 6$, leading to the chain 
\begin{equation}
\begin{array}{rl}
Z_{3(n+1)}:&\left\{ A_{15}\oplus\mathcal{P}_{321}A_{15}\oplus\mathcal{P}_{123}A_{15}
\oplus A_{12}\mathcal{P}_{13}\right.\\
&\quad\left.\oplus\mathcal{P}_{23}A_{12}\mathcal{P}_{13}
\oplus\mathcal{P}_{13}A_{12}\mathcal{P}_{13}\right\}\mathcal{P}\,.
\end{array}
\end{equation}
The associated charge vector is
\begin{equation}
\left\{
\begin{array}{l}
\left(1,\omega_{3(n+1)}^{2(n+1)},\omega_{3(n+1)}^{n+1},\omega_{3(n+1)}^{-k_1+2(n+1)},\omega_{3(n+1)}^{-k_1+n+1},\omega_{3(n+1)}^{-k_1}\right)\,.\\
\pm k_{L1}=\mp k_{L2}=(n+1)
\end{array}
\right.
\end{equation}

\item[(2)] \underline{$k_{L1}=k_1-k_{L2},\, k_{L2}=k_1-k_{L1}$}

This implies $k_1=k_{L1}+k_{L2}$ and $k_{L1}\neq\{2k_{L2},\, n/2\}$, $k_{L2}\neq\{-k_{L1},\,2k_{L1},\, n/2\}$ and $2k_{L1}\neq 2k_{L2}$. For $n=6$ we have four possible charges for $k_{L1}$ and $k_{L2}$; however, there are five constraints between the $k$'s. The order of the group has to be $n\geq 7$, leading to the chain
\begin{equation}
\begin{array}{rl}
Z_{n\geq 7}:& \left\{A_{15}\oplus\mathcal{P}_{12}A_{15}\mathcal{P}_{13}\oplus\mathcal{P}_{13}A_{15}\mathcal{P}_{12}
\oplus \mathcal{P}_{13}A_{12}\mathcal{P}_{13}\right.\\
&\quad\left.\oplus A_{12}\mathcal{P}_{23}
\oplus\mathcal{P}_{23}A_{12}\oplus A_0\right\}\mathcal{P}\,.
\end{array}
\end{equation}
The associated charge vector is
\begin{equation}
\begin{array}{l}
\left(1,\omega_n^{k_{L2}-k_1},\omega_n^{-k_{L2}},\omega_n^{-k_1},\omega_n^{k_{L2}-k_{L1}},\omega_n^{k_{L1}-k_{L2}},\omega_n^{k}\right)\,.
\end{array}
\end{equation}

\item[(3)] \underline{$k_1-k_{L1}=k_{L1}-k_{L2},\, k_{L2}-k_{L1}=k_1-k_{L2}$}

This implies $3k_{L1}=3k_{L2}$ (or $k_{L2}=k_{L1}\pm n/3$). The order of the group has to be $n\in\, 3\mathbb{Z}$ with $n\geq 6$, leading to the chain
\begin{equation}
\begin{array}{rl}
Z_{3(n+1)}:&\left\{A_{15}\oplus A_{15}\mathcal{P}_{123}
\oplus A_{15}\mathcal{P}_{321}\oplus\mathcal{P}_{13}A_{12}\right.\\
&\quad\left.
\oplus\mathcal{P}_{13}A_{12}\mathcal{P}_{23}
\oplus\mathcal{P}_{13}A_{12}\mathcal{P}_{13}\right\}\mathcal{P}\,.
\end{array}
\end{equation}
The associated charge vector is 
\begin{equation}
\left\{
\begin{array}{l}
\left(1,\omega_{3(n+1)}^{\mp 2(n+1)},\omega_{3(n+1)}^{\mp(n+1)},\omega_{3(n+1)}^{-k_{L1}\mp(n+1)},\omega_{3(n+1)}^{-k_{L1}},\right.\\
\quad\left.\omega_{3(n+1)}^{-k_{L1}\mp2(n+1)}\right)\,,\\
k_{L2}=k_{L1}\pm (n+1)
\end{array}\right.
\end{equation}

\item[(4)] \underline{$k_{L1}=k_1-k_{L1},\,k_{L2}=k_{L1}-k_{L2}$}

This implies $k_1=4k_{L2}$ and $k_{L1}=2k_{L2}$. The order of the group has to be $n\geq 6$, leading to the chain
\begin{equation}
\begin{array}{rl}
Z_{n\geq 6}:&\left\{ A_{15}\oplus \mathcal{P}_{123}A_{15}\oplus
\mathcal{P}_{13}A_{15}\mathcal{P}_{123}\right.\\
&\quad\left.\oplus
\mathcal{P}_{13}A_{12}\mathcal{P}_{13}\oplus\mathcal{P}_{23}A_{12}
\oplus A_{12}\mathcal{P}_{13}\right\}\mathcal{P}\,.
\end{array}
\end{equation}
The associated charge vector is
\begin{equation}
\left\{
\begin{array}{l}
\left(1,\omega_{n}^{-k_{L2}},\omega_{n}^{-2k_{L2}},\omega_{n}^{-4k_{L2}},\omega_{n}^{k_{L2}},\omega_{n}^{-3k_{L2}}\right)\,,\\
k_{L1}=2k_{L2}
\end{array}\right.
\end{equation}

\item[(5)] \underline{$k_{L1}=k_{L2}-k_{L1},\,k_{L2}=k_{1}-k_{L2}$}

This implies $k_1=4k_{L1}$ and $k_{L2}=2k_{L1}$. The order of the group has to be $n\geq 6$, leading to the chain
\begin{equation}
\begin{array}{rl}
Z_{n\geq 6}:&\left\{ A_{15}\oplus \mathcal{P}_{321}A_{15}\oplus
\mathcal{P}_{123}A_{15}\mathcal{P}_{12}\right.\\
&\quad\left.\oplus
\mathcal{P}_{13}A_{12}\mathcal{P}_{13}\oplus A_{12}\mathcal{P}_{23}
\oplus \mathcal{P}_{23}A_{12}\mathcal{P}_{13}\right\}\mathcal{P}\,.
\end{array}
\end{equation}
The associated charge vector is
\begin{equation}
\left\{
\begin{array}{l}
\left(1,\omega_{n}^{-k_{L1}},\omega_{n}^{-2k_{L1}},\omega_{n}^{-4k_{L1}},\omega_{n}^{k_{L1}},\omega_{n}^{-3k_{L1}}\right)\,,\\
k_{L1}=2k_{L2}
\end{array}\right.
\end{equation}
It is obtained from case $(4)$ with $\mathcal{P}_L=\mathcal{P}=\mathcal{P}_{23}$.
\item[(6)] \underline{$k_1=k_{L2}-k_{L1},\,k_{L2}=k_{L1}-k_{L2}$}

This implies $k_1=-k_{L2}$ and $k_{L1}=2k_{L2}$. The order of the group has to be $n\geq 6$, leading to the chain
\begin{equation}
\begin{array}{rl}
Z_{n\geq 6}:&\left\{ A_{15}\oplus \mathcal{P}_{321}A_{15}\mathcal{P}_{12}\oplus
\mathcal{P}_{123}A_{15}\right.\\
&\quad\left.\oplus
A_{12}\mathcal{P}_{13}\oplus \mathcal{P}_{13}A_{12}\mathcal{P}_{23}
\oplus \mathcal{P}_{23}A_{12}\mathcal{P}_{13}\right\}\mathcal{P}\,.
\end{array}
\end{equation}
The associated charge vector is
\begin{equation}
\left\{
\begin{array}{l}
\left(1,\omega_{n}^{k_{L2}},\omega_{n}^{2k_{L2}},\omega_{n}^{-k_{L2}},\omega_{n}^{-2k_{L2}},\omega_{n}^{3k_{L2}}\right)\,,\\
k_{L1}=2k_{L2}
\end{array}\right.
\end{equation}
It is obtained from case $(4)$ with $\mathcal{P}_L=\mathcal{P}_{12}$ and $\mathcal{P}=\mathcal{P}_{23}$.

\item[(7)] \underline{$k_1=k_{L1}-k_{L2},\,k_{L1}=k_{L2}-k_{L1}$}

This implies $k_1=-k_{L1}$ and $k_{L2}=2k_{L1}$. The order of the group has to be $n\geq 6$, leading to the chain
\begin{equation}
\begin{array}{rl}
Z_{n\geq 6}:&\left\{ A_{15}\oplus \mathcal{P}_{12}A_{15}\mathcal{P}_{123}\oplus
\mathcal{P}_{23}A_{12}\mathcal{P}_{13}\right.\\
&\quad\left.\oplus
\mathcal{P}_{321}A_{15}\oplus \mathcal{P}_{13}A_{12}
\oplus A_{12}\mathcal{P}_{13}\right\}\mathcal{P}\,.
\end{array}
\end{equation}
The associated charge vector is
\begin{equation}
\left\{
\begin{array}{l}
\left(1,\omega_{n}^{k_{L1}},\omega_{n}^{2k_{L1}},\omega_{n}^{-k_{L1}},\omega_{n}^{-2k_{L1}},\omega_{n}^{3k_{L1}}\right)\,,\\
k_{L2}=2k_{L1}
\end{array}\right.
\end{equation}
It is obtained from case $(4)$ with $\mathcal{P}_L=\mathcal{P}_{321}$.

\item[(8)] \underline{$k_1=k_{L2}-k_{L1},\,k_{L2}=k_{1}-k_{L2}$}

This implies $k_1=-2k_{L1}$ and $k_{L2}=-k_{L1}$. The order of the group has to be $n\geq 6$, leading to the chain
\begin{equation}
\begin{array}{rl}
Z_{n\geq 6}:&\left\{ A_{15}\oplus \mathcal{P}_{123}A_{15}\mathcal{P}_{12}\oplus
\mathcal{P}_{321}A_{15}\mathcal{P}_{12}\right.\\
&\quad\left.\oplus
\mathcal{P}_{13}A_{12}\mathcal{P}_{23}\oplus A_{12}\mathcal{P}_{23}
\oplus \mathcal{P}_{23}A_{12}\mathcal{P}_{13}\right\}\mathcal{P}\,.
\end{array}
\end{equation}
The associated charge vector is
\begin{equation}
\left\{
\begin{array}{l}
\left(1,\omega_{n}^{k_{L1}},\omega_{n}^{2k_{L1}},\omega_{n}^{-k_{L1}},\omega_{n}^{-2k_{L1}},\omega_{n}^{3k_{L1}}\right)\,,\\
k_{L2}=-k_{L1}
\end{array}\right.
\end{equation}
It is obtained from case $(4)$ with $\mathcal{P}_L=\mathcal{P}_{13}$ and $\mathcal{P}=\mathcal{P}_{321}$.

\item[(9)] \underline{$k_1=k_{L1}-k_{L2},\,k_{L1}=k_{1}-k_{L1}$}

This implies $k_1=2k_{L1}$ and $k_{L2}=-k_{L1}$. The order of the group has to be $n\geq 6$, leading to the chain
\begin{equation}
\begin{array}{rl}
Z_{n\geq 6}:&\left\{ A_{15}\oplus \mathcal{P}_{321}A_{15}\mathcal{P}_{13}\oplus
\mathcal{P}_{12}A_{15}\mathcal{P}_{123}\right.\\
&\quad\left.\oplus
\mathcal{P}_{13}A_{12}\oplus \mathcal{P}_{23}A_{12}
\oplus A_{12}\mathcal{P}_{13}\right\}\mathcal{P}\,.
\end{array}
\end{equation}
The associated charge vector is
\begin{equation}
\left\{
\begin{array}{l}
\left(1,\omega_{n}^{-k_{L1}},\omega_{n}^{-2k_{L1}},\omega_{n}^{k_{L1}},\omega_{n}^{2k_{L1}},\omega_{n}^{-3k_{L1}}\right)\,,\\
k_{L2}=-k_{L1}
\end{array}\right.
\end{equation}
It is obtained from case $(4)$ with $\mathcal{P}_L=\mathcal{P}_{123}$ and $\mathcal{P}=\mathcal{P}_{13}$.

\item[(10)] \underline{$k_{L1}=k_{1}-k_{L1},\, k_{L2}-k_{L1}=k_{1}-k_{L2}$}

This implies $k_1=2k_{L1}$ and $2k_{L2}=3k_{L1}$. The order of the group has to be $n\geq 6$, leading to the chain
\begin{equation}
\begin{array}{rl}
Z_{n\geq 6}:&\left\{ A_{15}\oplus A_{15}\mathcal{P}_{321}\oplus
\mathcal{P}_{13}A_{15}\mathcal{P}_{123}\right.\\
&\quad\left.\oplus
A_{12}\mathcal{P}_{23}\oplus\mathcal{P}_{13}A_{12}
\oplus\mathcal{P}_{13}A_{12}\mathcal{P}_{13}\right\}\mathcal{P}
\end{array}
\end{equation}
The associated charge vector is
\begin{equation}
\left\{
\begin{array}{l}
\left(1,\omega_n^{k_{L1}-k_{L2}},\omega_n^{-k_{L1}},\omega_n^{k_{L2}-k_{L1}},\omega_n^{-k_{L2}},\omega_{n}^{-2k_{L1}}\right)\,,\\
3k_{L1}=2k_{L2}
\end{array}
\right.
\end{equation}

\item[(11)] \underline{$k_1-k_{L1}=k_{L1}-k_{L2},\, k_{L2}=k_1-k_{L2}$}

This implies $k_1=2k_{L2}$ and $3k_{L2}=2k_{L1}$. The order of the group has to be $n\geq 6$, leading to the chain
\begin{equation}
\begin{array}{rl}
Z_{n\geq 6}:&\left\{ A_{15}\oplus A_{15}\mathcal{P}_{123}\oplus
\mathcal{P}_{123}A_{15}\mathcal{P}_{12}\right.\\
&\quad\left.\oplus
\mathcal{P}_{23}A_{12}\oplus\mathcal{P}_{13}A_{12}\mathcal{P}_{23}
\oplus\mathcal{P}_{13}A_{12}\mathcal{P}_{13}\right\}\mathcal{P}
\end{array}
\end{equation}
The associated charge vector is
\begin{equation}
\left\{
\begin{array}{l}
\left(1,\omega_n^{k_{L2}-k_{L1}},\omega_n^{-k_{L2}},\omega_n^{k_{L1}-k_{L2}},\omega_n^{-k_{L1}},\omega_{n}^{-2k_{L2}}\right)\,,\\
2k_{L1}=3k_{L2}
\end{array}
\right.
\end{equation}
It is obtained from $(10)$ with $\mathcal{P}_L=\mathcal{P}=\mathcal{P}_{23}$.

\item[(12)] \underline{$k_1=k_{L1}-k_{L2},\, k_{L2}-k_{L1}=k_1-k_{L2}$}

This implies $k_1=k_{L1}-k_{L2}$ and $2k_{L1}=3k_{L2}$. The order of the group has to be $n\geq 6$, leading to the chain
\begin{equation}
\begin{array}{rl}
Z_{n\geq 6}:&\left\{ A_{15}\oplus \mathcal{P}_{12}A_{15}\mathcal{P}_{123}\oplus
\mathcal{P}_{13}A_{12}\oplus
A_{15}\mathcal{P}_{321}\right.\\
&\quad\left.\oplus\mathcal{P}_{13}A_{12}\mathcal{P}_{23}
\oplus\mathcal{P}_{23}A_{12}\mathcal{P}_{13}\right\}\mathcal{P}
\end{array}
\end{equation}
The associated charge vector is
\begin{equation}
\left\{
\begin{array}{l}
\left(1,\omega_n^{k_{L2}-k_{L1}},\omega_n^{-k_{L2}},\omega_n^{k_{L1}-k_{L2}},\omega_n^{-k_{L1}},\omega_{n}^{k_{L2}}\right)\,,\\
2k_{L1}=3k_{L2}
\end{array}
\right.
\end{equation}
It is obtained from $(10)$ with $\mathcal{P}_L=\mathcal{P}_{23}$ and $\mathcal{P}=\mathcal{P}_{12}$.

\item[(13)] \underline{$k_1=k_{L2}-k_{L1},\,k_1-k_{L1}=k_{L1}-k_{L2}$}

This implies $k_1=k_{L2}-k_{L1}$ and $3k_{L1}=2k_{L2}$. The order of the group has to be $n\geq 6$, leading to the chain
\begin{equation}
\begin{array}{rl}
Z_{n\geq 6}:&\left\{ A_{15}\oplus \mathcal{P}_{321}A_{15}\mathcal{P}_{12}\oplus
\mathcal{P}_{13}A_{12}\mathcal{P}_{23}\right.\\
&\quad\left.\oplus
A_{15}\mathcal{P}_{123}\oplus\mathcal{P}_{13}A_{12}
\oplus A_{12}\mathcal{P}_{13}\right\}\mathcal{P}
\end{array}
\end{equation}
The associated charge vector is
\begin{equation}
\left\{
\begin{array}{l}
\left(1,\omega_n^{k_{L1}-k_{L2}},\omega_n^{-k_{L1}},\omega_n^{k_{L2}-k_{L1}},\omega_n^{-k_{L2}},\omega_{n}^{k_{L1}}\right)\,,\\
3k_{L1}=2k_{L2}
\end{array}
\right.
\end{equation}
It is obtained from $(10)$ with $\mathcal{P}=\mathcal{P}_{123}$.

\item[(14)] \underline{$k_1=k_{L1}-k_{L2},\, k_{L2}=k_1-k_{L2}$}

This implies $k_1=2k_{L2}$ and $k_{L1}=3k_{L2}$. The order of the group has to be $n\geq 6$, leading to the chain
\begin{equation}
\begin{array}{rl}
Z_{n\geq 6}:&\left\{ A_{15}\oplus \mathcal{P}_{12}A_{15}\mathcal{P}_{123}\oplus
\mathcal{P}_{123}A_{15}\mathcal{P}_{12}\right.\\
&\quad\left.\oplus
\mathcal{P}_{23}A_{12}\mathcal{P}_{13}\oplus\mathcal{P}_{23}A_{12}
\oplus\mathcal{P}_{13}A_{12}\mathcal{P}_{23}\right\}\mathcal{P}
\end{array}
\end{equation}
The associated charge vector is
\begin{equation}
\left\{
\begin{array}{l}
\left(1,\omega_n^{2k_{L2}},\omega_n^{-k_{L2}},\omega_n^{k_{L2}},\omega_n^{2k_{L2}},\omega_{n}^{-3k_{L2}}\right)\,,\\
k_{L1}=3k_{L2}
\end{array}
\right.
\end{equation}
It is obtained from $(10)$ with $\mathcal{P}_L=\mathcal{P}_{123}$ $\mathcal{P}=\mathcal{P}_{13}$.

\item[(15)] \underline{$k_1=k_{L2}-k_{L1},\,k_{L2}=k_1-k_{L1}$}

This implies $k_1=2k_{L1}$ and $k_{L2}=3k_{L1}$. The order of the group has to be $n\geq 6$, leading to the chain
\begin{equation}
\begin{array}{rl}
Z_{n\geq 6}:&\left\{ A_{15}\oplus \mathcal{P}_{321}A_{15}\mathcal{P}_{12}\oplus
\mathcal{P}_{321}A_{15}\mathcal{P}_{13}\right.\\
&\quad\left.\oplus
A_{12}\mathcal{P}_{13}\oplus A_{12}\mathcal{P}_{23}
\oplus \mathcal{P}_{13}A_{12}\right\}\mathcal{P}
\end{array}
\end{equation}
The associated charge vector is
\begin{equation}
\left\{
\begin{array}{l}
\left(1,\omega_n^{-2k_{L1}},\omega_n^{-k_{L1}},\omega_n^{k_{L1}},\omega_n^{2k_{L1}},\omega_{n}^{-3k_{L1}}\right)\,,\\
k_{L2}=3k_{L1}
\end{array}
\right.
\end{equation}
It is obtained from $(10)$ with $\mathcal{P}_L=\mathcal{P}_{13}$ $\mathcal{P}=\mathcal{P}_{321}$.

\end{itemize}

$\bullet$\, Two matrices from $\mathcal{P}^\prime A_{12}\mathcal{P}$.

In this case we have the following possibilities
\begin{itemize}
\item[(1)] \underline{$k_{L1}=k_1-k_{L1}$}

This implies $k_1=2k_{L1},\, k_{L1}\neq n/2$, $k_{L2}\neq\left\{-k_{L1},\,2k_{L1},\,3k_{L1},\,k_{L1}+n/2,n/2\right\}$, and $3k_{L1}\neq 2k_{L2} $. The order of the group has to be $n\geq 7$; however, for $n=7$ we get five possibilities for $k_{L2}$ given a $k_{L1}$. But these are the different forbidden $k_{L2}$; therefore, $n=7$ is not possible, leading to $n\geq 8$. The chain is given by 
\begin{equation}
\begin{array}{rl}
Z_{n\geq 8}:&\left\{A_{15}\oplus\mathcal{P}_{321}A_{15}\mathcal{P}_{13}\oplus
\mathcal{P}_{13}A_{12}\mathcal{P}_{13}\oplus\mathcal{P}_{23}A_{12}\right.\\
&\quad\left.\oplus\mathcal{P}_{13}A_{12}\oplus A_{12}\mathcal{P}_{23}
\oplus A_{12}\mathcal{P}_{13}\oplus A_0\right\}\mathcal{P}
\end{array}
\end{equation}
The associated charge vector is
\begin{equation}
\begin{array}{l}
\left(1,\omega_n^{-k_{L1}},\omega_n^{-2k_{L1}},\omega_n^{k_{L1}-k_{L2}},\omega_n^{-k_{L2}},\right.\\\left.\quad\omega_n^{k_{L2}-k_{L1}},\omega_{n}^{k_{L2}-2k_{L1}},\omega_n^{\tilde{k}}\right)
\end{array}
\end{equation}

\item[(2)] \underline{$k_{L2}=k_{1}-k_{L2}=k$}

This implies $k_1=2k_{L2},\, k_{L2}\neq n/2$, $k_{L1}\neq\left\{-k_{L2},\,2k_{L2},\,3k_{L2},\,k_{L2}+n/2,n/2\right\}$, and $3k_{L2}\neq 2k_{L1} $.The order of the group is $n\geq 8$. The chain is given by 
\begin{equation}
\begin{array}{rl}
Z_{n\geq 8}:&\left\{A_{15}\oplus\mathcal{P}_{123}A_{15}\mathcal{P}_{12}\oplus
\mathcal{P}_{13}A_{12}\mathcal{P}_{13}\oplus A_{12}\mathcal{P}_{23}\right.\\
&\quad\left.\oplus\mathcal{P}_{13}A_{12}\mathcal{P}_{23}\oplus \mathcal{P}_{23}A_{12}
\oplus \mathcal{P}_{23}A_{12}\mathcal{P}_{13}\oplus A_0\right\}\mathcal{P}
\end{array}
\end{equation}
The associated charge vector is
\begin{equation}
\begin{array}{l}
\left(1,\omega_n^{-k_{L2}},\omega_n^{-2k_{L2}},\omega_n^{k_{L2}-k_{L1}},\omega_n^{-k_{L1}},\right.\\\left.\quad\omega_n^{k_{L1}-k_{L2}},\omega_{n}^{k_{L1}-2k_{L2}},\omega_n^{\tilde{k}}\right)
\end{array}
\end{equation}
It is obtained from $(1)$ with $\mathcal{P}_L=\mathcal{P}_R=\mathcal{P}_{23}$.

\item[(3)] \underline{$k_1=k_{L2}-k_{L1}$}

This implies $k_{L1}\neq \{2k_{L2},\,n/2\}$, $k_{L2}\neq\left\{-k_{L1},\,2k_{L1},\,3k_{L1},\,k_{L1}+n/2,n/2\right\}$, and $3k_{L1}\neq 2k_{L2} $.The order of the group is $n\geq 8$. The
\begin{equation}
\begin{array}{rl}
Z_{n\geq 8}:&\left\{A_{15}\oplus\mathcal{P}_{321}A_{15}\mathcal{P}_{12}\oplus
\mathcal{P}_{13}A_{12}\mathcal{P}_{23}\oplus \mathcal{P}_{13}A_{12}\right.\\
&\quad\left.\oplus A_{12}\mathcal{P}_{13}\oplus  A_{12}\mathcal{P}_{23}
\oplus \mathcal{P}_{23}A_{12}\mathcal{P}_{13}\oplus A_0\right\}\mathcal{P}
\end{array}
\end{equation}
The associated charge vector is
\begin{equation}
\begin{array}{l}
\left(1,\omega_n^{k_{L1}-k_{L2}},\omega_n^{-k_{L1}},\omega_n^{-k_{L2}},\omega_n^{k_{L1}},\right.\\\left.\quad\omega_n^{k_{L2}-k_{L1}},\omega_{n}^{2k_{L1}-k_{L2}},\omega_n^{\tilde{k}}\right)
\end{array}
\end{equation}
It is obtained from $(1)$ with $\mathcal{P}_L=\mathcal{P}_{13}$ and $\mathcal{P}_R=\mathcal{P}_{123}$.

\item[(4)] \underline{$k_1=k_{L1}-k_{L2}$}

This implies $k_{L2}\neq \{2k_{L1},\,n/2\}$, $k_{L1}\neq\left\{-k_{L2},\,2k_{L2},\,3k_{L2},\,k_{L2}+n/2,n/2\right\}$, and $3k_{L2}\neq 2k_{L1} $.The order of the group is $n\geq 8$. The
\begin{equation}
\begin{array}{rl}
Z_{n\geq 8}:&\left\{A_{15}\oplus\mathcal{P}_{12}A_{15}\mathcal{P}_{123}\oplus
\mathcal{P}_{13}A_{12}\oplus \mathcal{P}_{13}A_{12}\mathcal{P}_{23}\right.\\
&\quad\left.\oplus \mathcal{P}_{23}A_{12}\mathcal{P}_{13}\oplus  \mathcal{P}_{23}A_{12}
\oplus A_{12}\mathcal{P}_{13}\oplus A_0\right\}\mathcal{P}
\end{array}
\end{equation}
The associated charge vector is
\begin{equation}
\begin{array}{l}
\left(1,\omega_n^{k_{L2}-k_{L1}},\omega_n^{-k_{L2}},\omega_n^{-k_{L1}},\omega_n^{k_{L2}},\right.\\\left.\quad\omega_n^{k_{L1}-k_{L2}},\omega_{n}^{2k_{L2}-k_{L1}},\omega_n^{\tilde{k}}\right)
\end{array}
\end{equation}
It is obtained from $(1)$ with $\mathcal{P}_L=\mathcal{P}_{321}$ and $\mathcal{P}_R=\mathcal{P}_{13}$.
\item[(5)] \underline{$k_{L1}=k_{L2}-k_{L1}$}

This implies $k_{L2}=2k_{L1},\, k_{L1}\neq n/2,\, n/3$, and $k_1\neq\left\{k_{L1},\,-k_{L1},\,2k_{L1},\,3k_{L1},\,4k_{L1}\right\}$. The order of the group has to be $n\geq 7$, leading to the chain
\begin{equation}
\begin{array}{rl}
Z_{n\geq 7}:&\left\{A_{15}\oplus\mathcal{P}_{321}A_{15}\oplus
\mathcal{P}_{13}A_{12}\mathcal{P}_{13}\oplus\mathcal{P}_{13}A_{12}\right.\\
&\left.\oplus A_{12}\mathcal{P}_{23}\oplus A_{12}\mathcal{P}_{13}
\oplus\mathcal{P}_{23}A_{12}\mathcal{P}_{13}\right\}\mathcal{P}
\end{array}
\end{equation}
The associated charge vector is 
\begin{equation}
\left\{
\begin{array}{l}
\left(1,\omega_n^{-k_{L1}},\omega_{n}^{-k_1},\omega_n^{-2k_{L1}},\omega_n^{k_{L1}},\omega_n^{2k_{L1}-k_1},\omega_n^{k_{L1}-k_1}\right)\\
k_{L2}=2k_{L1}
\end{array}\right.
\end{equation}

\item[(6)] \underline{$k_{L2}=k_{L1}-k_{L2}$}

This implies $k_{L1}=2k_{L2},\, k_{L2}\neq n/2,\, n/3$, and $k_1\neq\left\{k_{L2},\,-k_{L2},\,2k_{L2},\,3k_{L2},\,4k_{L2}\right\}$. The order of the group has to be $n\geq 7$, leading to the chain
\begin{equation}
\begin{array}{rl}
Z_{n\geq 7}:&\left\{A_{15}\oplus\mathcal{P}_{123}A_{15}\oplus
\mathcal{P}_{13}A_{12}\mathcal{P}_{13}\oplus\mathcal{P}_{13}A_{12}\mathcal{P}_{23}\right.\\
&\left.\oplus \mathcal{P}_{23}A_{12}\oplus \mathcal{P}_{23}A_{12}\mathcal{P}_{13}
\oplus A_{12}\mathcal{P}_{13}\right\}\mathcal{P}
\end{array}
\end{equation}
The associated charge vector is 
\begin{equation}
\left\{
\begin{array}{l}
\left(1,\omega_n^{-k_{L2}},\omega_{n}^{-k_1},\omega_n^{-2k_{L2}},\omega_n^{k_{L2}},\omega_n^{2k_{L2}-k_1},\omega_n^{k_{L2}-k_1}\right)\\
k_{L1}=2k_{L2}
\end{array}\right.
\end{equation}
It is obtained from $(5)$ with $\mathcal{P}_L=\mathcal{P}_R=\mathcal{P}_{23}$.

\item[(7)] \underline{$k_1-k_{L1}=k_{L1}-k_{L2}$}

This implies $k_1=2k_{L1}-k_{L2}$, $k_{L1}\neq\left\{k_{L2},2k_{L2}, k_{L2}+n/2\right\}$, $3k_{L1}\neq 2k_{L2}$, $3k_{L2}\neq 2k_{L1}$. The order of the group has to be $n\geq 7$, leading to the chain
\begin{equation}
\begin{array}{rl}
Z_{n\geq 7}:&\left\{ A_{15}\oplus A_{15}\mathcal{P}_{123}\oplus
A_{12}\mathcal{P}_{13}\right.\\
&\quad \oplus\mathcal{P}_{23}A_{12}
\oplus\mathcal{P}_{13}A_{12}\oplus\mathcal{P}_{13}A_{12}\mathcal{P}_{23}\\
&\quad\left.\oplus\mathcal{P}_{13}A_{12}\mathcal{P}_{13}\right\}\mathcal{P}
\end{array}
\end{equation}
The associated charge vector is
\begin{equation}
\begin{array}{l}
\left(1,\omega_n^{k_{L2}-k_{L1}},\omega_n^{2(k_{L2}-k_{L1})},\omega_n^{k_{L1}-k_{L2}},\omega_n^{-k_{L2}},\right.\\
\quad\left.\omega_n^{-k_{L1}},\omega_n^{k_{L2}-2k_{L1}}\right)
\end{array}
\end{equation}

\item[(8)] \underline{$k_{L2}-k_{L1}=k_1-k_{L2}$}

This implies $k_1=2k_{L2}-k_{L1}$, $k_{L2}\neq\left\{k_{L1},2k_{L1}, k_{L1}+n/2\right\}$, $3k_{L1}\neq 2k_{L2}$, $3k_{L2}\neq 2k_{L1}$. The order of the group has to be $n\geq 7$, leading to the chain
\begin{equation}
\begin{array}{rl}
Z_{n\geq 7}:&\left\{ A_{15}\oplus A_{15}\mathcal{P}_{321}\oplus
\mathcal{P}_{23}A_{12}\mathcal{P}_{13}\right.\\
&\quad \oplus A_{12}\mathcal{P}_{23}
\oplus\mathcal{P}_{13}A_{12}\mathcal{P}_{23}\oplus\mathcal{P}_{13}A_{12}\\
&\quad\left.\oplus\mathcal{P}_{13}A_{12}\mathcal{P}_{13}\right\}\mathcal{P}
\end{array}
\end{equation}
The associated charge vector is
\begin{equation}
\begin{array}{l}
\left(1,\omega_n^{k_{L1}-k_{L2}},\omega_n^{2(k_{L1}-k_{L2})},\omega_n^{k_{L2}-k_{L1}},\omega_n^{-k_{L1}},\right.\\
\quad\left.\omega_n^{-k_{L2}},\omega_n^{k_{L1}-2k_{L2}}\right)
\end{array}
\end{equation}
It is obtained from $(7)$ with $\mathcal{P}_R=\mathcal{P}_{123}$
\item[(9)] \underline{$k_{L2}-k_{L1}=k_{L1}-k_{L2}$}

This implies $2k_{L1}=2k_{L2}$ (or $k_{L2}=k_{L1}+n/2$). The order of the group has to be $n\in 2\,\mathbb{Z}$ with $n>7 $, leading to the chain
\begin{equation}
\begin{array}{rl}
Z_{2(3+n)}:&\left\{A_{15}\oplus\mathcal{P}_{23}A_{15}\oplus
\mathcal{P}_{13}A_{12}\mathcal{P}_{13}\right.\\
&\quad \oplus\mathcal{P}_{13}A_{12}\oplus\mathcal{P}_{13}A_{12}\mathcal{P}_{23}\oplus A_{12}\mathcal{P}_{13}\\
&\quad\left.\oplus\mathcal{P}_{23}A_{12}\mathcal{P}_{13}\oplus A_0\right\}\mathcal{P}
\end{array}
\end{equation}
The associated charge vector is 
\begin{equation}
\left\{
\begin{array}{l}
\left(1,\omega_{2(3+n)}^{3+n},\omega_{2(3+n)}^{k_1},\omega_{2(3+n)}^{-k_{L1}+3+n},\omega_{2(3+n)}^{-k_{L1}},\omega_{2(3+n)}^{k_{L1}-k_{1}+3+n}\right.\\
\quad\left.\omega_{2(3+n)}^{k_{L1}-k_1},\omega_{2(3+n)}^{k}\right)\\
k_{L2}=k_{L1}+3+n
\end{array}\right.
\end{equation}

\end{itemize}

$\bullet$\, One matrix form $\mathcal{P}^\prime A_{15}\mathcal{P}$

In this case we only have one possibility and the constraints $2k_{L1}\neq 2k_{L2}$, $k_{L1}\neq 2k_{L2}$, $k_{L2}\neq 2k_{L1}$ and $k_{1}\neq\{\pm(k_{L1}-k_{L2}),\,2k_{L1,2},k_{L1}+k_{L2},2k_{L1}-k_{L2},2k_{L2}-k_{L1}\}$, leading to the chain
\begin{equation}
\begin{array}{rl}
Z_{n\geq 8}:&\left\{ A_{15}\oplus A_{12}\mathcal{P}_{23}
\oplus A_{12}\mathcal{P}_{13}\oplus\mathcal{P}_{23}A_{12}\right.\\
&\oplus\mathcal{P}_{23}A_{12}\mathcal{P}_{13}\oplus\mathcal{P}_{13}A_{12}
\oplus\mathcal{P}_{13}A_{12}\mathcal{P}_{23}\\
&\quad\left.\oplus\mathcal{P}_{13}A_{12}\mathcal{P}_{13}\right\}\mathcal{P}
\end{array}
\end{equation}
The associated charge vector is
\begin{equation}
\left(1,\omega_n^{k_{L2}-k_{L1}},\omega_n^{k_{L2}-k_1},\omega_n^{k_{L1}-k_{L2}},\omega_n^{k_{L1}-k_1}\right.\\
\quad\left.\omega_n^{-k_{L2}},\omega_n^{-k_{L1}},\omega_n^{k_{1}}\right)
\end{equation}

\subsubsection{Only matrices from $\mathcal{P}^\prime A_{12}\mathcal{P}$}

The symmetry generators are given by
\begin{equation}
\left\{
\begin{array}{l}
\mathcal{S}_R=\mathcal{P}^T\text{diag}\left(\omega_n^{k_1},\, \omega_n^{k_2},\,\omega_n^{k_{L2}}\right)\mathcal{P}\,,\\ \mathcal{S}_L=\text{diag}\left(1,\, \omega_n^{k_{L1}},\, \omega_n^{k_{L2}}\right)
\end{array}\right.
\end{equation}
leading to the phase transformation matrix $\Theta_{A_{12}\mathcal{P}}$
\begin{equation}
\frac{2\pi}{n}
%
\mathcal{P}
\end{equation}
There is only one chain possible when $k's$ are all different. We get
\begin{equation}
\begin{array}{rl}
Z_{n\geq 9}:&\mathcal{P}^\prime\left\{ A_{12}\oplus\mathcal{P}_{23}A_{12}\mathcal{P}_{23}\oplus A_{12}\mathcal{P}_{23}\oplus A_{12}\mathcal{P}_{13}\right.\\
&\quad\oplus\mathcal{P}_{23}A_{12}\oplus\mathcal{P}_{23}A_{12}\mathcal{P}_{13}
\oplus\mathcal{P}_{13}A_{12}\\
&\quad\left.\oplus\mathcal{P}_{13}A_{12}\mathcal{P}_{23}
\oplus\mathcal{P}_{13}A_{12}\mathcal{P}_{13}\right\}\mathcal{P}
\end{array}
\end{equation}
The associated charge vector is
\begin{equation}
\begin{array}{l}
\left(1,\omega_n^{k_{L1}-k_2},\omega_n^{k_{L2}-k_2},\omega_n^{k_{L2}-k_1},\omega_n^{k_{L1}-k_{L2}},\omega_n^{k_{L1}-k_1},\right.\\
\quad\left.\omega_n^{-k_{L2}},\omega_n^{-k_{2}},\omega_n^{-k_1}\right)
\end{array}
\end{equation}

\section{Models belonging to classes $\mathbf{(2,i)}$}\label{21N3}
Explicit representation of the available models for classes $\mathbf{2,i}$.

\begin{subequations}
\begin{align}
&
%
\label{C523C623}
\end{align}
\end{subequations}

\section{Tables}\label{Tables}
In this appendix Tables X, XI, XII, XIII, and XIV, useful in different stages of the paper, are presented.

\begin{table}[H]
\begin{tabular}{rl}
\hline\hline
$C_1^{\mathbf{(1,1)}}:$&None\\
$C_1^{\mathbf{(1,2)}}:$&$Z_{n\geq 2}$\\
$C_3^{\mathbf{(1,3)}}:$&$Z_{n\geq3}$\\
\hline\hline
\end{tabular}
\begin{tabular}{rl}
\hline\hline
$C_1^{\mathbf{(2,1)}}:$&$Z_{n\geq2}$\\
$C_1^{\mathbf{(2,2)}}:$&$Z_{2n}$\\
$C_{2,3}^{\mathbf{(2,2)}}:$&$Z_{n\geq 3}$\\
$C_4^{\mathbf{(2,2)}}:$&$Z_{n\geq 4}$\\
$C_1^{\mathbf{(2,3)}}:$&$Z_{3n}$\\
$C_2^{\mathbf{(2,3)}}:$&$Z_{2(n+1)}$\\
$C_{3,4}^{\mathbf{(2,3)}}:$&$Z_{n\geq4}$\\
$C_5^{\mathbf{(2,3)}}:$&$Z_{n\geq5}$\\
$C_6^{\mathbf{(2,3)}}:$&$Z_{n\geq6}$\\
\hline\hline
\end{tabular}
\begin{tabular}{rl}
\hline\hline
$C_1^{\mathbf{(3,1)}}:$&$Z_{n\geq3}$\\
$C_1^{\mathbf{(3,2)}}:$&$Z_{3n}$\\
$C_2^{\mathbf{(3,2)}}:$&$Z_{2(n+1)}$\\
$C_{3,4}^{\mathbf{(3,2)}}:$&$Z_{n\geq4}$\\
$C_5^{\mathbf{(3,2)}}:$&$Z_{n\geq5}$\\
$C_6^{\mathbf{(3,2)}}:$&$Z_{n\geq6}$\\
$C_1^{\mathbf{(3,3)}}:$&$Z_{3n}$\\
$C_2^{\mathbf{(3,3)}}:$&$Z_{4n}$\\
$C_3^{\mathbf{(3,3)}}:$&$Z_{n\geq5}$\\
$C_4^{\mathbf{(3,3)}}:$&$Z_{2(n+2)}$\\
$C_5^{\mathbf{(3,3)}}:$&$Z_{n\geq7}$\\
$C_6^{\mathbf{(3,3)}}:$&$Z_{2(n+2)}$\\
$C_7^{\mathbf{(3,3)}}:$&$Z_{5n}$\\
$C_8^{\mathbf{(3,3)}}:$&$Z_{3(n+1)}$\\
$C_9^{\mathbf{(3,3)}}:$&$Z_{n\geq7}$\\
$C_{10}^{\mathbf{(3,3)}}:$&$Z_{3(n+1)}$\\
$C_{11}^{\mathbf{(3,3)}}:$&$Z_{n\geq6}$\\
$C_{12}^{\mathbf{(3,3)}}:$&$Z_{n\geq6}$\\
$C_{13}^{\mathbf{(3,3)}}:$&$Z_{n\geq8}$\\
$C_{14}^{\mathbf{(3,3)}}:$&$Z_{n\geq7}$\\
$C_{15}^{\mathbf{(3,3)}}:$&$Z_{n\geq7}$\\
$C_{16}^{\mathbf{(3,3)}}:$&$Z_{2(n+3)}$\\
$C_{17}^{\mathbf{(3,3)}}:$&$Z_{n\geq9}$\\
$C_{18}^{\mathbf{(3,3)}}:$&$Z_{n\geq9}$\\
\hline\hline
\end{tabular}
\caption{\label{SymTab}Symmetry groups that implement a given chain.}
\end{table}

\begin{table}[H]
\begin{center}
\begin{tabular}{r||c}
&Classes\\
\hline\hline
$A_9\oplus \mathcal{P}_{23}A_9\oplus \mathcal{P}_{13}A_9$&$C^{\mathbf{(3,1)}}_1$\\
$A_{14}\oplus \mathcal{P}_{123}A_{14}\oplus A_{0}$&$ ^0C^{\mathbf{(3,2)}}_{1,3}$\\
$A_{14}\oplus  \mathcal{P}_{321}A_{14}\oplus A_0$&$ ^0C^{\mathbf{(3,2)}}_1$\\
$ \mathcal{P}_{123}A_{14}\oplus \mathcal{P}_{321}A_{14}\oplus A_0$&$ ^0C^{\mathbf{(3,2)}}_1$\\
$A_{14}\oplus \mathcal{P}_{123}A_{14}\oplus \mathcal{P}_{321}A_{14}$&$ C^{\mathbf{(3,2)}}_1$\\
$A_{14}\oplus\mathcal{P}_{23}A_{14}
\oplus\mathcal{P}_{13}A_{11}$&$C_2^\mathbf{(3,2)}$\\
$A_{14}\oplus\mathcal{P}_{23}A_{14}
\oplus\mathcal{P}_{13}A_{12}$&$C_2^\mathbf{(3,2)}$\\
\hline\hline
\end{tabular}
\caption{\label{tab3iI} Combinations with $N=3$ which lead to 3 mixing angles for classes $\mathbf{(3,i)}$. (Part I)}
\end{center}
\end{table}

\begin{table}[H]
\begin{center}
\begin{tabular}{r||c}
&Classes\\
\hline\hline
$\mathcal{P}_{23}A_{15}
\oplus\mathcal{P}_{321}A_{15}\mathcal{P}_{13}\oplus\mathcal{P}_{13}A_{12}\mathcal{P}_{13}$&$C_{6}^\mathbf{(3,3)}$\\
$\mathcal{P}_{23}A_{15}
\oplus\mathcal{P}_{123}
A_{15}\mathcal{P}_{12}\oplus\mathcal{P}_{13}A_{12}\mathcal{P}_{13}$&$C_{6}^\mathbf{(3,3)}$\\
$A_{15}\oplus A_{15}\mathcal{P}_{123}\oplus
\mathcal{P}_{321}A_{15}\mathcal{P}_{12}$&$C_{7}^\mathbf{(3,3)}$\\
$A_{15}\oplus A_{15}\mathcal{P}_{123}\oplus
\mathcal{P}_{123}A_{15}\mathcal{P}_{12}$&$C_{7}^\mathbf{(3,3)}$\\
$A_{15}\oplus A_{15}\mathcal{P}_{123}\oplus
\mathcal{P}_{13}
A_{12}\mathcal{P}_{23}$&$C_{7,10,15}^\mathbf{(3,3)}$\\
$A_{15}\oplus
\mathcal{P}_{321}A_{15}\mathcal{P}_{12}\oplus
\mathcal{P}_{123}A_{15}\mathcal{P}_{12}$&$C_{7}^\mathbf{(3,3)}$\\
$A_{15}\oplus
\mathcal{P}_{321}A_{15}\mathcal{P}_{12}\oplus\mathcal{P}_{13}
A_{12}\mathcal{P}_{23}$&$C_{7}^\mathbf{(3,3)}$\\
\hline\hline
\end{tabular}
\caption{\label{tab3iII} Combinations with $N=3$ which lead to 3 mixing angles for classes $\mathbf{(3,i)}$. (Part II)}
\end{center}
\end{table}

\begin{widetext}
\begin{table}[H]
\begin{center}
\begin{tabular}{r|ll}
\hline\hline
$C_1^{\mathbf{(3,1)}}$&$(1,\omega_n^{k_{L1}-k_{L2}},\omega_n^{-k_{L2}})$&\\
&\\
$C_1^{\mathbf{(3,2)}}$&$(1,\omega_{3n}^{2n},\omega_{3n}^n)$&$k_{L1}=2k_{L2}=2n$\\
$\mathcal{P}_{12}\times\quad$&$(1,\omega_{3n}^{2n},\omega_{3n}^n)$&$k_{L2}=2k_{L1}=2n$\\
$C_2^{\mathbf{(3,2)}}$&$(1,\omega_{2(n+1)}^{n+1},\omega_{2(n+1)}^{-k_{L1}},\omega_{2(n+1)}^{-k_{L1}+n+1})$&$k_{L2}=k_{L1}+n+1$\\
$\mathcal{P}_{12}\times\quad$&$(1,\omega_{2(n+1)}^{n+1},\omega_{2(n+1)}^{k_{L1}},\omega_{2(n+1)}^{k_{L1}+n+1})$&$ k_{L2}=n+1$\\
$\mathcal{P}_{13}\times\quad$&$(1,\omega_{2(n+1)}^{n+1},\omega_{2(n+1)}^{k_{L2}+n+1},\omega_{2(n+1)}^{k_{L2}})$&$ k_{L1}=n+1$\\
$C_3^{\mathbf{(3,2)}}$&$(1,\omega_n^{-k_{L2}},\omega_n^{-2k_{L2}},\omega_n^{k_{L2}})$&$k_{L1}=2k_{L2}$\\
\end{tabular}
\end{center}
\end{table}

\begin{table}[H]
\begin{center}
\begin{tabular}{r|ll}
$C_{13}^{\mathbf{(3,3)}}$&$(1,\omega_n^{-k_{L1}},\omega_n^{-2k_{L1}},\omega_n^{k_{L1}-k_{L2}},\omega_n^{-k_{L2}},\omega_n^{k_{L2}-k_{L1}},\omega_n^{k_{L2}-2k_{L1}},\omega_n^{k^\prime})$&\\
$\mathcal{P}_{12}\times\quad$&$(1,\omega_n^{k_{L1}},\omega_n^{2k_{L1}},\omega_n^{-k_{L2}},\omega_n^{k_{L1}-k_{L2}},\omega_n^{k_{L2}},\omega_n^{k_{L2}+k_{L1}},\omega_n^{k^\prime})$&\\
$\mathcal{P}_{13}\times\quad$&$(1,\omega_n^{k_{L2}-k_{L1}},\omega_n^{2(k_{L2}-k_{L1})},\omega_n^{k_{L1}},\omega_n^{k_{L2}},\omega_n^{-k_{L1}},\omega_n^{k_{L2}-2k_{L1}},\omega_n^{k^\prime})$&\\
$\mathcal{P}_{23}\times\quad$&$(1,\omega_n^{-k_{L2}},\omega_n^{-2k_{L2}},\omega_n^{k_{L2}-k_{L1}},\omega_n^{-k_{L1}},\omega_n^{k_{L1}-k_{L2}},\omega_n^{k_{L1}-2k_{L2}},\omega_n^{k^\prime})$&\\
$\mathcal{P}_{123}\times\quad$&$(1,\omega_n^{k_{L1}-k_{L2}},\omega_n^{2(k_{L1}-k_{L2})},\omega_n^{k_{L2}},\omega_n^{k_{L1}},\omega_n^{-k_{L2}},\omega_n^{k_{L1}-2k_{L2}},\omega_n^{k^\prime})$&\\
$\mathcal{P}_{321}\times\quad$&$(1,\omega_n^{k_{L2}},\omega_n^{2k_{L2}},\omega_n^{-k_{L1}},\omega_n^{k_{L2}-k_{L1}},\omega_n^{k_{L1}},\omega_n^{k_{L1}+k_{L2}},\omega_n^{k^\prime})$&\\
$C_{14}^{\mathbf{(3,3)}}$&$(1,\omega_n^{-k_{L1}},\omega_n^{-k_1},\omega_n^{-2k_{L1}},\omega_n^{k_{L1}},\omega_n^{2k_{L1}-k_1},\omega_n^{k_{L1}-k_1})$&$k_{L2}=2k_{L1}$\\
$\mathcal{P}_{12}\times\quad$&$(1,\omega_n^{k_{L1}},\omega_n^{-k_1},\omega_n^{2k_{L1}},\omega_n^{-k_{L1}},\omega_n^{-2k_{L1}-k_1},\omega_n^{-k_{L1}-k_1})$&$k_{L2}=-k_{L1}$\\
\hline\hline
\end{tabular}
\end{center}
\caption{\label{CV3ifull} Vector charge for the classes $\mathbf{(3,i)}$.}
\end{table}

\begin{table}[H]
\begin{center}
\begin{tabular}{c|c|c|c}
\hline\hline
Classes&$N=1$&$N=2$&$N=3$\\
\hline
$C_1^{\mathbf{(1,1)}}$&All&All&All\\
$C_1^{\mathbf{(1,2)}}$&&All&All\\
$C_1^{\mathbf{(1,3)}}$&&&All\\
&&&\\
$C_1^{\mathbf{(2,1)}}$&&All&All\\
$C_1^{\mathbf{(2,2)}}$&$(1)$&All&All\\
$C_2^{\mathbf{(2,2)}}$&$(1)$&$(1,2);(1,3)$&All\\
\hline\hline
\end{tabular}
\caption{\label{detup3} Combinations, up to $N=3$, of textures that lead to a non-singular mass matrix.}
\end{center}
\end{table}

\end{widetext}

\end{document}